%% file: Kerr_etal_AlfvenWaves_2016_ToSubmit.tex
\documentclass[iop, apj]{emulateapj}
\usepackage[colorlinks=true, citecolor = blue, urlcolor = blue, linkcolor = blue]{hyperref}
\usepackage{natbib, twoopt}
\usepackage{amsmath}
\usepackage{float}
\usepackage[caption=false]{subfig}
\usepackage{graphicx}
\usepackage{txfonts}

\usepackage{color} \input{rgb}

\shorttitle{Alfv\'en Wave Flare Heating}
\shortauthors{Kerr et al}

\begin{document}


\title{Simulations of the  Mg~\textsc{ii} k  and Ca \textsc{ii} 8542 lines from an Alfv\'en Wave-heated flare chromosphere}
\author{Graham~S.~Kerr$^{1}$, Lyndsay. Fletcher$^{1}$, Alexander J.B. Russell$^{2}$, and Joel C. Allred$^{3}$}
\affil{$^{1}$ SUPA, School of Physics and Astronomy, University of Glasgow, G12 8QQ, Scotland, U.K.\\ \url{g.kerr.2@research.gla.ac.uk} ; \url{Lyndsay.Fletcher@glasgow.ac.uk} \\
$^{2}$ School of Science and Engineering, University of Dundee, Nethergate, Dundee, DD1 4HN, Scotland, U.K. \\ \url{arussell@maths.dundee.ac.uk}\\
$^{3}$ NASA Goddard Space Flight Center, Heliophysics Science Division, Code 671, 8800 Greenbelt Rd., Greenbelt, MD 20771, U.S.A \\ \url{joel.c.allred@nasa.gov}}

\keywords{Sun: atmosphere, Sun: chromosphere, Sun: flares, Sun: waves, Sun: UV radiation, methods: numerical}

\begin{abstract}

We use radiation hydrodynamic simulations to examine two models of solar flare chromospheric heating:  Alfv\'en wave dissipation and electron beam collisional losses.  Both mechanisms are capable of strong chromospheric heating, and we show that the distinctive atmospheric evolution in the mid-to-upper chromosphere results in  Mg~\textsc{ii} k-line emission that should be observably different between wave-heated and beam-heated simulations. We also present Ca \textsc{ii} 8542\AA\ profiles which are formed slightly deeper in the chromosphere. The Mg~\textsc{ii} k-line profiles from our wave-heated simulation are quite different from those from a beam-heated model and are more consistent with IRIS observations. The predicted differences between the Ca \textsc{ii} 8542\AA\  in the two models are small. We conclude that careful observational and theoretical study of lines formed in the mid-to-upper chromosphere holds genuine promise for distinguishing between competing models for chromospheric heating in flares. 
\end{abstract}


\section{Introduction}\label{sec:intro}

In a solar flare, up to $10^{32-33}$~ergs is released when the coronal magnetic field reconfigures. It is transported along the coronal magnetic field to be deposited in the dense chromosphere. This results in plasma heating and broadband enhancement to the solar radiative output. The sudden temperature increase also causes strong upflows (chromospheric  `evaporation')  and down flows (`condensations') observed via Doppler shifts in spectral lines. Hard X-ray (HXR) footpoint sources indicating the presence of high-energy electrons  are closely associated with enhancements to optical radiation, the locations of which are a subset of the UV and EUV ribbons that delineate the endpoints of just-reconnected field. For an overview of flare observations, see e.g. \cite{2011SSRv..159...19F} and \cite{2015SoPh..290.3399M}.

How energy is transported from the corona, where it is stored, to the chromosphere where it is dissipated is still a matter of debate. The transport must be fast (sub-second in the flare impulsive phase), ducted along the coronal magnetic field and must involve the acceleration of electrons out of the thermal background. It is possible that different flares have different combinations of transport by particle beams, conduction or MHD waves, and the challenge is to use observations and modelling to distinguish between these. Coronal observations will tend to be ambiguous, as thermal conduction, strong flows, and optically-thin radiation from the hot, ionised plasma smear out and superpose, and hence obscure signatures of the energy transport. The chromosphere, which produces the majority of flare radiation, has its own complications: it is optically thick in some lines, partially ionised, and highly structured both vertically and horizontally in temperature and density. However, chromospheric line profiles in principle allow us to probe the conditions at different atmospheric depths, meaning there is the prospect of making progress in distinguishing different energy transport models on the basis of how they affect the chromosphere. It is the purpose of this paper to model the profiles of some important diagnostic spectral lines generated in a chromosphere heated as a result of energy transport from the corona by electron beams and by Alfv\'en waves.

In the electron beam (EB) model \citep{1971SoPh...18..489B,1972SoPh...24..414H}, electrons, accelerated to mildly relativistic speeds in or close to the coronal energy release site, stream along the field. They deposit their energy  via Coulomb collisions, primarily in the chromosphere which presents a collisional thick target. This results in heating and radiation, including non-thermal hard X-ray (HXR) bremsstrahlung.  Using the collisional thick-target model (CTTM) the energy and number spectrum of the non-thermal coronal beam can be inferred from the HXR emission (e.g. \citealt{2003ApJ...595L..97H}) as has frequently been done using data from the RHESSI satellite \citep{2002SoPh..210....3L}. The resulting electron spectra are frequently used as input to numerical codes for simulating the response of the chromosphere to flare energy input. 

Flare energy transport by Alfv\'en waves (AWs) was first proposed by \cite{1982SoPh...80...99E} to explain TMR heating and \cite{2008ApJ...675.1645F} restarted the discussion on Alfv\'enic perturbations as a potential flare energy transport mechanism. Since a solar flare is, fundamentally, a large scale restructuring of the solar magnetic field, it is very reasonable to expect that AWs are produced. \cite{2008ApJ...675.1645F} argued that a 5-10\% Alfv\'enic perturbation to a 500G coronal field could supply adequate power to the flare chromosphere. The dissipation of the wave, by various damping mechanisms, in the highly structured chromosphere would lead to heating and potentially also to \emph{in situ} electron acceleration, but the details of electron acceleration have to be worked out.

The CTTM is attractive in that it neatly combines energy transport, HXR generation and footpoint heating, but it has difficulty explaining  the low depths of some flare optical emission \citep[e.g.][]{2012ApJ...753L..26M} and can also imply high coronal EB densities. A high coronal EB density, given the inferred ambient densities suggest that the beam-return current relative speed is high enough that instabilities may follow.

Flare line and continuum radiation show that the temperature minimum region (TMR) is heated by up to a few hundred kelvin during flares \citep{1975SoPh...42..395M, 1978SoPh...58..363M,1990ApJ...365..391M}. Additionally, white light flare (WLF) images also show that the optical continuum enhancement originates low in the solar atmosphere; WL and HXR emission were found to be co-spatial and located at $<$ $0.5$~Mm above the photosphere in an event studied by \cite{2012ApJ...753L..26M}. To reach and heat this depth requires significantly higher power in high energy electrons than is observed \citep[e.g.][]{1978SoPh...58..363M,1989SoPh..121..261N,1990ApJ...365..391M}. For this reason radiative backwarming \citep{1989SoPh..124..303M} is often used to explain WLF observations in the EB model.

The CTTM implies up to 10$^{36}~\rm{electrons}~s^{-1}$ \citep{1976SoPh...48..197H,2003ApJ...595L..97H} which, for a typical coronal density of $10^9~\rm{cm}^{-3}$ is equivalent to accelerating all the electrons in a coronal volume of $(10,000~\rm{km})^3$ each second. A continual resupply of electrons for acceleration is needed, which could happen via the return current that is established by the ambient plasma. As discussed by \cite{2008ApJ...675.1645F} and \cite{2011ApJ...739...96K}, observations imply coronal beam densities comparable to or greater than the ambient coronal density meaning a return current speed that is comparable to the beam speed. In this case, the beam and its return current are likely to be unstable in the corona, and to dissipate a large fraction of their energy in turbulence and heating. We note that in the scenario of a significantly greater ambient coronal density (on the order of $10^{11}$~\rm{cm}$^{-3}$) then the replenishment of the beam does not pose as much of a problem. Higher densities have been inferred in some flares \citep[e.g][]{2004ApJ...603L.117V, 2012ApJ...755...32G}.

Atmospheric heating via damping of AWs generated by photospheric drivers and propagating upwards into the chromosphere has been proposed as a quiet Sun chromospheric heating mechanism, e.g. by ion-neutral damping in the partially-ionised chromosphere. \citep{2001ApJ...558..859D, 2004A&A...422.1073K,2005A&A...442.1091L}.

Downwards-propagating AWs produced by flaring perturbations in the coronal magnetic field will also be damped in the chromospheric plasma, resulting in heating. \cite{1982SoPh...80...99E} suggested the  resistive (Joule) dissipation of the currents associated with AWs as a means of heating the TMR, and calculated that $\Delta T$ of 100-200~kelvin was possible for frequencies of order 1-10 Hz. More recently, simulations by \cite{2013ApJ...765...81R} of AWs traveling downwards through chromosphere with a realistic stratification of Alfv\'en speed and ionisation showed that for sufficiently high frequencies (around 1 Hz) a significant fraction of coronal AW energy can be transmitted to the deep chromosphere and damped by ion-neutral damping in the TMR  (and electron resistivity lower down).  \cite{2008ApJ...675.1645F} and \cite{2014SoPh..289..881M} propose that AWs are the dominant energy transport mechanism through the flaring corona, and discuss the viability of AWs in accelerating electrons to produce the observed HXR.  

In the simulations of \cite{2013ApJ...765...81R}, heating in the upper chromosphere was also observed, and \cite{2016ApJ...818L..20R} investigated this further. They updated the approach of \cite{1982SoPh...80...99E} for describing the energy deposition by waves to use ambipolar resistivity instead of classical resistivity and implemented this as an energy input in the HYDRAD code \citep{2003A&A...401..699B}. The result was that AW damping in the mid-upper chromosphere produced strong heating and evaporation, and looked very similar to what is found in electron-beam driven simulations. Heating was most efficient for perpendicular wave numbers $k_\perp > 1\times10^{-4}$~cm$^{-1}$ and frequencies around 10~Hz.

We expect that beam-driven and wave-driven models of energy input will have different heating profiles, and different time evolution, which will form the basis of discriminating between models. High spatial, spectral and temporal resolution data of chromospheric and transition region (TR) radiation in the near-UV (NUV) and far-UV (FUV) are now available from the Interface Region Imaging Spectrograph (IRIS; \citealt{2014SoPh..289.2733D}) spacecraft.  For example, \cite{2015A&A...582A..50K, 2015SoPh..290.3525L} and \cite{2015ApJ...807L..22G} discuss the complex chromospheric Mg~\textsc{ii} spectra observed during flares. The Daniel K. Inouye Solar Telescope (DKIST) will also provide high resolution chromospheric observations in the optical and infrared (IR). These resources provide the opportunity to probe models of energy transport in flares by comparing the synthetic spectra output by advanced models to observations. 

In this paper we use the radiation hydrodynamics code RADYN (\S~\ref{sec:model_descript}) to describe the chromospheric temperature, density and ionisation profiles resulting from numerical experiments that simulate chromospheric heating by high fluxes of flare-generated AWs, and compare with those from a standard EB simulation (\S~\ref{sec:hydro_results}). We use an approximated form of AW heating developed by \cite{1982SoPh...80...99E} and \cite{2016ApJ...818L..20R} as a heating term. Finally, we synthesise the observational signatures that result from these experiments (the Mg~\textsc{ii} h \& k lines, and the Ca~\textsc{ii} 8542\AA\ line) (\S~\ref{sec:rad_results}) and present discussion and conclusions (\S~\ref{sec:discussion}, \S~\ref{sec:conclusion})

\section{Implementing Alfv\'en Wave Heating in the RADYN code}\label{sec:model_descript}

\subsection{RADYN}
The radiation hydrodynamics code RADYN is a well established code for investigating chromospheric dynamics. Originally created by \cite{1995ApJ...440L..29C,1997ApJ...481..500C}, RADYN was used to study acoustic waves in the chromosphere, and was adapted by \cite{1999ApJ...521..906A} to simulate the chromospheric response to flare energy deposition by an EB. Later updates have included improved treatment of the EB, including a Fokker-Planck description, soft X-ray, extreme-UV (EUV) and UV radiation backwarming and photoionisation \citep{2005ApJ...630..573A,2015ApJ...809..104A}. We used the \cite{2015ApJ...809..104A} version of RADYN for results presented in this paper, with our own modifications described in \S~\ref{sec:alfwaves}. 

RADYN solves the plane-parallel, coupled, non-linear equations of hydrodynamics, radiation transfer, charge conservation and atomic level populations on a 1D grid that extends from the sub-photosphere to the corona, representing one leg of a symmetric flux tube. An adaptive grid \citep{1987JCoPh..69..175D} with 191 grid points resolves shocks and steep gradients.  Elements important for chromospheric energy balance are computed using non-Local Thermodynamic Equilibrium (nLTE) radiative transfer, with other atomic species included as background continuum opacity (assumed in LTE) using the Uppsala opacity package of \cite{Gustaffson_1973}. A radiative loss function approximates the optically thin coronal radiation transfer by summing all transitions in the CHIANTI database \citep{1997A&AS..125..149D,2013ApJ...763...86L}, apart from the transitions treated in detail (See \cite{2015ApJ...809..104A} for a line list). The atomic level populations are solved for a six-level-with-continuum hydrogen atom, a nine-level-with-continuum helium atom, a six-level-with-continuum Ca \textsc{ii} ion, and a four-level-with-continuum Mg~\textsc{ii} ion. Transitions (22 bound-bound transitions and 24 bound-free transitions) with up to 100 frequency points and 5 angular points are computed assuming complete redistribution (CRD), except the Lyman transitions which are truncated at 10 Doppler widths to approximate the effects of partial redistribution (PRD). We return to this issue when discussion the Mg~\textsc{ii} lines in \S~\ref{sec:mgii_k_profiles}. The product of the coronal/TR emission measure and emissivities (from CHIANTI) is integrated to find the XEUV spectrum, which is included as downward-directed incident radiation when solving the nLTE radiation transfer and ionisation equations.

Typically when simulating flares, the EB CTTM model has been used.  A non-thermal EB with a power law energy flux spectrum is introduced at the apex of the corona loop. It deposits energy as it travels, heating the plasma with heating rate $Q_{\rm BEAM}$ calculated from collisional losses.

\subsection{Alfv\'en Wave Dissipation and Heating}\label{sec:alfwaves}
We follow \cite{2016ApJ...818L..20R} to include an additional heating rate term due to AWs, $Q_{AW}$, in RADYN using the WKB approximation to obtain the period-averaged Poynting flux of the AW as a function of distance in a magnetic flux tube  \citep[as was described in][]{1982SoPh...80...99E}. Collisions between ions, electrons and neutrals damp the Poynting flux, and the dissipation of Poynting flux gives the heating term for the plasma. As noted by \cite{2016ApJ...818L..20R}, this approximation is accurate if the parallel wavelength is less than or comparable to the gradient length scale of the Alfv\'en speed, which also means that reflections are assumed negligible. As discussed in \S~\ref{sec:intro}, reflection at the corona-TR boundary has been shown to significant, so as in \cite{2016ApJ...818L..20R} we choose an initial Poynting flux giving a reasonable flux at the top of the chromosphere. 

In the following, $i$, $n$, $e$ and $t$ subscripts refer to ions, neutrals, electrons and total. The collisional frequencies are computed as follows. The formula for $\nu_{e,n}$ is quoted in \cite{1986A&A...159....1G} as

	\begin{equation}\label{eq:ven}
		\nu_{e,n} = 6.97\times10^{-14}~T^{0.1}~n_{H},
	\end{equation}
where $T$ is temperature, and $n_H$ is the number density of neutral hydrogen. \cite{2012ApJ...745...52H} gives an expression for the electron-ion collision time, $\tau_{e}$

	$$
		\tau_e = \frac{3}{4}\left(\frac{m_e}{2\pi}\right)^{1/2}\left(\frac{k_b~T}{n_e~\lambda~e^4}\right)^{3/2}
	$$

	\begin{equation}\label{eq:vei}
		\nu_{e,i} = 1/\tau_{e},
	\end{equation}
where $m_e$ is the electron mass, $k_b$ is Boltzmann's constant, $n_e$ is electron number density, $e$ is electron charge and and $\lambda$ is the Coulomb logarithm. Finally, $\nu_{n,i}$ is discussed in \cite{2013ApJ...765...81R} (noting their typo in the first $T$) and \cite{schunk_nagy}

	\begin{align}\label{eq:vni}
		\nu_{n,i} = 2.65\times10^{-16}~T^{1/2}~(1-&0.083~{\rm{log_{10}}}~T)^2~n_p   \notag \\ 
			& + 2.11\times10^{-15}~(n_e-n_p),
	\end{align}
where $n_p$ is the proton number density.

The parallel (to the field) and perpendicular resistivities of the plasma are defined as

	\begin{equation}\label{eta_par}
		\eta_{||} = \frac{m_e~(\nu_{e,i} + \nu_{e,n})}{n_e~e^2}
	\end{equation}
	
	\begin{align}\label{eq:eta_perp}
		\eta_{\perp} & =  \eta_{||} + \eta_C  \notag \\
		  & = \eta_{||} + \frac{B^2~\rho_n}{c^2~\nu_{n,i}~\rho_t^2~(1+\xi^2~\theta^2)}, 
	\end{align}
where $\rho$ is mass density, $c$ the speed of light, $\eta_C$ the Cowling resistivity, $\xi$ the hydrogen ionisation fraction and $\theta = \omega/\nu_{n,i}$ for $\omega = 2\pi f$.

\cite{1982SoPh...80...99E} derived an expression for the effective damping length of Alfv\'en waves, $L_D(z)$, with height along the modelled flux tube. We modify the \cite{1982SoPh...80...99E} $L_D(z)$ to use ambipolar resistivity as proposed by \cite{2016ApJ...818L..20R}

	\begin{align}\label{eq:effdamplen}
		L_D(z) & =  \left( \frac{1}{L_{\perp}(z)} + \frac{1}{L_{||}(z)}\right)^{-1} \notag \\
		            & = \left( \frac{\eta_{||}~k_x^2~c^2}{4~\pi~v_A} + \frac{\eta_{\perp}~w^2~c^2}{4~\pi~v_A^3}\right)^{-1} \notag \\
			    & = \frac{4~\pi~v_A^3}{c^2~(\eta_{||~}k_x^2~v_A^2 + \eta_{\perp}~w^2)}, 
	\end{align}
where $v_A(z)$ is the Alfv\'en speed. As in \cite{2016ApJ...818L..20R} we modify the Aflv\'en speed for the presence of neutrals,
	
	\begin{equation}\label{alfv_sp}
		v_{A}(z) = \frac{B}{\sqrt{4\pi \rho_{t}}}\left(\frac{1+\xi\theta^{2}}{1+\xi^{2}\theta^{2}}\right)^{1/2}.
	\end{equation}

 The period-averaged Poynting flux injected at the loop apex, $S_a$, is then damped to give the flux as a function of height

	\begin{equation}\label{eq:poyntflux}
		S(z) = S_a~{\rm{exp}}~\left(- \int_0^z \frac{dz^\prime}{L_D(z^\prime)}\right).
	\end{equation}

\noindent The volumetric heating rate is then the change in Poynting flux with distance along the flux tube:

	\begin{equation}\label{eq:q_alf}
		Q_{AW} = \frac{dS}{dz}.
	\end{equation}

A magnetic field strength $B(z)$ is imposed 
which depends on height as a function of pressure, $P(z)$ \citep{1983ApJ...264..648Z}, with $B_{0}$ defined as the photospheric value (note, this is only used in calculating the wave damping, and is not updated in the hydrodynamic or radiation transfer solutions):

	\begin{equation}\label{eq:b_field} 
		B(z) = B_{0}\left(\frac{P(z)}{P_{0}}\right)^\alpha.
	\end{equation}
We choose $\alpha = 0.139$ as in \cite{2013ApJ...765...81R} and \cite{2016ApJ...818L..20R}. $B(z)$ is constant with time, and at each timestep is interpolated to the updated grid. 

The perpendicular wavenumber varies as a function of $B(z)$ due to variations in the cross-section of the flux tube. In this work we use the relation:
	\begin{equation}\label{eq:wavenumber}
		k_x(z) = k_{x,a}\left(\frac{B(z)}{B_a}\right).
	\end{equation}
where subscript $a$ denotes values at the loop apex. This linear scaling is found when the magnetic field expands in one dimension, as in a magnetic arcade. An alternative two-dimensional expansion leads to a square root dependence however comparison of simulations for both geometries \citep[by][]{2016ApJ...818L..20R} suggests that the conclusions of AW heating studies depend only weakly on the choice of geometrical scaling. 

Flares are simulated with user inputs: $f$, $k_{x,a}$, $B_0$ \& $S_{a}$. Currently $S_a$ can be varied as a function of time, and future work will allow $f$, and $k_{x,a}$ to vary in time also.

\subsection{Simulations}\label{sec:simulations}
We model a 10~Mm flux tube extending from below the photosphere ($z=0$ defined where $\tau_{5000} = 1$) into the corona at temperature $T=1$~MK. The pre-flare atmosphere is the QS.SL.LT model atmosphere discussed in \cite{2015ApJ...809..104A}. This is the PF2 atmosphere used by \cite{1999ApJ...521..906A} and \cite{2005ApJ...630..573A}, modified to include the XEUV backwarming of \cite{2015ApJ...809..104A}. The PF2 atmosphere was originally created by adding a TR and corona to the \cite{1997ApJ...481..500C} radiative equilibrium atmospheric model. Non-radiative heating is applied to maintain the photospheric and coronal energy balance in grid cells with column mass greater than $7.6$~g~cm$^{-2}$ (photosphere) and less than $1\times10^{-6}$~g~cm$^{-2}$ (corona). We use a fixed boundary condition in the sub-photosphere and a reflecting boundary condition at the top of the loop, to mimic the effect of disturbances from the other half of the flux tube. 

Two simulations are compared here, one in which the flare energy transport mechanism is a non-thermal EB (F11) and one in which the energy transport is via AW dissipation (S11). Both have the same injected energy flux of $10^{11}$~ergs~cm$^{-2}$~s$^{-1}$, which is constant for $t =10$~s, which is representative of the `dwell time' of a flare footpoint at a particular chromospheric position. The additional EB simulation parameters are:  $\delta = 5$ and $E_c = 25$~keV. The additional AW simulation parameters are: $f = 10$~Hz, $k_{x,a} = 4\times10^{-4}$~cm$^{-1}$ and $B_0 = 1000$~G.

\section{Atmospheric Response}\label{sec:hydro_results}

\begin{figure*}
	\centering
	\hbox{
		\hspace{-0.25in}
		\subfloat{\includegraphics[width = 0.35\textwidth]{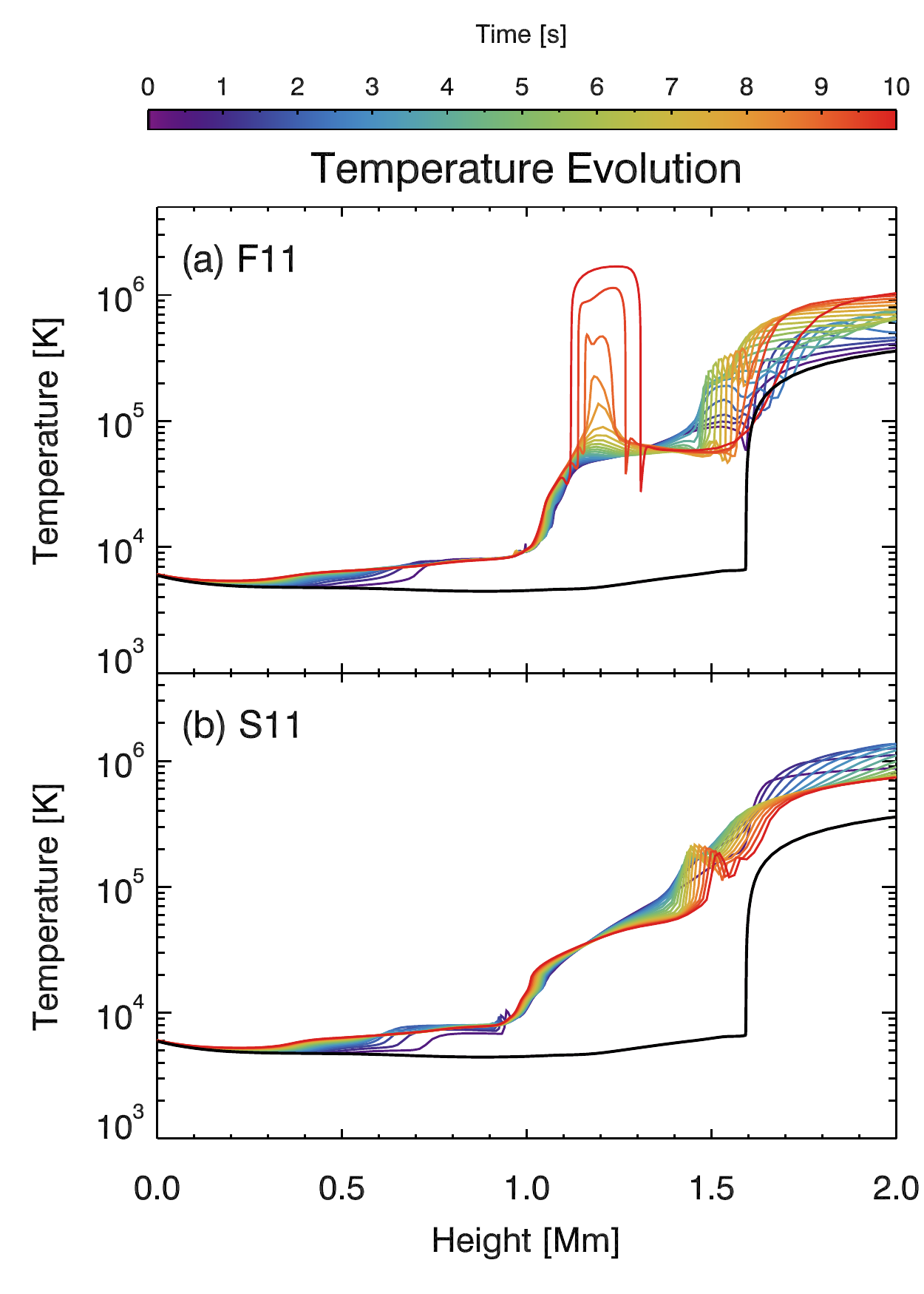}}
		\subfloat{\includegraphics[width = 0.35\textwidth]{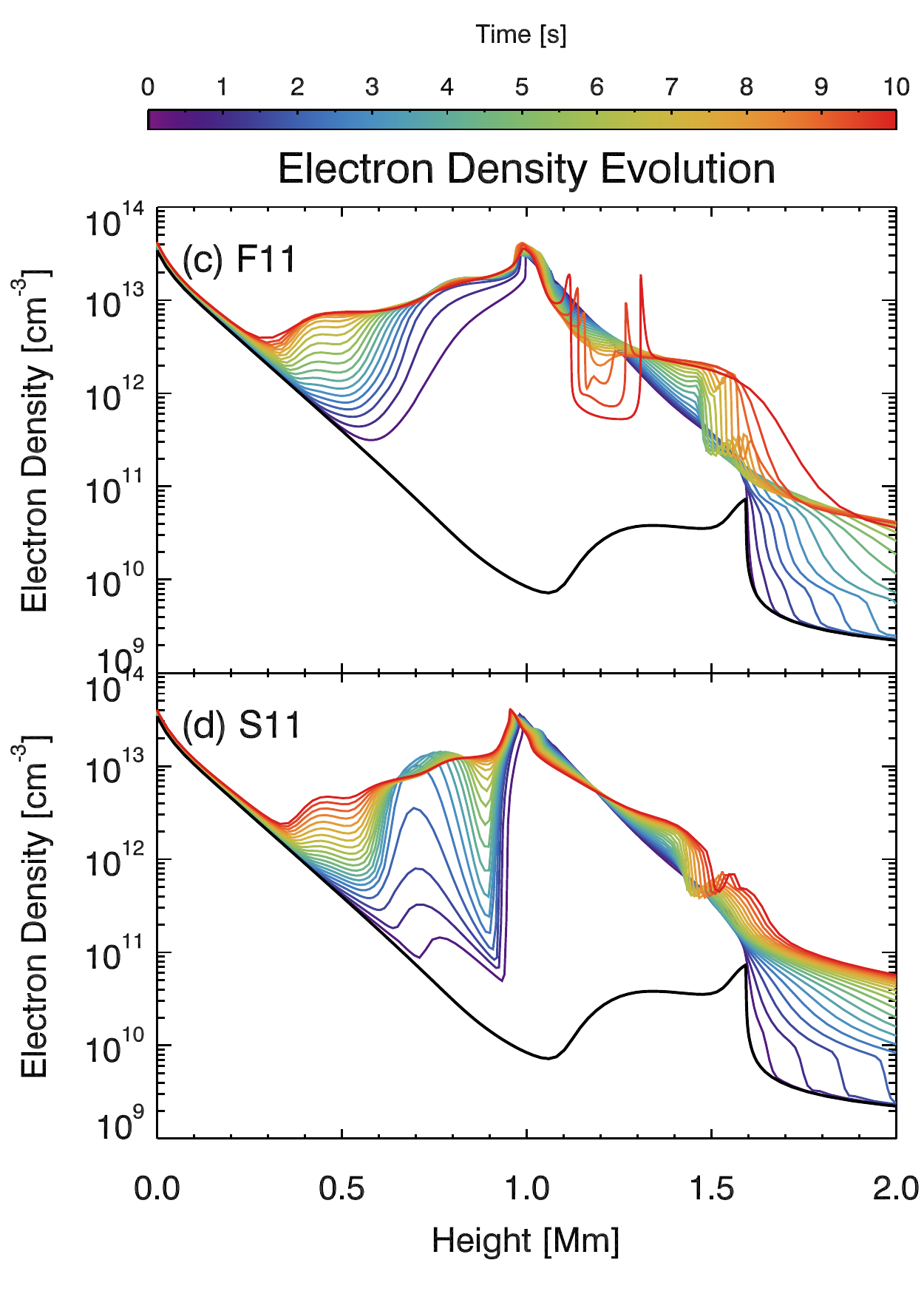}}
		\subfloat{\includegraphics[width = 0.35\textwidth]{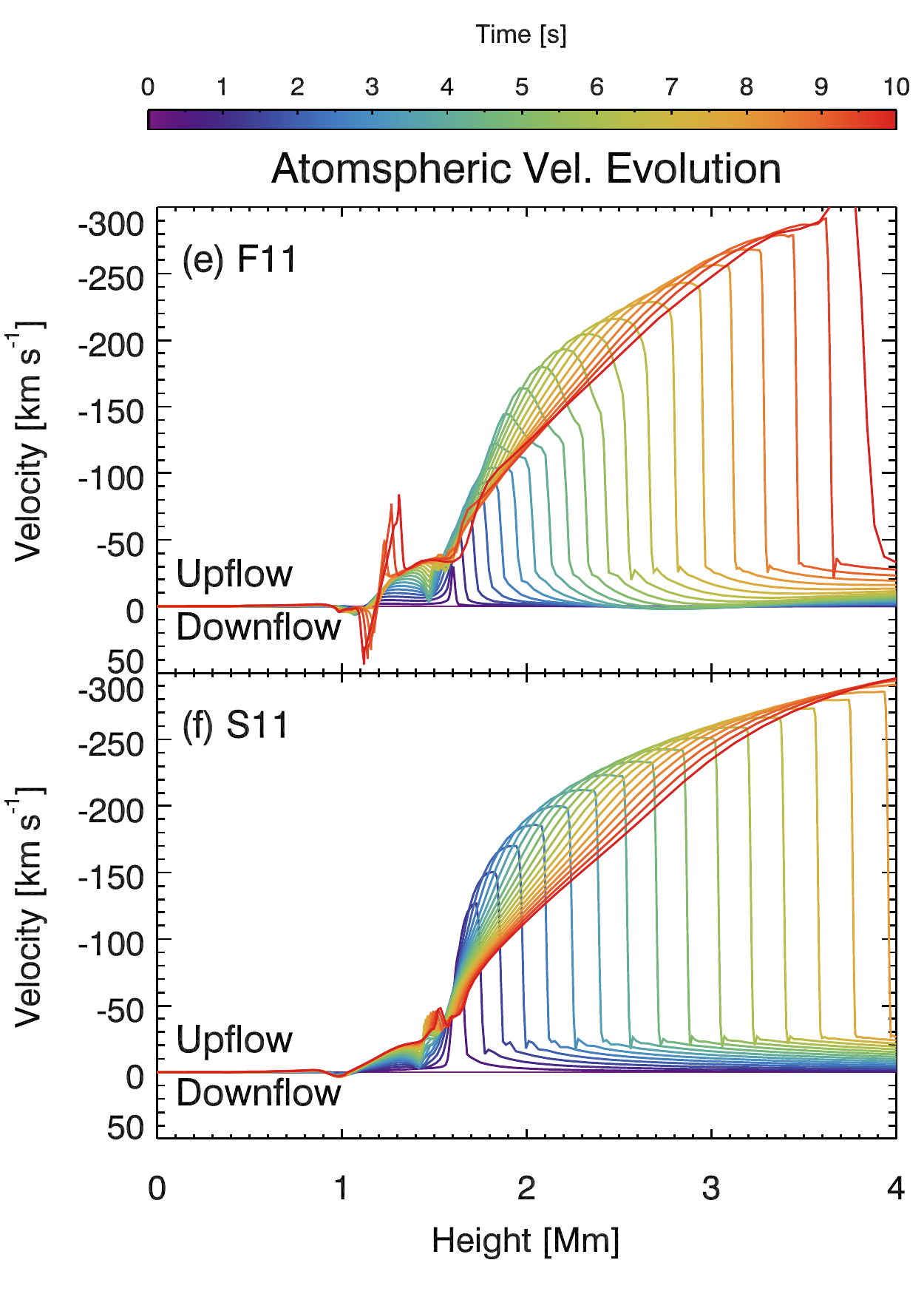}}
		}
	\vspace{-0.2in}
	\hbox{
		\hspace{-0.25in}
		\subfloat{\includegraphics[width = 0.35\textwidth]{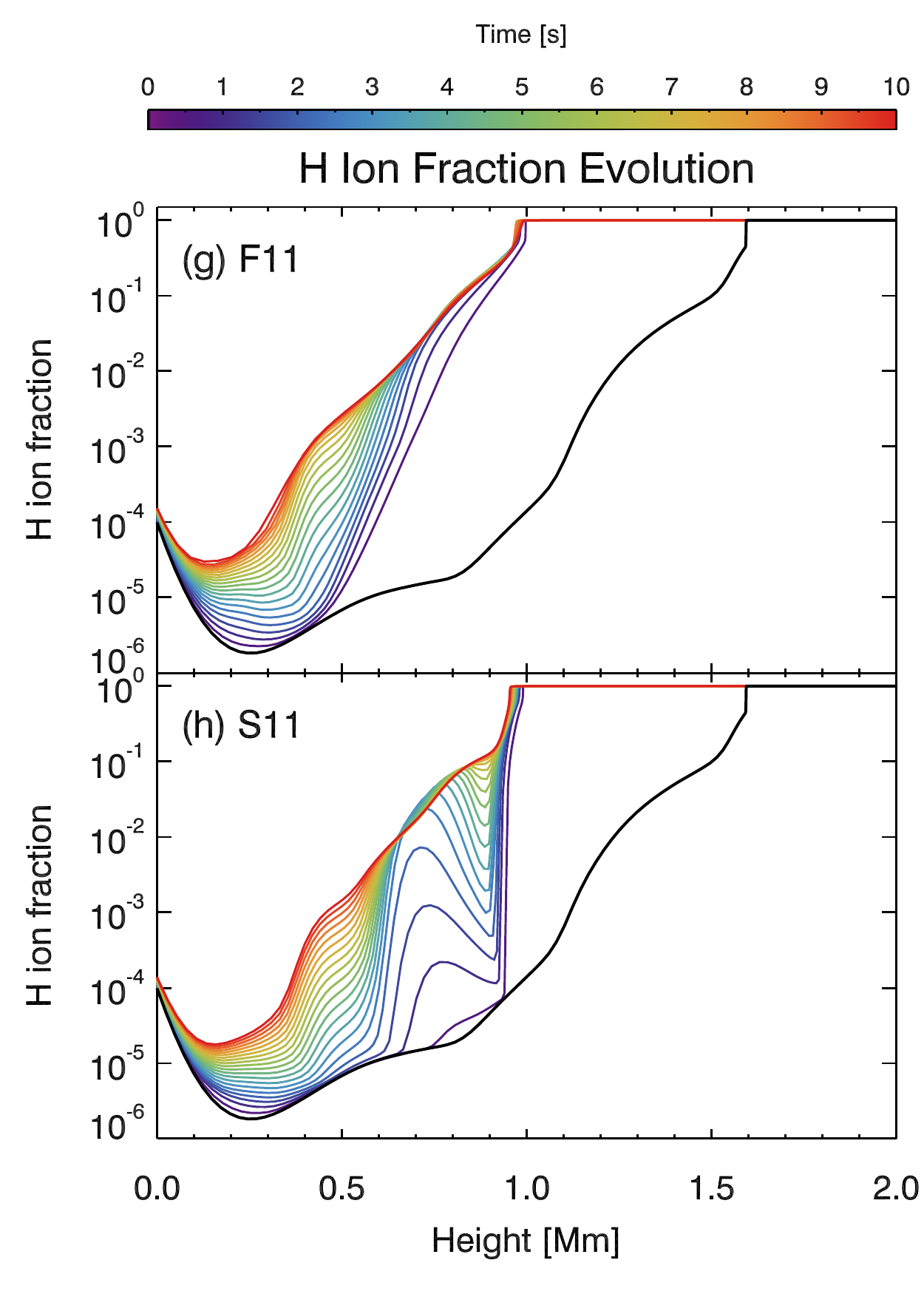}}
		\subfloat{\includegraphics[width = 0.35\textwidth]{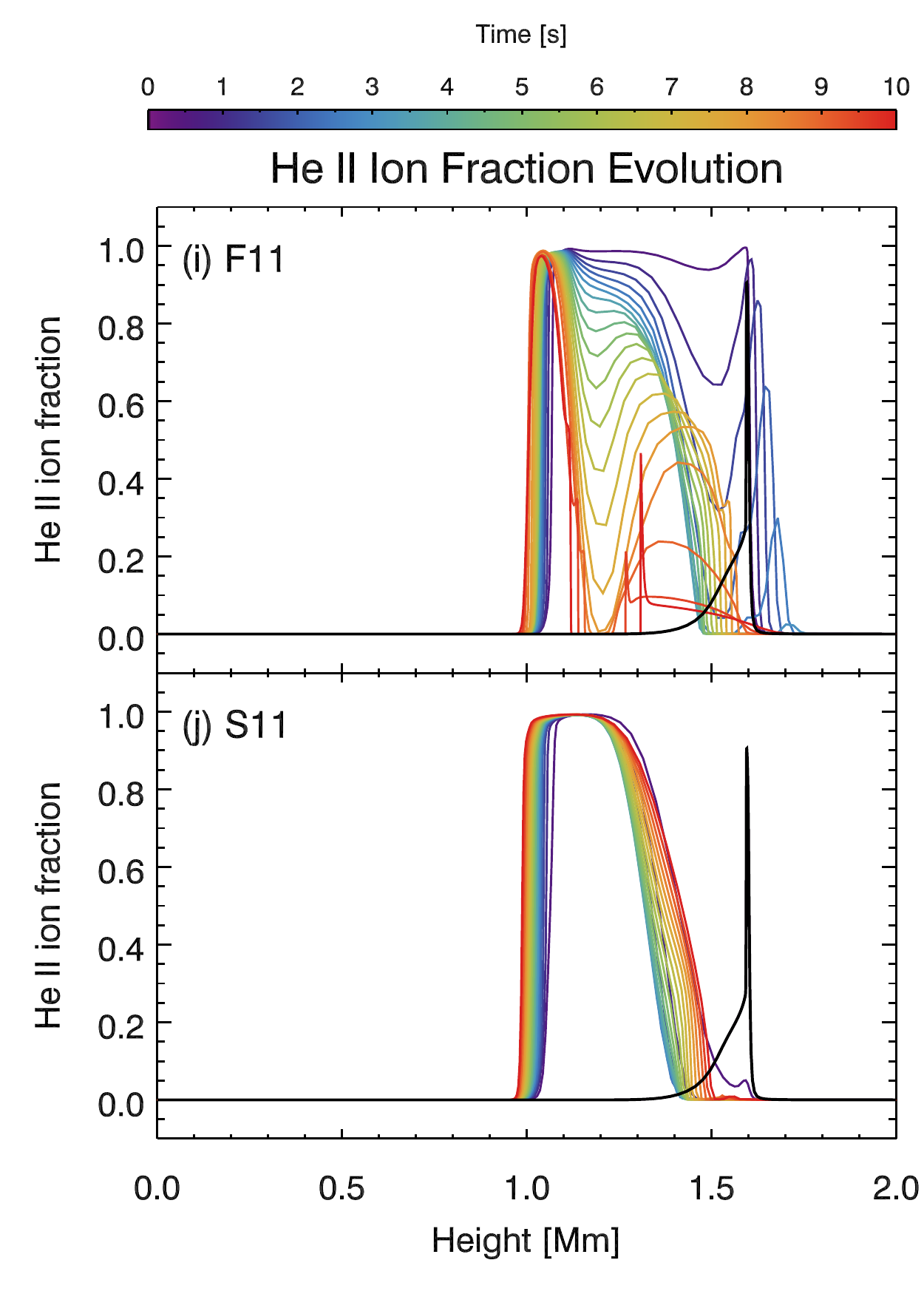}}
		\subfloat{\includegraphics[width = 0.35\textwidth]{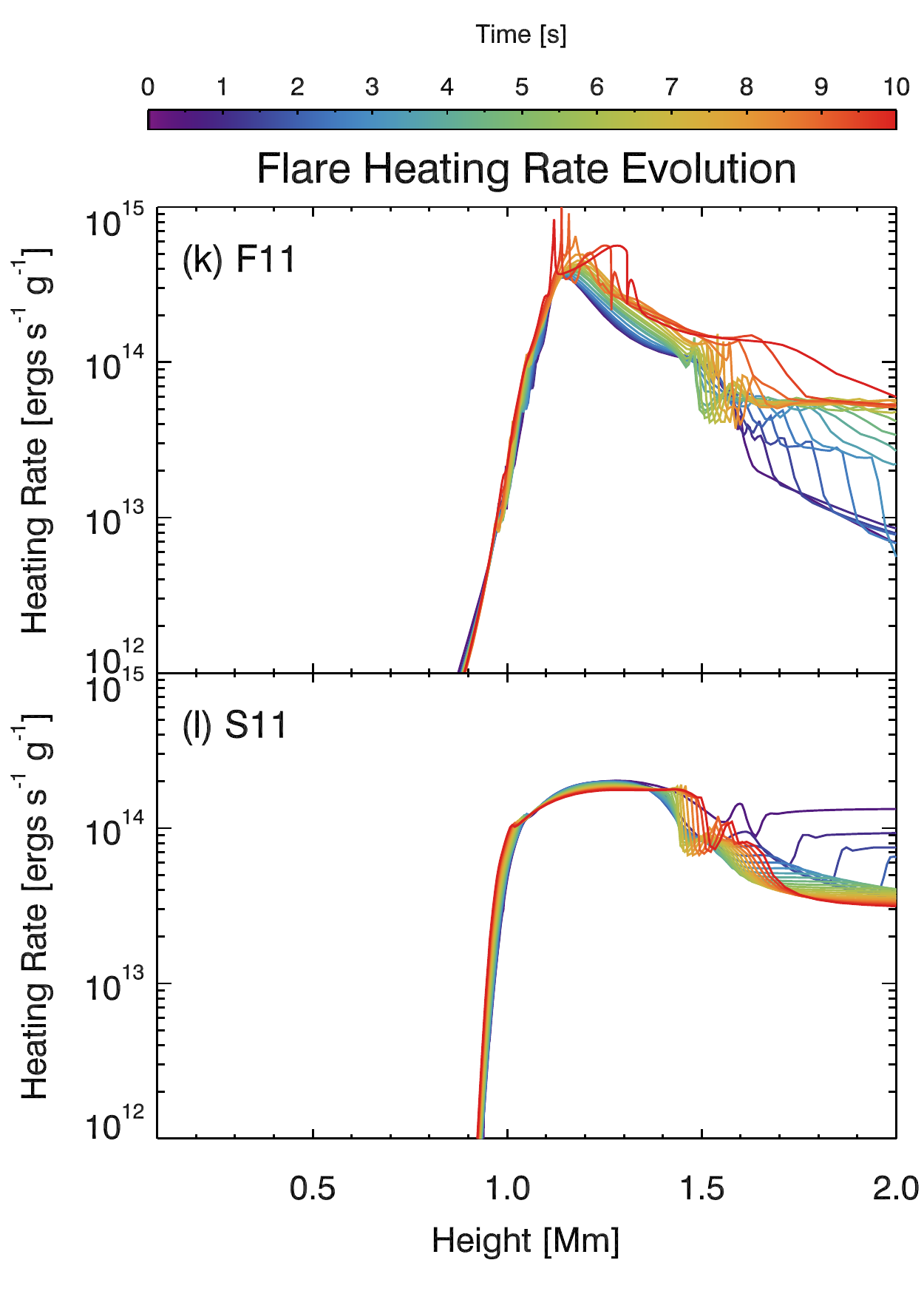}}		
	}
	\caption{\textsl{The evolution of the atmosphere for different energy transport mechanisms F11 (EB) and S11 (AW dissipation) where (a) \& (b) show temperature, (c) \& (d) show electron density, (e) \& (f) show velocity with upflow negative, (g) \& (h) show hydrogen ionisation fraction, (i) \& (j) show He \textsc{ii} ionisation fraction, and (k) \& (l) show flare heating rate per mass. Note that the x-scale of the atmospheric velocity is larger than for the other panels, so as to show the large velocity achieved high in the loop. Colour represents time with output plotted at 0.5~s intervals.}}
	\label{fig:comps}
\end{figure*}

Figure~\ref{fig:comps} shows the evolution of the atmosphere in each of the simulations, where F11 refers to the EB simulation and S11 the AW simulation. The colour of the lines represents the time in the simulation, where we plot from $t = [0, 10]$~s in 0.5~s intervals. The temperature, electron density, velocity, H ion fraction, He~\textsc{ii} ion fraction and flare heating rate are shown. In each case the lower panel shows the S11 atmosphere and the upper panel shows the F11 atmosphere. We discuss features of the dynamics below. We also show the energetics of the simulations at various times in Figure~\ref{fig:ebal_f11} (F11) and Figure~\ref{fig:ebal_s11} (S11), where positive quantities are heating and negative are cooling. 

\subsection{EB Simulation (F11)}
\begin{figure*}
	\centering
	\includegraphics[width = \textwidth]{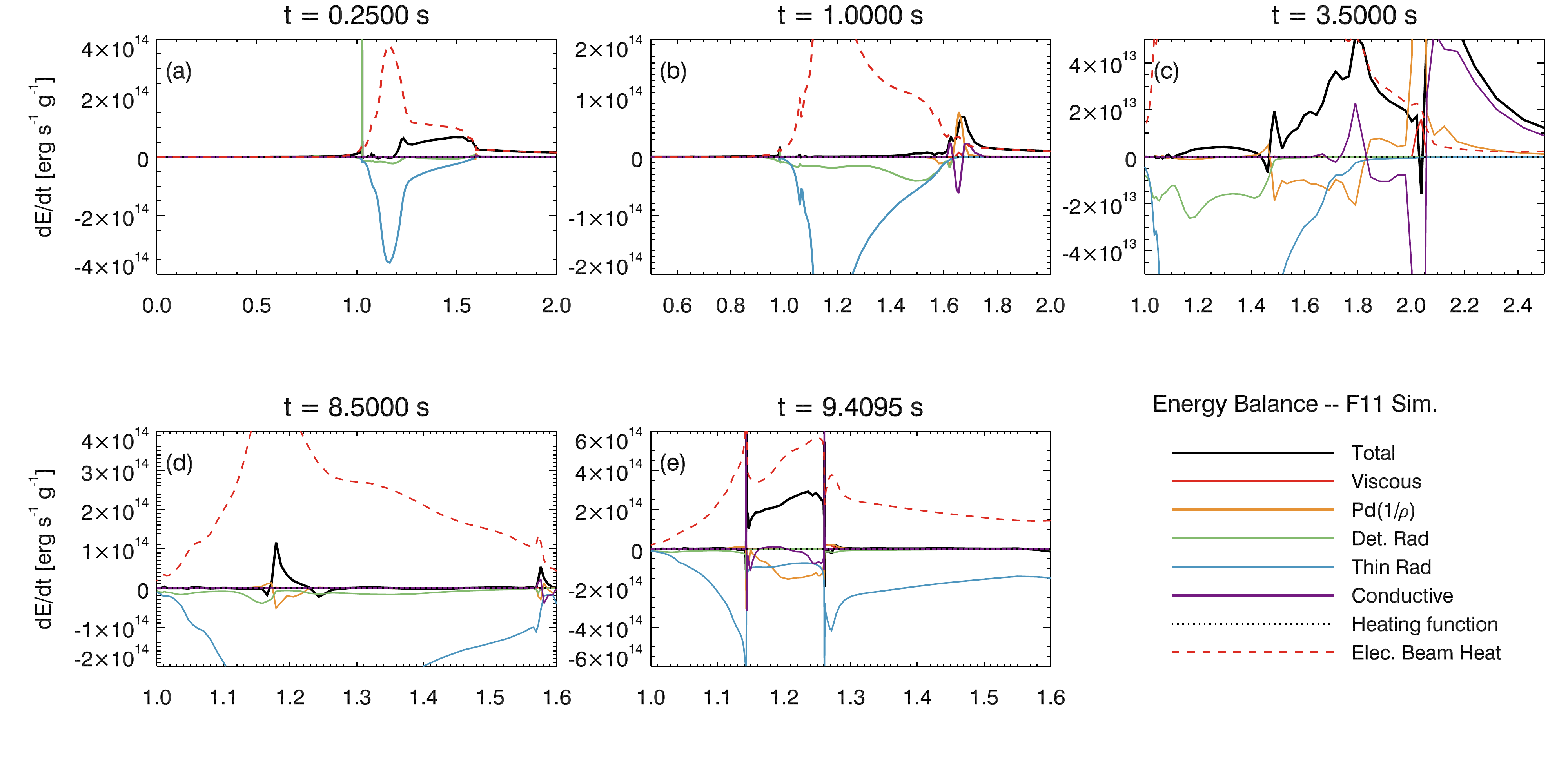}		
	\caption{\textsl{Energy balance in the EB simulation. Contributions to the energy balance are shown: total (black), viscous heating (red), work done by pressure (yellow), optically thick radiation computed in detail (green), optically thin radiation (blue), conductive flux (purple), the background heating function (black, dotted) and the flare heating rate (red, dashed). Positive represents heating, and negative cooling. Panel (a) shows that by $t = 0.25$~s radiative losses effectively balance energy input in the lower atmosphere, but are unable to balance beam heating in the mid-upper chromosphere. By $t = 1$~s, however, strong optically thick losses balance beam energy. At this time the pressure from enhanced temperature in the upper chromosphere results in upflows $\sim1.7$~Mm. Panel (c) illustrates the decrease in temperature around $\sim1.4$~Mm at $t= 3.5$~s and that conductive flux helps to increase the temperature in the upper atmosphere. Panel (d) shows the onset of the high temperature bubble in the mid-chromosphere. Radiative losses limit the temperature rise at this time, but as panel (e) shows, high temperatures increases ionisation in this region removing losses from He~\textsc{ii} allowing the explosive temperature rise to $>1$~MK.}}
	\label{fig:ebal_f11}
\end{figure*}

\begin{figure*}
	\centering
	\includegraphics[width = \textwidth]{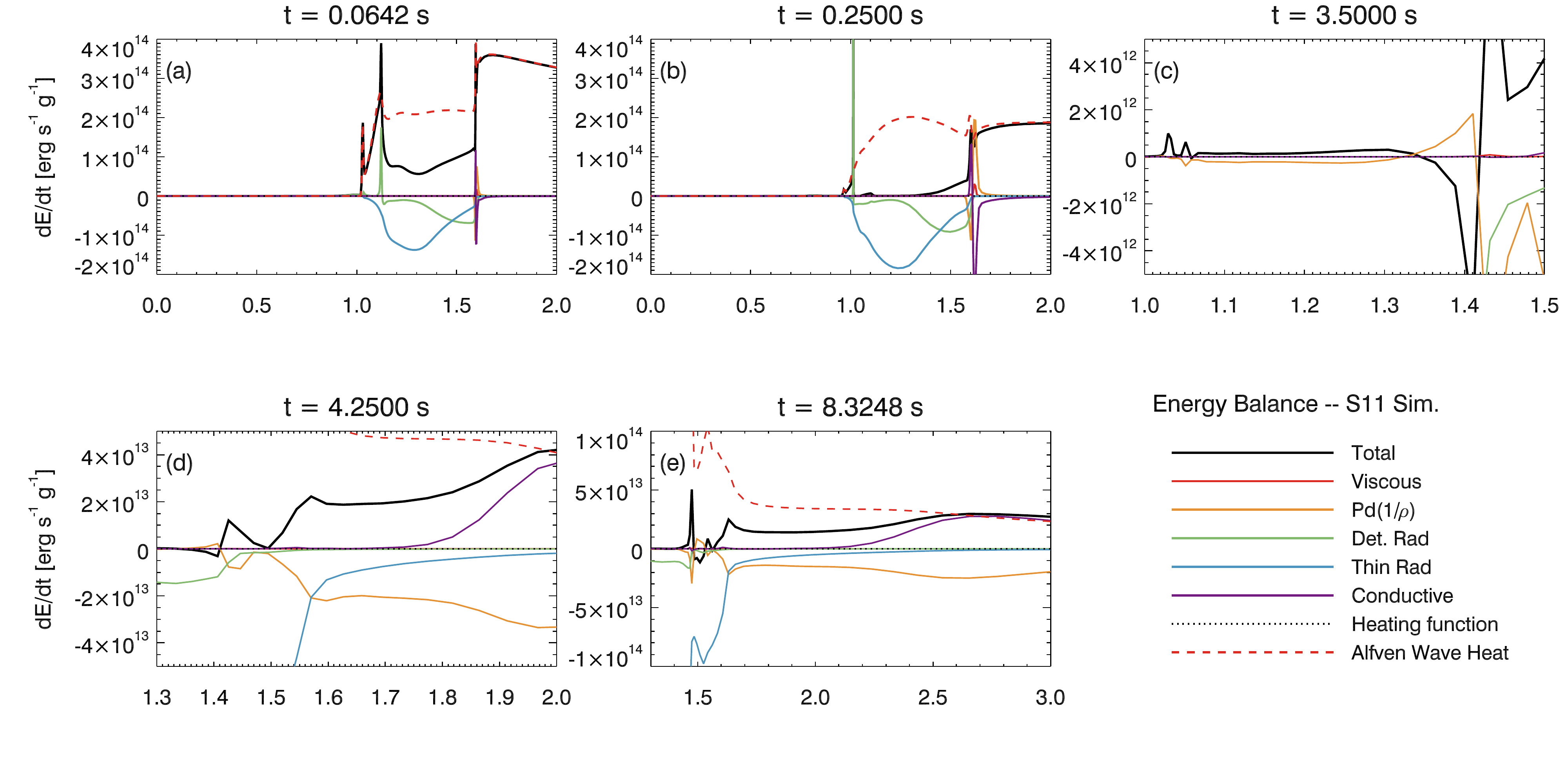}		
	\caption{\textsl{Energy balance in the AW simulation. Lines are as described in Figure~\ref{fig:ebal_f11}. Panel (a) shows the sudden energy input results in a large amount of unbalanced energy to heat and ionise the plasma. Panel (b) shows that very quickly this energy input is well balanced by radiative losses in the lower atmosphere, and that the large temperature enhancement results in a strong pressure wave leading to high velocity upflows. Panel (c) shows that losses exceed energy input at $\sim1.4$~Mm, leading to a decrease in temperature, and that a pressure wave pushes hot material upwards. Panel (d) shows that a hot bubble begins to form, with the energy balanced immediately above the high temperature region at $\sim1.5$~Mm, and that a strong conductive flux is present in the upper atmosphere. Panel (e) shows the complex dynamics of the temperature bubble. The upward propagating pressure wave has made the bubble under-dense, decreasing the heating rate, but also reducing radiative losses so that temperature increases greatly here. Immediately above the bubble is a local temperature minimum.}}
	\label{fig:ebal_s11}
\end{figure*}

{\bf t$<$1~s}: The temperature in the mid-upper chromosphere increases significantly over the background, to $T\approx[6000-7000]$~K between $0.6-1$~Mm and $\approx[40,000-85,000]$~K over $1.15-1.5$~Mm. In the lower atmosphere, flare energy largely goes into ionisation of hydrogen, which becomes completely ionised at $>1$~Mm, and partially ionised between $0.5-1$~Mm. Enhanced ionisation means a significantly increased electron density between $0.5-1.6$~Mm (and by more than three orders of magnitude at $1$~Mm). Helium ionisation also occurs at greater depths, with He~\textsc{ii} quickly forming between $\sim1.05-1.6$~Mm. In the upper atmosphere, just below the original TR position, T increases to $\approx[85,000-90,000]$~K, so that the He~\textsc{ii} fraction decreases again at $\sim1.5$~Mm as He~\textsc{iii} starts to form. A pressure wave starts at the TR ($1.6$~Mm) resulting from the sudden temperature increase to approximately $10^5$K, producing an upward mass motion with a velocity of more than $50$~km~s$^{-1}$, increasing with height. Figure~\ref{fig:ebal_f11}(a) \& (b) show that beam energy input is mostly balanced by radiative losses.

{\bf t = 1-4~s}:  Energy input into the lower chromosphere at $0.6-1$~Mm largely results in increased hydrogen ionisation causing the temperature plateau to only very slowly increase in temperature. The plateau extends to deeper layers, and electron density increases with further ionisation.  The transition from $T\approx[7000-40,000]$~K, at $1-1.15$~Mm, steadily steepens. At $1.15$~Mm the temperature increases by a few $\times 10^4$K to $T\approx60,000$~K, but radiative losses largely balance (and occasionally exceed) energy input between $1.15-1.4$~Mm meaning that the temperature changes little, and actually decreases at $\sim$1.4~Mm. Radiative losses decrease with time above this height and are no longer able to balance energy input, resulting in a temperature bubble in excess of $T=200,000$~K. Figure~\ref{fig:ebal_f11}(c) illustrates the energetics at this time. Within this bubble temperatures are high enough to almost completely ionise He to  He~\textsc{iii}. Above $1.6$~Mm the temperature continues to increase but not smoothly. Loop density is enhanced there by strong upflows ($v\sim150$~km~s$^{-1}$), so the beam deposits more energy at greater height. A strong conductive flux helps to increase temperature $> 2$~Mm.

{\bf t = 4-7.5~s}: Conditions at heights $<~1.15$~Mm continue to evolve in a similar manner to previously. The peak of the EB heating rate moves slightly higher, to $1.18$~Mm. Losses are just unbalanced at this point allowing temperature to rise to $T=85,000$~K. Losses are able to balance, and at times exceed, energy input between $\sim1.2-1.35$~Mm resulting in a drop in temperature. There is a corresponding drop in electron density as recombinations to He~\textsc{ii} take place. Note also at this time the amount of He~\textsc{iii} in the mid-chromosphere around $1.18$~Mm increases due to high temperatures, so that a narrow region of almost fully ionised He begins to form. Initially, the hot bubble at heights $> 1.4$~Mm is smoothed out as it is heated to  $>400,000$~K, due to a conductive flux into the cooler material ahead of the bubble, which increases the temperature in those regions. However, increased temperature at $\sim1.5$~Mm leads to an increased pressure which drives material away, making a narrow, under-dense region. Radiative losses decrease as a result of decreased density allowing temperature to increase further. Immediately ahead of this under-dense region is a locally over-dense region which, due to increased radiative losses, forms a local temperature minimum. 

{\bf t$>$7.5~s}: The final stage in the evolution is the formation of the large temperature bubble in the mid-upper chromosphere. High chromospheric temperatures ionise a large proportion of the He~\textsc{ii} to He~\textsc{iii} at $\sim~1.2$~Mm. Decreasing radiative losses from He~\textsc{ii} can no longer balance the beam energy deposition, which produces an ever-increasing temperature at that location (in excess of $1.5$~MK), and further ionisation. Figure~\ref{fig:ebal_f11}(d) shows the decrease in radiative losses allowing temperature to quickly rise. This high temperature bubble is very under-dense as material is pushed away by the strong pressure difference between the bubble and surrounding plasma, increasing the size of the high temperature region (Figure~\ref{fig:ebal_f11}(e)). Chromospheric condensations are much stronger at these times, reaching up to $v\sim45$~km~s$^{-1}$. Since more mass is evaporated into the loop the heating rate again increases at greater heights, increasing the temperature of the corona and pushing the TR upwards. The increase in density on either side of the bubble (material evacuated from the high pressure region) results in strong radiative losses that exceed energy input, creating very narrow, cool regions that permit recombination to He~\textsc{ii}. These regions propagate away from the shock.  

\subsection{Alfv\'en Wave Simulation (S11)}
The lower panel of each atmospheric property in Figure~\ref{fig:comps} shows the evolution of the atmosphere in response to AW dissipation. Immediately clear from these figures is that the atmosphere evolves in a largely similar manner to the EB flare simulation, but also that the extreme temperature, low density bubble does not form in the mid chromosphere. 

{\bf t $<$ 1~s:} The location of the peak of the heating rate (Figure~\ref{fig:comps}(l)) is within $\sim0.1$~Mm of the peak in the EB heating rate, but the peak is broader. By $t=1$~s the temperature between $0.7-0.95$~Mm rises to $T\approx[6000 - 7500]$~K. Mid-upper chromospheric temperatures show enhancements over pre-flare values, rising over a shallow gradient from $T\sim7500$~K at 0.95~Mm to $T\sim200,000$~K at 1.55~Mm. The TR temperature is increased to $>1$~MK. 

This initial enhancement to the mid-upper chromospheric temperature occurs rapidly, with radiative losses almost completely balancing energy input up to $\sim1.3$~Mm by $t=0.25$~s. Figure~\ref{fig:ebal_s11}(a,b) illustrates the energy balance at these times showing that following the rapid ionisation the radiative losses increases sufficiently to mostly balance flare energy input. Hydrogen is almost entirely  ionised above $1$~Mm, and ionisation continues gradually to greater depth (Figure~\ref{fig:ebal_s11}(h)). The elevated temperature in the mid-chromosphere results in ionisation of He, which is mostly ionised to He~\textsc{ii} between $1.05-1.3$~Mm and to He~\textsc{iii} above $1.4$~Mm, leading to an increase in the electron density between $0.9-1.6$~Mm. The peak in $n_e$ occurs at the hydrogen ionisation boundary (between $0.9-1$~Mm), with $n_e\sim4\times10^{13}$~cm$^{-3}$. By $t=0.25-0.5$~s a strong pressure wave at $1.65$~Mm pushes chromospheric material into the corona with velocities in excess of $v\sim130$~km~s$^{-1}$, significantly higher than in the EB simulation. Since the AW heating rate decreases sharply around 0.9~Mm, the ionisation of hydrogen below this occurs at later times in comparison to the EB simulation. 

{\bf t = 1-5~s:} Between $0.6-0.9$~Mm radiative losses almost completely balance the energy input, and so temperatures rise only modestly, to $T\sim8000$~K. Hydrogen ionisation also increases, creating a small region of locally high electron density $n_e = 1-1.5\times10^{13}$~cm$^{-3}$ at $\sim 0.75$~Mm. In the mid-chromosphere a temperature `pivot' point forms at $\sim1.15$~Mm, with temperature decreasing with time above this point, and increasing below. As the temperature increases in the deep atmosphere, ionisation to He~\textsc{ii} follows producing a small electron density increase at $1-1.05$~Mm. The associated pressure changes results in upflows of a few $\times 10$~km~s$^{-1}$. The initial high velocity upflow reaches heights above $2.5$~Mm, with $v\sim200$~km~s$^{-1}$.

The temperature gradient between $1.15-1.45$~Mm flattens slightly from $T\approx[35,000 - 90,000]$~K  to $T\approx[30,000 - 70,000]$~K. This occurs because hot plasma pushed up into the loop at a few $\times 10$~km~s$^{-1}$, due to an increase in pressure above 1.15~Mm, leaves cooler material in its place, and because radiative losses above $1.25$~Mm begin to exceed the energy input which leads to cooling (see Figure\ref{fig:ebal_s11}(c)). A narrow high temperature ($T\sim100,000$~K) bubble begins to form at $\sim1.4-1.5$~Mm. The TR heats as a result of energy deposition and via a conductive flux from below. Figure~\ref{fig:ebal_s11}(d) shows the decrease in radiative losses that allow the formation of the high temperature at $1.4$~Mm, and the upwards propagating pressure wave.

{\bf t = 5-10~s:} For the remainder of the simulation the dynamics evolves as before. The atmosphere cools slightly between $1.25-1.4$~Mm,  narrowing the high temperature bubble around $1.4-1.5$~Mm, making it more pronounced. This results in a greater pressure difference that increases the flows (see Figure~\ref{fig:comps}(f)), leading to an under-dense region (similar to the process that resulted in the high-altitude temperature bubble in the EB simulation). The bubble cools slightly, but remains hot since, despite the heating rate in the bubble being reduced because of the lower density, the density change also significantly reduces radiative losses around 1.5~Mm. Increased density ahead of the bubble leads to enhanced radiative losses and decreasing temperature, producing a local temperature minimum. Figure~\ref{fig:ebal_s11}(e) illustrates the energetics at this time.\\

\section{Radiation Results}\label{sec:rad_results}

\subsection{Ca~\textsc{ii} 8542 \AA\ Line Profiles}\label{sec:caii_8542_profiles}

The Ca~\textsc{ii} 8542 \AA\ line is part of the Ca~\textsc{ii} subordinate infrared (IR) triplet, which are sensitive to the  temperature at their formation height in the low chromosphere, and to magnetic structures, making them good tracers of solar and stellar activity \citep[e.g.][]{1972SoPh...25..357S, 1979ApJS...41..481L, 2006ApJ...639..516U}. Since this line is so sensitive to lower chromospheric temperature, and will be observed with high spatial and spectral resolution by DKIST, we investigate the differences in the spectral line profiles, and their formation properties, between our EB and AW simulations.  
 
\begin{figure}
	\centering
	\vbox{
		\subfloat{\includegraphics[width = 0.5\textwidth, clip = true, trim = 0.85cm 0cm 0cm 0.2cm]{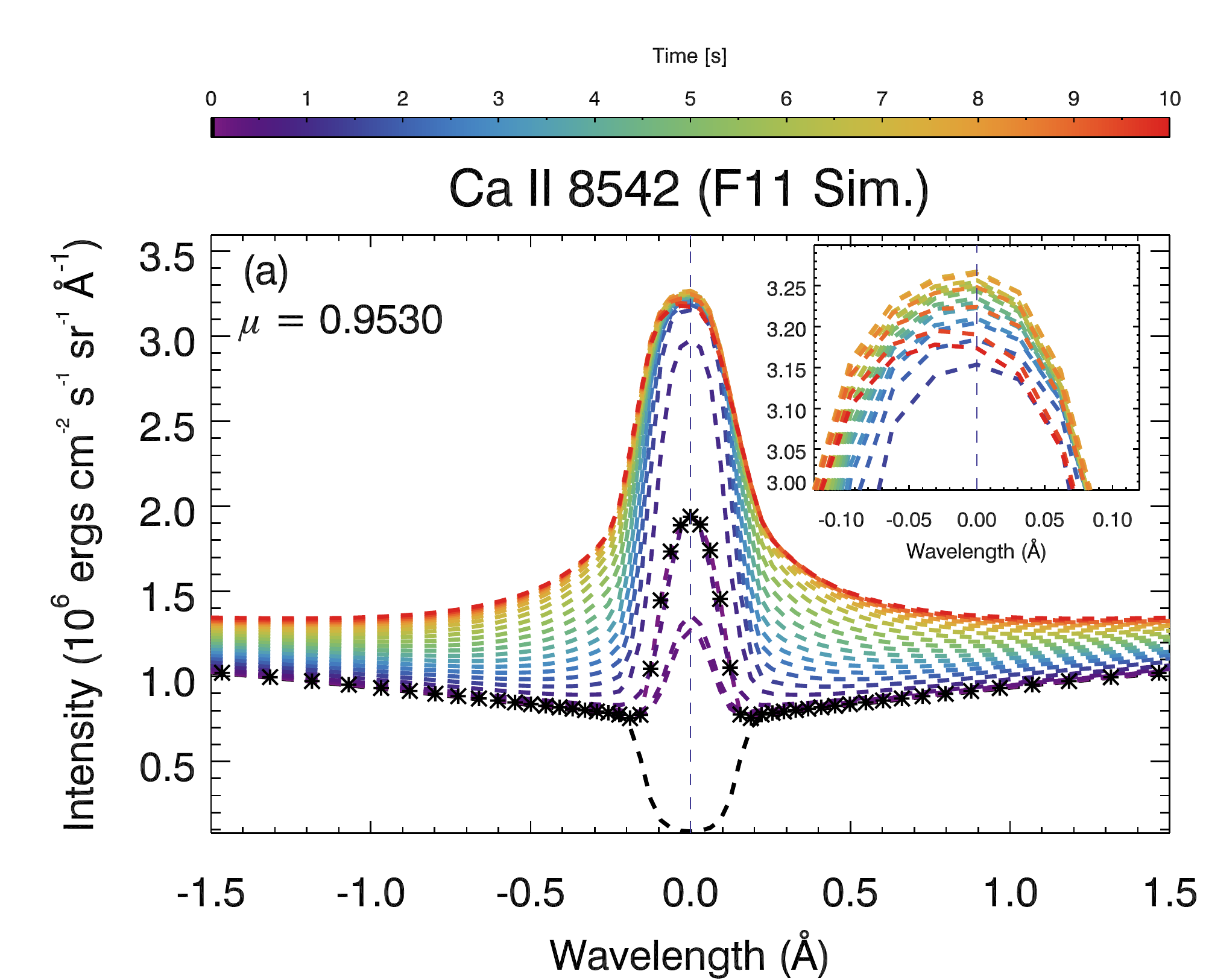}}		
		}
	\vbox{
		\subfloat{\includegraphics[width = 0.5\textwidth, clip = true, trim = 0.85cm 0cm 0cm 0.2cm]{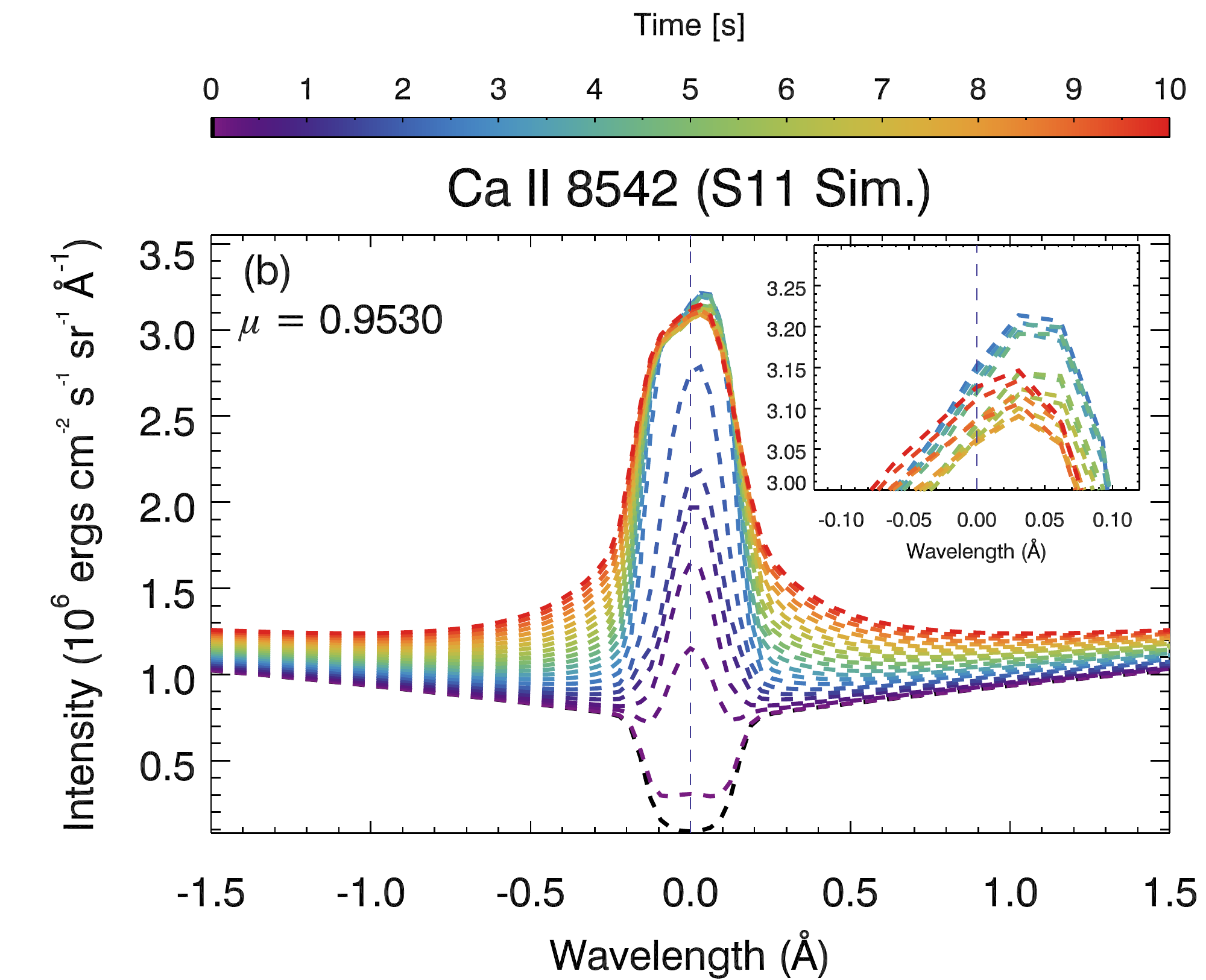}}	
		}
	\caption{\textsl{The Ca~\textsc{ii} 8542 \AA\ line, computed in (a) the EB simulation and (b) the AW simulation. In both cases colour represents time, and the inset shows a closer view of the line core. In panel (a) symbols are over-layed on the profile at $t=0.072$~s to help it stand out against profiles from $t<1$~s  }}
	\label{fig:caii_8542_lines}
\end{figure}

Figure~\ref{fig:caii_8542_lines}(a) shows how the Ca~\textsc{ii} 8542 \AA\ line responds to flare energy input in the EB (F11) simulation. Colour refers to simulation time (note that we plot $t = [0,~0.072,~0.25,~0.302,~0.5]$~s and then every 0.5~s thereafter). The inset shows a zoom of the core, and the vertical dashed line indicates the rest wavelength. The quiet Sun profile is in absorption but the line core immediately goes into emission in response to beam energy input. Between $t=0.072$ and $t=0.25$~s the core intensity drops significantly, but the far wing intensity continues to rise (note that on Figure~\ref{fig:caii_8542_lines}(a) we plot symbols for the $t=0.072$~s profile, since the colours at early times are very similar). Over the next few seconds the core intensity increases again, reaching a peak of $\sim3.27\times10^{6}$~ergs~cm$^{-2}$~s$^{-1}$~sr$^{-1}$~\AA$^{-1}$ at $t = 7.6$~s followed by a small decrease.
The core intensity changes little after $t\sim3$~s, but the wing intensity shows a strong enhancement. The far wing intensity is initially $\sim1\times10^6$~ergs~cm$^{-2}$~s$^{-1}$~sr$^{-1}$~\AA$^{-1}$, rising to $\sim1.35\times10^6$~ergs~cm$^{-2}$~s$^{-1}$~sr$^{-1}$~\AA$^{-1}$. The line was initially narrow and symmetric but over time a slight blueshift develops in the line core and the profile widens.

The equivalent is shown for the AW (S11) simulation in Figure~\ref{fig:caii_8542_lines}(b). We plot $t= 0, 0.064, 0.25, 0.5$~s and then every 0.5~s thereafter, and include an inset of the line core. it is clear that the AW simulation produces profiles with a stronger asymmetry. The line is slower to go into emission than in the EB simulation, and the intensity increases slowly over time.

Since the line is optically thick the peak intensity may not be the core of the line. Instead we consider the line centroid (the centre of mass of the line). The intensity of the line is lower than in the EB simulation, reaching a maximum line centroid intensity of 
$\sim3.16\times10^{6}$~ergs~cm$^{-2}$~s$^{-1}$~sr$^{-1}$~\AA$^{-1}$. The peak intensity of the profile is located redward of the line centroid and peaks at $t\sim3.5$ with a value of  $\sim3.22\times10^{6}$~ergs~cm$^{-2}$~s$^{-1}$~sr$^{-1}$~\AA$^{-1}$.The final intensity in the far wings is also lower than the EB simulation. The line appears redshifted initially, but this decreases with time. The final state is a small red asymmetry.

\subsection{Ca~\textsc{ii} 8542 \AA\ Line Formation}\label{sec:caii_8542_form}
We can study the formation properties of the line in each simulation, by writing the formal solution of the radiative transfer equation for the emergent intensity as in \cite{1994chdy.conf...47C}:	
	\begin{equation}\label{eq:cont_fn} 
		I_{\nu} = \frac{1}{\mu}\int_{z_0}^{z_1}S_{\nu}~\tau_{\nu}e^{-\tau_{\nu}/\mu}~\frac{\chi_{\nu}}{\tau_{\nu}}~dz = \frac{1}{\mu}\int_{z_0}^{z_1} C_{\rm{I}}~dz,
	\end{equation}
	
\noindent where $C_{\rm{I}}$ is  the contribution function to the emergent intensity and indicates how much emergent intensity originates from a certain height. In Eq~\ref{eq:cont_fn}, ${\mu}$ is the viewing angle, and the terms are a function of frequency $\nu$. Since CRD is assumed, the source function, $S_{\nu}$ (the ratio of emissivity to opacity), is independent of frequency across the line. The term  $\tau_{\nu}e^{-\tau_{\nu}/\mu}$ describes the attenuation by the optical depth, $\tau_{\nu}$. The monochromatic opacity per unit volume, $\chi_{\nu}$, is proportional to the density of emitting particles, so that the term~${\chi_{\nu}}/{\tau_{\nu}}$ is higher when there are large number of emitters at low optical depth (i.e photons are produced and can escape). This is sensitive to mass motions, and shows velocity gradients. 
 
\begin{figure*}
	\centering
	\vbox{
	\hbox{
		\subfloat[]{\includegraphics[width = 0.5\textwidth, clip = true, trim = 0.0cm 0.0cm 0cm 0.cm]{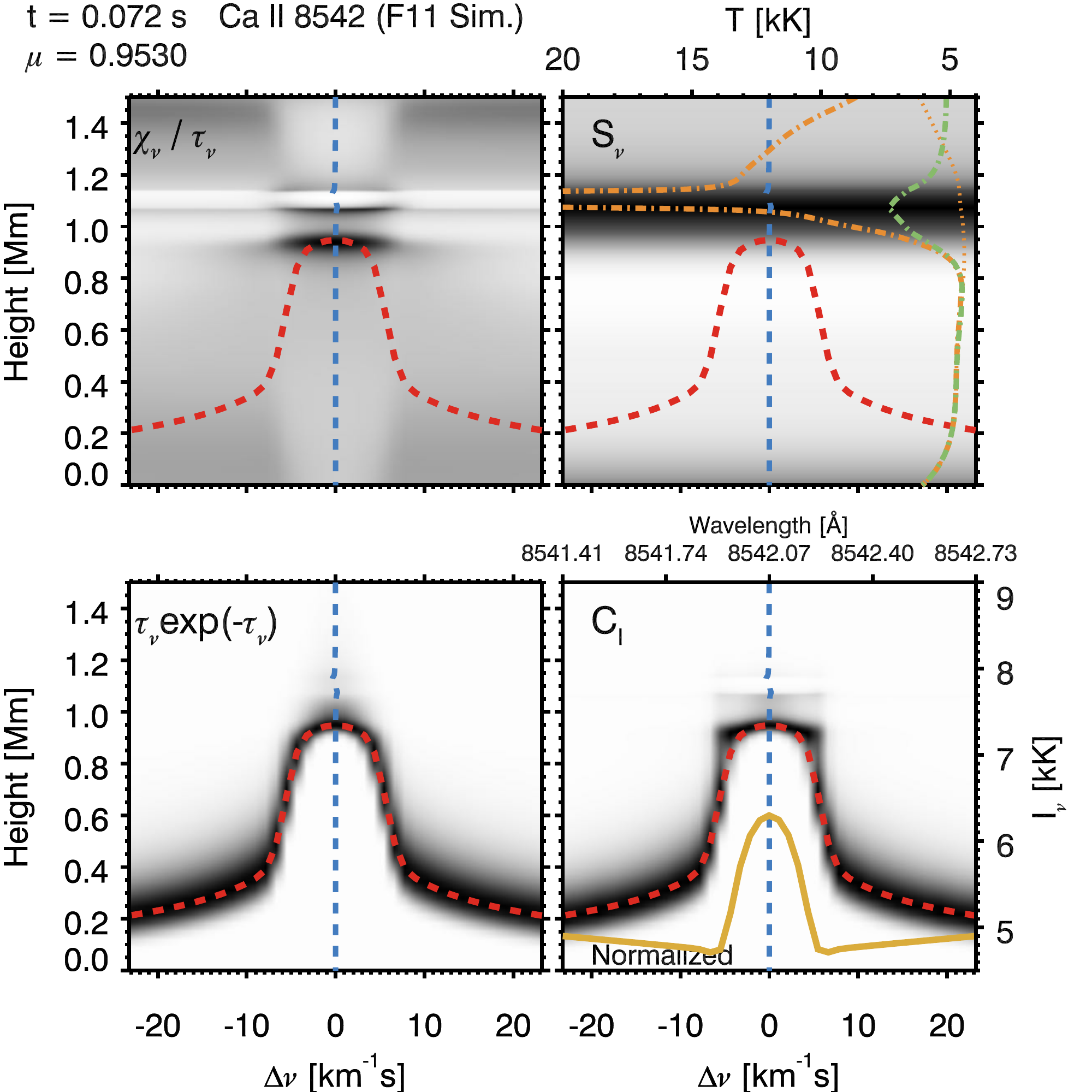}}		
		\subfloat[]{\includegraphics[width = 0.5\textwidth, clip = true, trim = 0.0cm 0cm 0cm 0.cm]{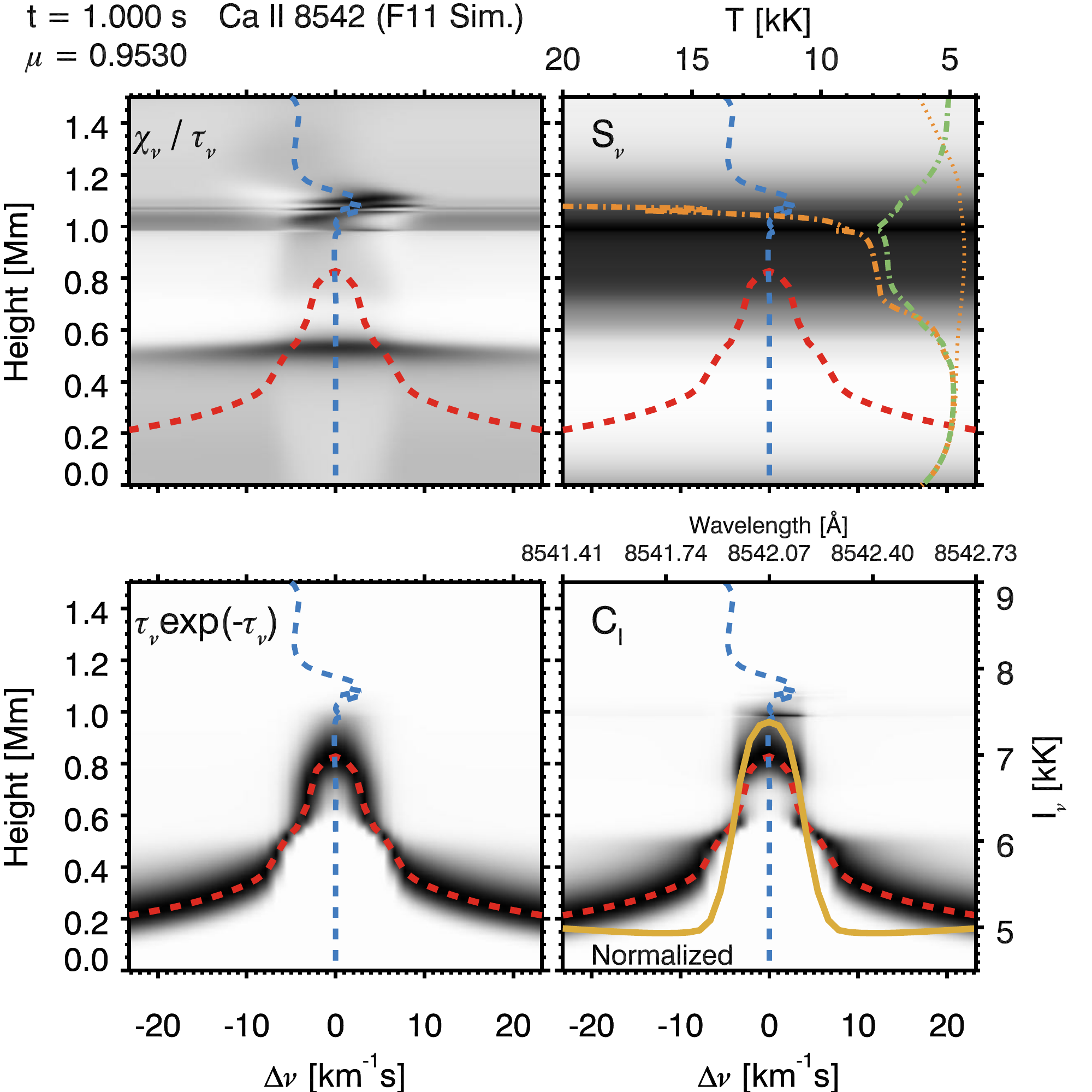}}		
		}
		}
		\vbox{
		\hbox{
		\subfloat[]{\includegraphics[width = 0.5\textwidth, clip = true, trim = 0.0cm 0cm 0cm 0.cm]{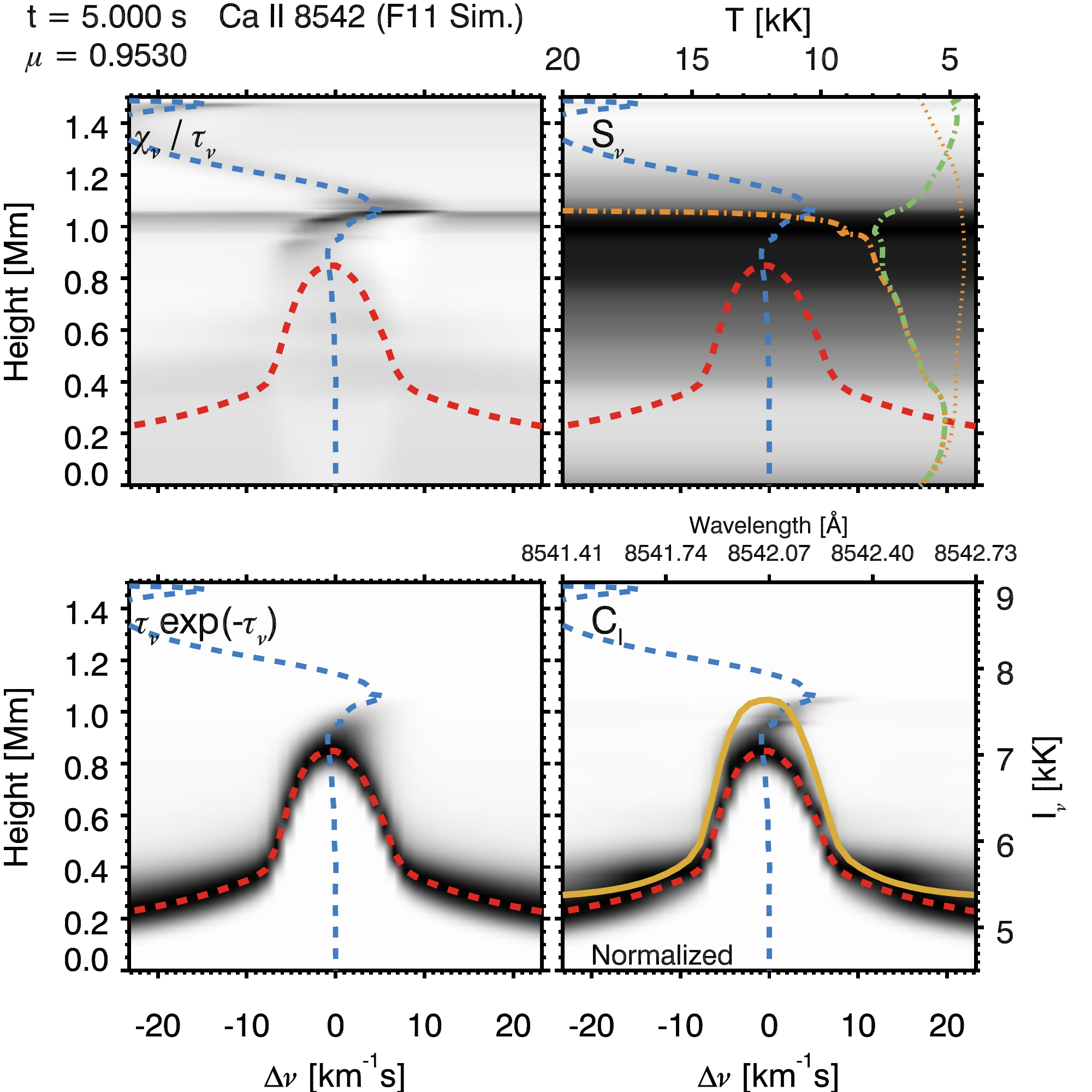}}		
		\subfloat[]{\includegraphics[width = 0.5\textwidth, clip = true, trim = 0.0cm 0cm 0cm 0.cm]{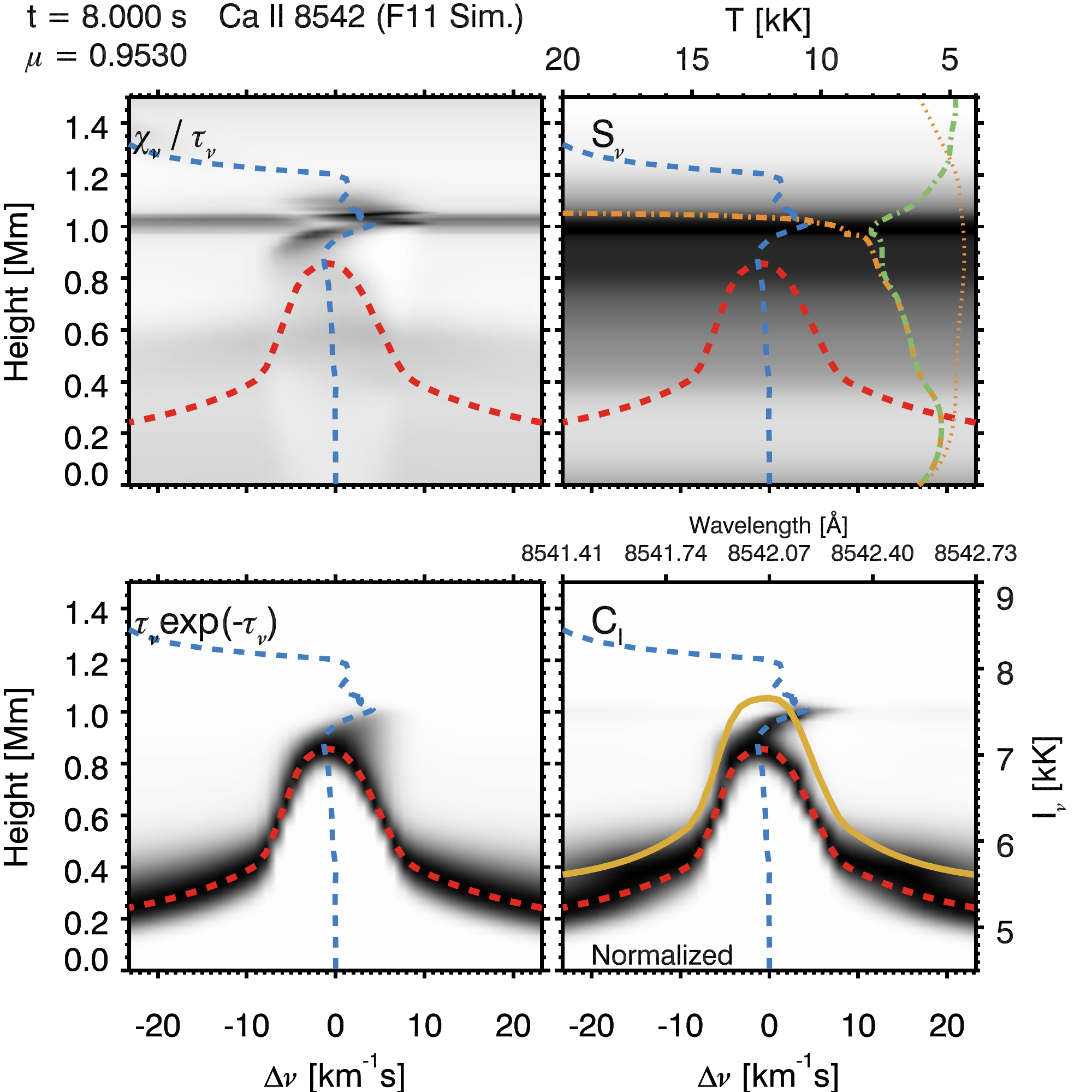}}		
		}
		}
			\caption{\textsl{Ca~\textsc{ii} 8542\AA\ line formation in the EB (F11) simulation at different times as indicated. Each panel shows the image of the quantity labelled in the corner of the image. Images are inverse scale. The atmospheric velocity (blue, dashed), $\tau_{\nu}=1$ curve (red, dashed), line source function (green, dot-dashed), Planck function (orange, dot-dashed), Planck function at $t=0$~s (orange, dotted), and emergent intensity (yellow, solid) are also plotted. Positive velocity is redshift/downflow, and intensity is expressed in units of radiation temperature. In the bottom right panels we have normalised the contribution function at each wavelength so that the image is not dominated by the line core, allowing details from the wings to be visible.}}
	\label{fig:caii_elec_contfn}
\end{figure*}

\begin{figure*}
	\centering
	\vbox{
	\hbox{
		\subfloat[]{\includegraphics[width = 0.5\textwidth, clip = true, trim = 0.0cm 0cm 0cm 0.cm]{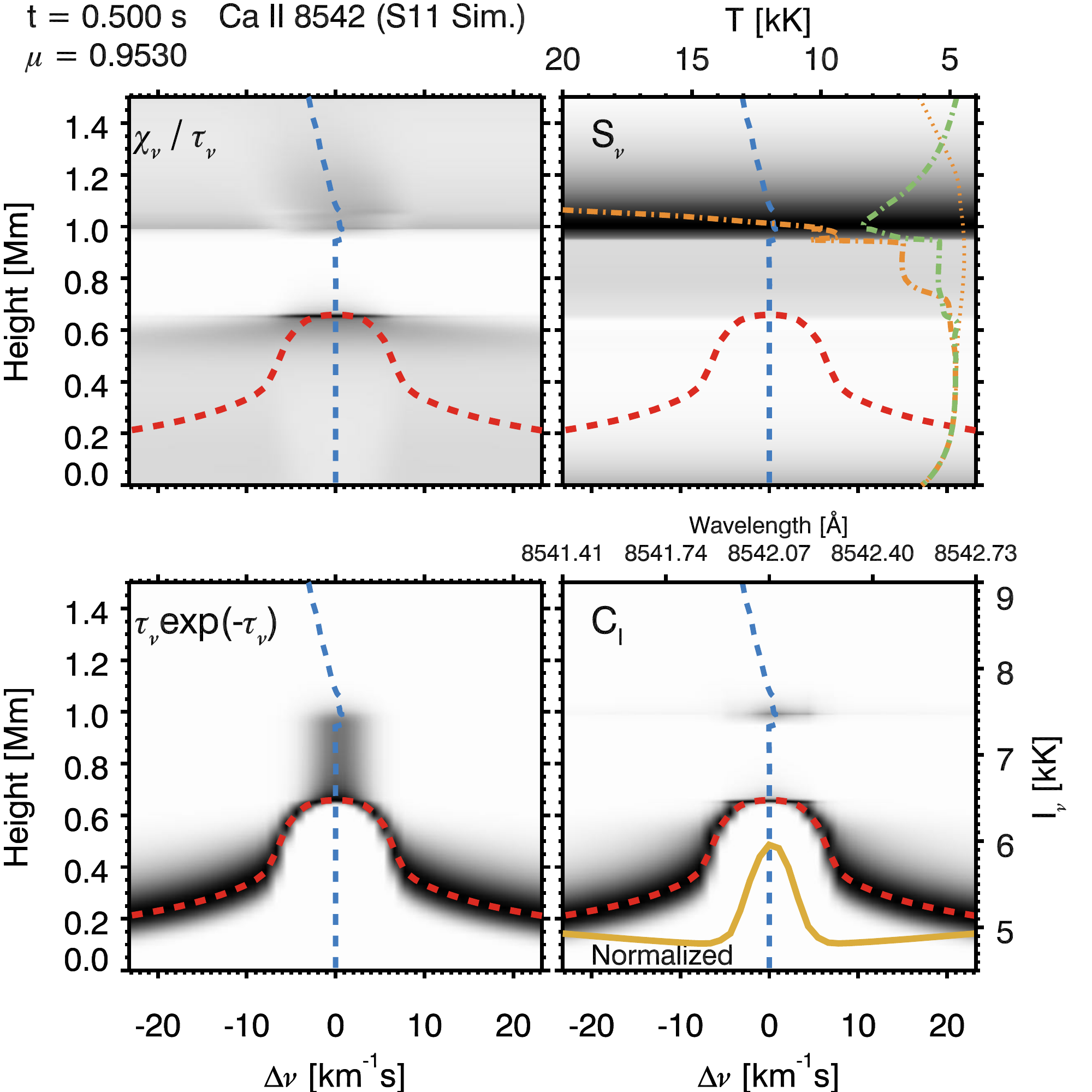}}		
		\subfloat[]{\includegraphics[width = 0.5\textwidth, clip = true, trim = 0.0cm 0cm 0cm 0.cm]{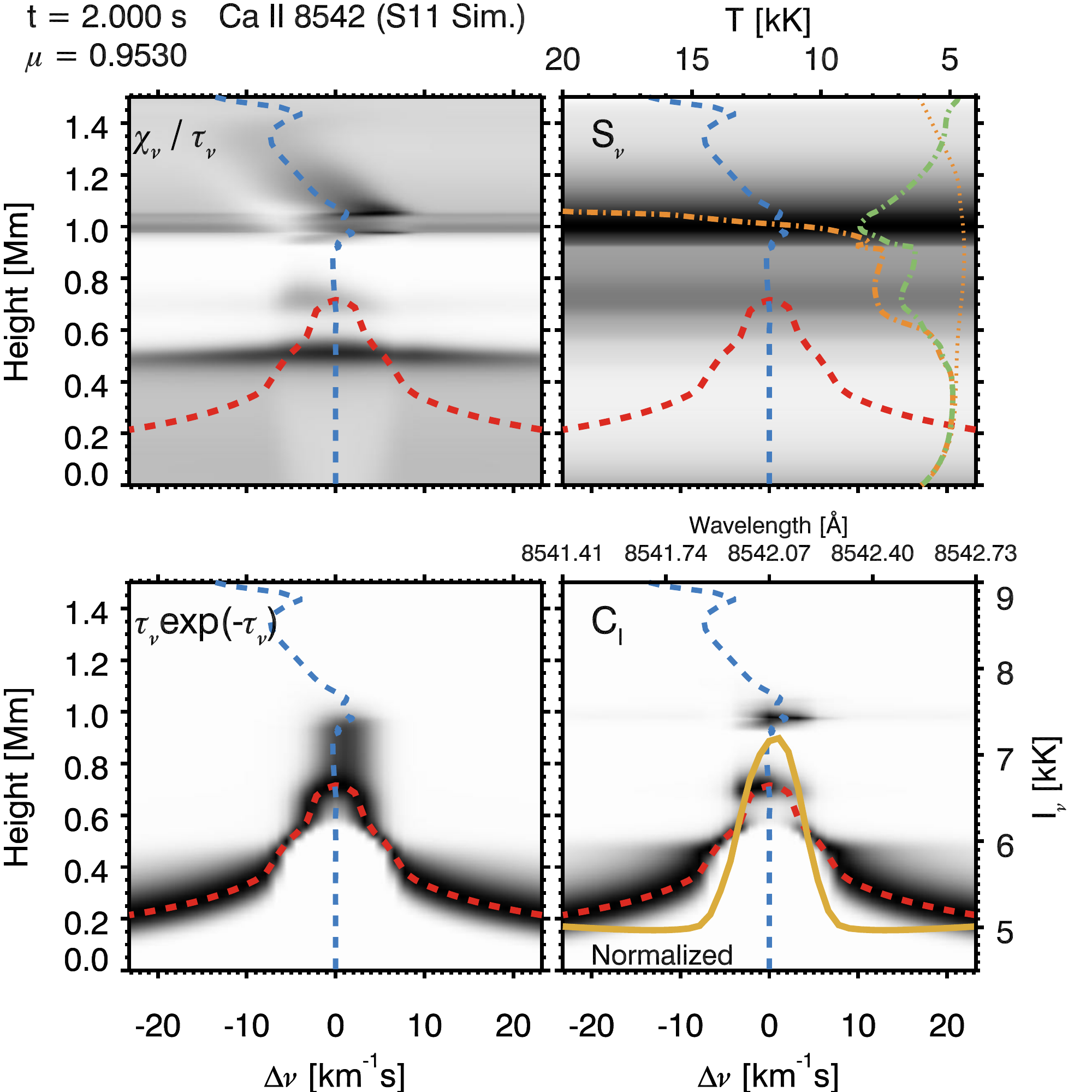}}		
		}
		}
		\vbox{
		\hbox{
		\subfloat[]{\includegraphics[width = 0.5\textwidth, clip = true, trim = 0.0cm 0cm 0cm 0.cm]{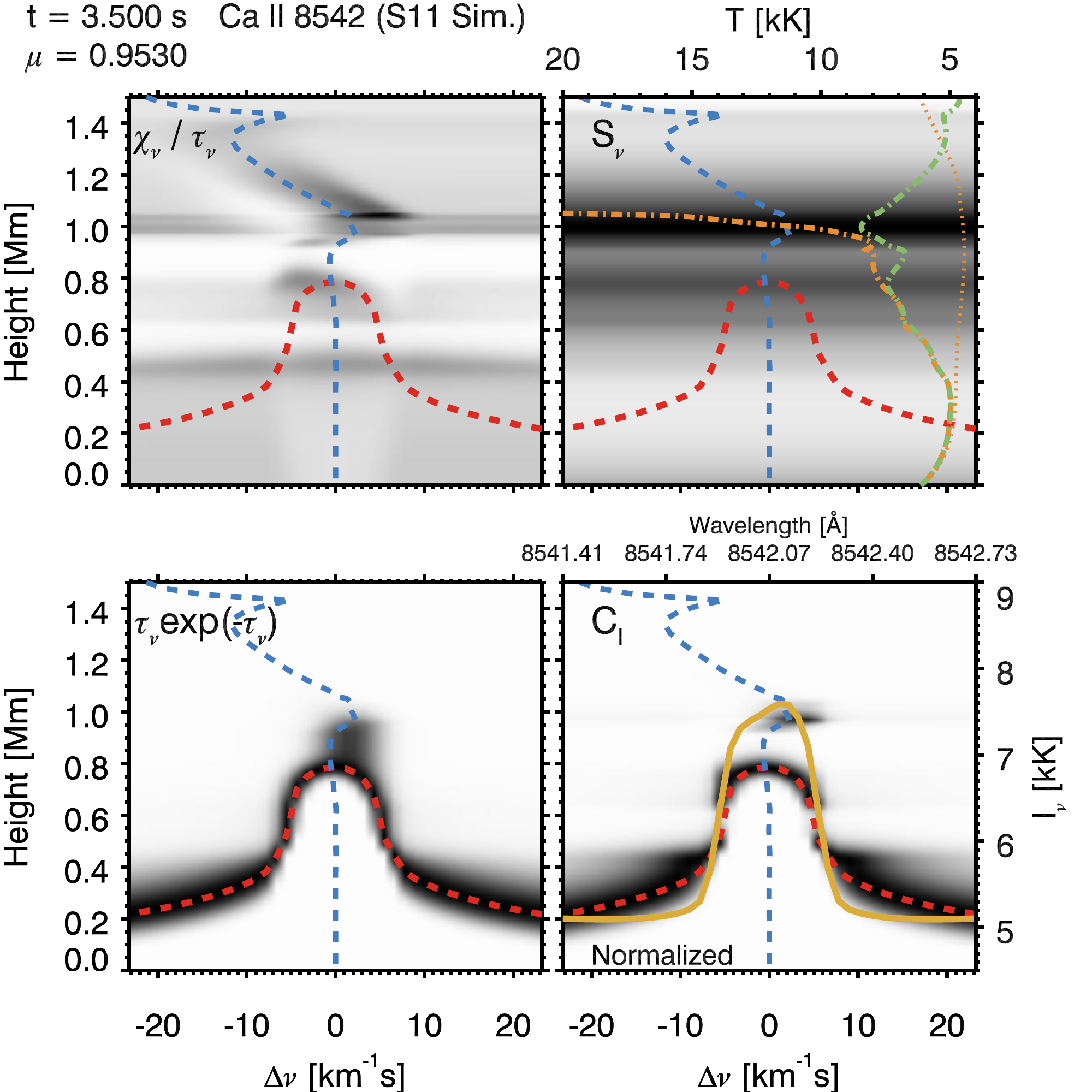}}		
		\subfloat[]{\includegraphics[width = 0.5\textwidth, clip = true, trim = 0.0cm 0cm 0cm 0.cm]{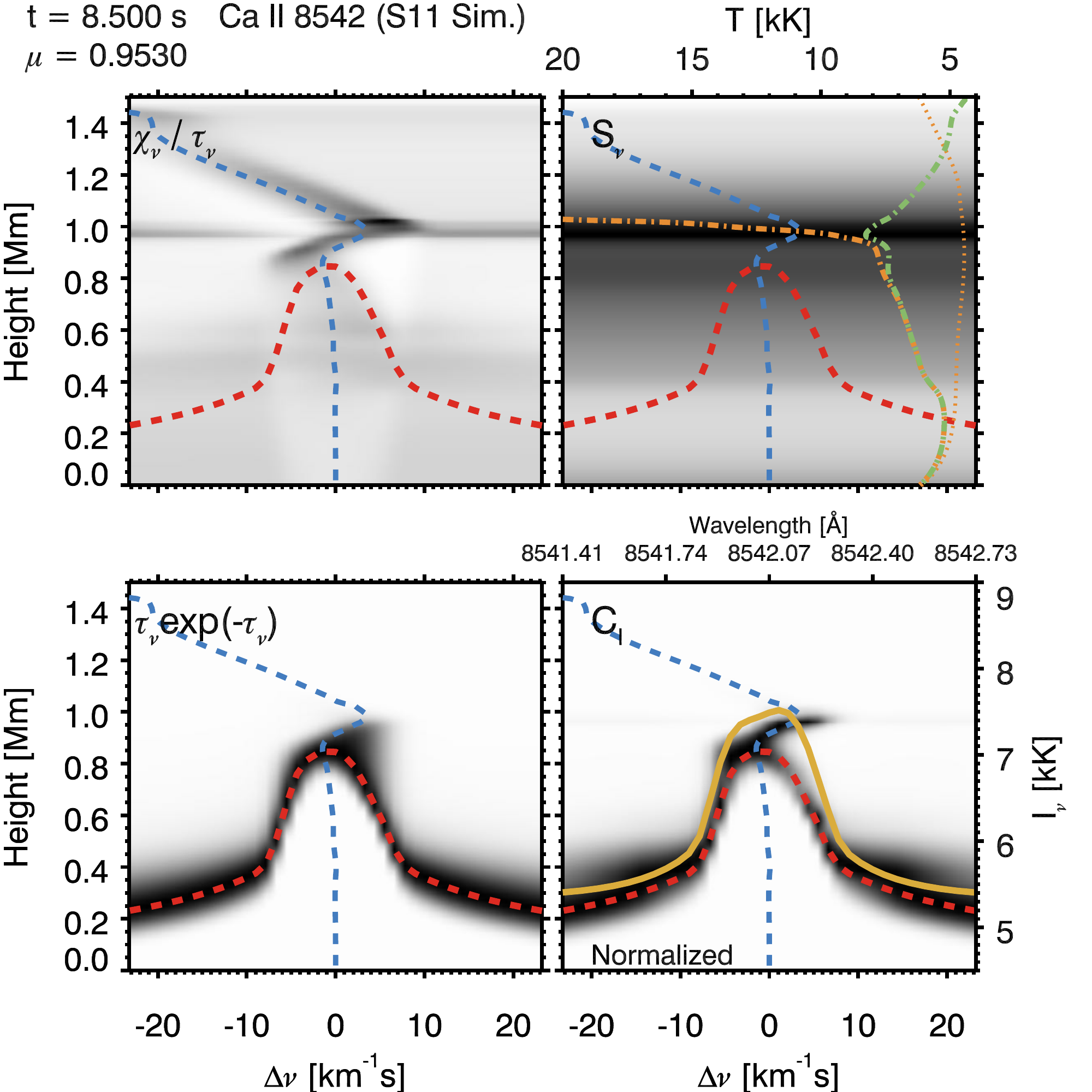}}		
		}
		}
			\caption{\textsl{Ca~\textsc{ii} 8542\AA\ line formation in the AW beam (S11) simulation at different times as indicated. Lines are as described in Figure~\ref{fig:caii_elec_contfn}}}
	\label{fig:caii_aw_contfn}
\end{figure*}

Some example snapshots are shown in Figures~\ref{fig:caii_elec_contfn} \& \ref{fig:caii_aw_contfn} to illustrate the line formation. In these figures the background images show the components of the contribution function, ${\chi_{\nu}}/{\tau_{\nu}}$ (top left), $S_{\nu}$ (top right) \& $\tau_{\nu}e^{-\tau_{\nu}/\mu}$ (bottom left), and the contribution function itself, $C_{\rm{I}}$ (bottom right). These are inverse scale so that dark regions show large values of each term. The $C_{\rm{I}}$ images are normalised within each wavelength bin to better show the contribution to the wing intensity (which is much lower intensity than the line core). These are shown as a function of wavelength and height, with wavelength expressed as a Doppler velocity, redshifts being positive. The lines shown on each figure are the atmospheric velocity (blue, dashed), where positive is downflow, the $\tau_{\nu}=1$ curve (red, dashed), the line source function (green, dot-dashed), the Planck function (purple, dot-dashed), the Planck function at $t=0$~s (purple, dotted), and the emergent intensity (yellow, solid). The source function, Planck function and emergent intensity are expressed in units of radiation temperature. If the line is optically thick then the contribution to the emergent intensity originates from near the $\tau_{\nu}=1$ height.\\

{\bf \textsc{Electron Beam model}}

{\bf t = 0-0.25~s}: see Figure~\ref{fig:caii_elec_contfn}(a). The very high temperature where the beam energy is deposited causes a reduction in the population of the Ca~\textsc{ii} 8542\AA\ upper levels above $1$~Mm. The $\tau_{\nu} = 1$ layer forms below this at $\sim0.9$~Mm where there is an increase in Ca~\textsc{ii} 8542\AA\ upper level populations. The contribution function peaks here with an additional small (optically thin) contribution to the line core, from between $0.9 - 1.1$~Mm. The line source function is strongly coupled to the Planck function up to the core formation height (Figure~\ref{fig:caii_elec_contfn}(a), top left panel), which has increased due to enhanced temperature, and so the line core is in emission.  Moving away from the core, the wings are formed progressively deeper in the atmosphere.  Far wings are formed at only 0.2~Mm above the photosphere. As time proceeds (not shown) the rapid increase in temperature at even greater depths starts to depopulate the upper levels further, driving down the formation height of the line core to $\sim0.75$~Mm. The line source function at this altitude is still strongly coupled to the Planck function, which is smaller, so the emergent intensity decreases. The line core still has an optically thin contribution from $0.75-1.1$~Mm. 

{\bf t = 0.25-1.25~s}: see Figure~\ref{fig:caii_elec_contfn}(b). In the region $\sim0.8-1$~Mm the upper level of Ca~\textsc{ii} begins to repopulate, and the $\tau = 1$ height moves upward to $\sim0.85$~Mm where it remains for the rest of the simulation. The line source function couples to the Planck function at ever increasing heights (up to 1~Mm), increasing the intensity of the optically thin component. The source function at the core formation height increases significantly compared to only a small change in the  wings, so the profiles look very narrow at these times.  A small downflow develops at~$1.1$~Mm,  causing an increase to the $\chi_{\nu}/\tau_{\nu}$ term redward of the line core. This provides an additional small, optically thin contribution to the red wing, and marks the start of the red asymmetry in the line profile.  

{\bf t = 1.5-6~s}: see  Figure~\ref{fig:caii_elec_contfn}(c).The upper level is repopulated across a wider range of heights so that emission from the near wings originates from higher layers, and intensity increases. The $\tau=1$ curve widens, increasing line width. The condensation has increased in magnitude and moved lower in the atmosphere, making the optically thin contribution to the red wing more pronounced. A small upflow at the line centre formation height ($\sim$ 0.85Mm) blueshifts the core.

 {\bf t = 6-8~s}: The peak formation height is still $\sim~0.85$~Mm and since the source function is still strongly coupled to the Planck function, it changes only little. Emergent intensity decreases slightly, largely due to the condensation which continues to move deeper in the atmosphere, adding more to the red wing and less to the line core as time progresses. 

{\bf t = 8-10~s}: As the under dense, high temperature bubble forms at $\sim 1.2$~Mm it expands, starting a second (much larger) condensation. There are not enough emitters at the height of the condensation to have much effect on the Ca~\textsc{ii} 8542\AA\ line, but the condensation does compress the atmosphere, resulting in a small increase to the optically thin emission in the red wing, increasing the asymmetry.  \\

{\bf \textsc{AW Model}}

{\bf t = 0-0.25~s}: (not shown) unlike in the EB simulation, at this time the line is still in absorption, as the source function is not strongly coupled to the Planck function, but decreases above $0.75$~Mm. The electron density is much lower compared to the EB simulation since there is initially less H ionisation below  0.95~Mm at early times. There is, however, a narrow region in the mid-chromosphere where there is a sufficient population of Ca~\textsc{ii} to produce an optically thin component to the contribution function near 1.1~Mm. 

{\bf t = 0.25-1.25~s}: see Figure~\ref{fig:caii_aw_contfn}(a). The temperature increase at greater depths depopulates the Ca~\textsc{ii} 8542\AA\ upper level between 0.65 - 0.9~Mm driving the formation height down to around 0.65~Mm, lower than in the EB simulation. The lower temperature at this height compared to the formation height of the line in the EB simulation means a smaller line intensity. The source function where the line forms is now more strongly coupled to the Planck function, so the line is in emission.  An upflow above $1$~Mm shifts the absorption profile to the blue, meaning more blue-wing than red-wing absorption. This creates  a small but growing asymmetry. In addition, the strength of the optically thin contribution increases, and is stronger on the red side of the profile due to a small condensation. Together, these features have the effect of making the emergent profile appear redshifted. Although the optically thin emission originates from a much narrower layer than in the EB simulation, the AW simulation has a somewhat higher temperature in this region, (the difference is $\approx 2000$~K). This raises the intensity of the optically thin component compared to in the EB simulation.

{\bf t = 1.25-6.5~s}:  see Figure~\ref{fig:caii_aw_contfn}(b,c).

Beginning around $t=2$~s the upper-level populations increase, raising the height of the $\tau_{\nu}=1$ layer. Figure~\ref{fig:caii_aw_contfn}(b) shows the start of this process. By $t=3.5$~s populations have increased enough that the location of the $\tau_{\nu}=1$ layer is raised and the line width increased. The line core is slightly blueshifted, but the optically-thin redshifted component arising from the condensation means that the line is further broadened and peaks in the red. By $t=6.5$~s the $\tau_{\nu} = 1$ height has risen to around 0.81~Mm, increasing the intensity of the whole line as it is formed in a region of higher temperature. It takes longer for upper level populations in the AW simulation to reach a similar state to the EB simulation, meaning that the intensity increase is slower. 

This occurs due to a difference in the length of time it takes for recombinations to increase the amount of Ca~\textsc{ii}. In both simulations the fraction of Ca~\textsc{iii} to Ca~\textsc{ii} increases significantly to large depth. The ratio of $n_{\rm{Ca~\textsc{iii}}}/n_{\rm{Ca~\textsc{ii}}}$ is $\sim50\%$ between 0.5-0.6~Mm, increasing quickly over a narrow height range so that calcium is almost all ionised to Ca~\textsc{iii} above 0.6~Mm. In the EB simulation, recombinations to Ca~\textsc{ii} between $z\sim0.7-0.9$~Mm take place early ($t\sim1$~s) so that the Ca~\textsc{ii} 8542\AA\ upper level is subsequently populated at that height. In contrast, the decreased electron density in the AW simulation relative to the EB simulation, at these heights, means that recombinations to Ca~\textsc{ii} do not occur as quickly. The electron density in the AW simulation begins to increase around $t\sim 2.25$~s, which in turn allows recombinations, and for the Ca~\textsc{ii} upper level to become populated over the next few seconds. 

{\bf t = 6.5-10~s}: There is little change over the remainder of the simulation, other than an intensity increase in the red wing as the density increases and the source function couples strongly to the Planck function. Similarly to the EB simulation, the small condensation extends further into the wing versus the line core towards the end of the simulation, reducing the extent of the red peak, and broadening more of the red wing (Figure~\ref{fig:caii_aw_contfn}(d)).\\

\subsection{Mg~\textsc{ii} k Line Profiles}\label{sec:mgii_k_profiles}

\begin{figure}
	\centering
	\vbox{
		\subfloat{\includegraphics[width = 0.5\textwidth, clip = true, trim = 1.05cm 0cm 0.15cm 0.2cm]{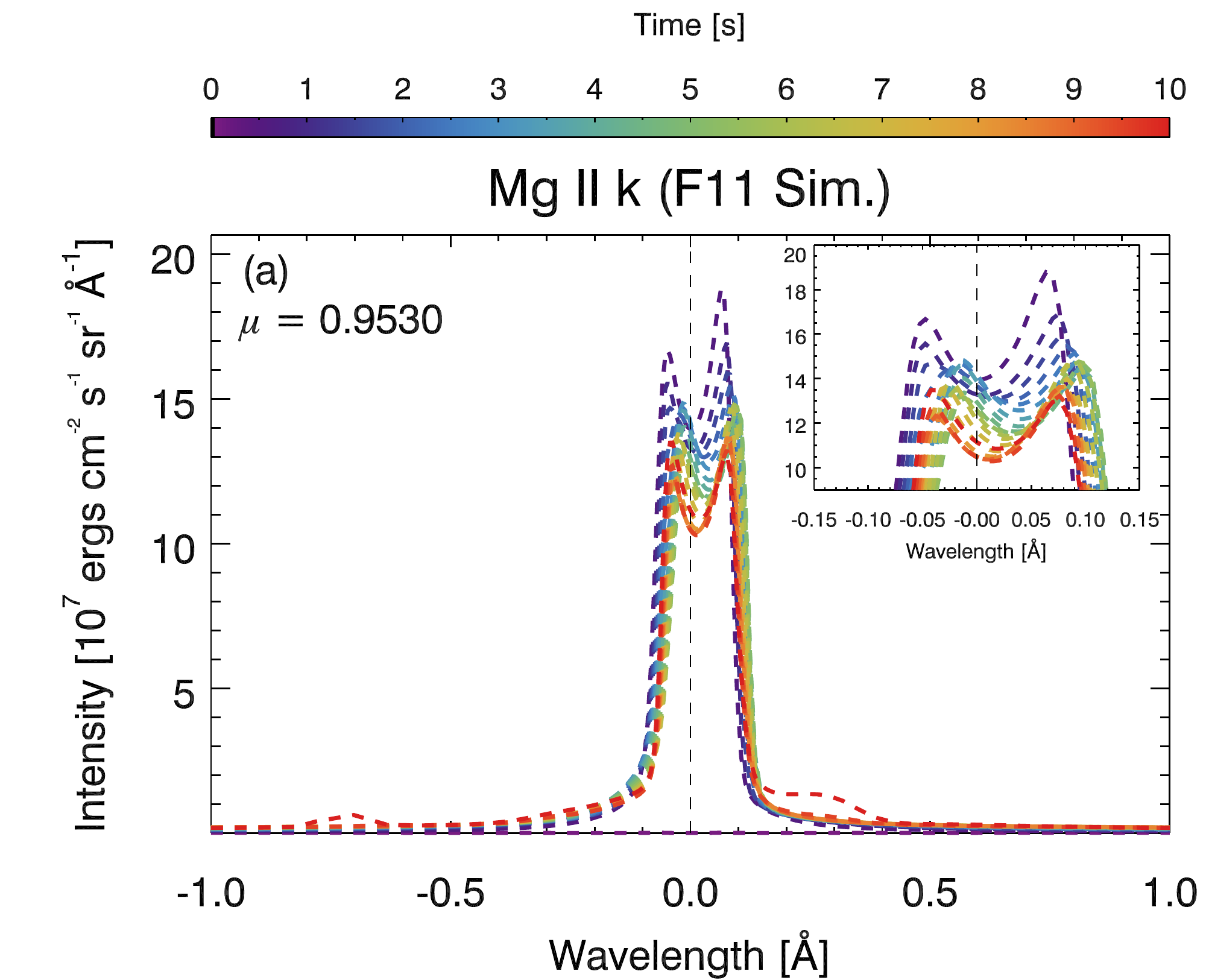}}
		}
	\vbox{
		\subfloat{\includegraphics[width = 0.5\textwidth, clip = true, trim = 1.05cm 0cm 0.15cm 0.2cm]{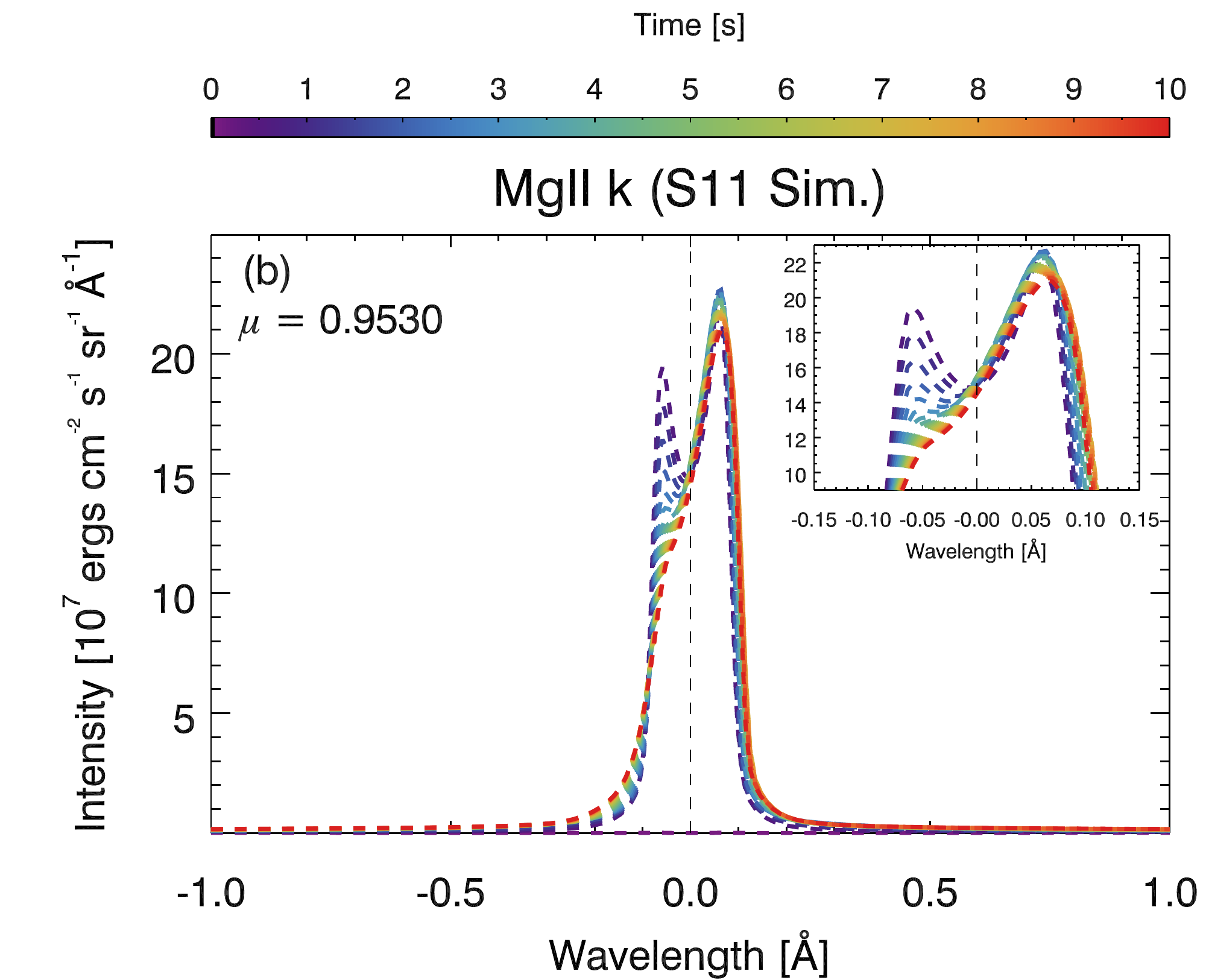}}	
		}
	\caption{\textsl{The Mg~\textsc{ii} k line computed in (a) the EB simulation and (b) the AW simulation. In both cases colour represents time, and the inset shows a closer view of the line core.}}
	\label{fig:mgii_k_lines}
\end{figure}

The Mg~\textsc{ii} h \& k resonance lines in the quiet Sun are formed over a wide range of chromospheric heights. They usually appear as doubly peaked profiles with a central reversal. The red and blue peaks of the k line are referred to as the k2r and k2v component, and the centrally reversed core as the k3 component. \cite{2015A&A...582A..50K} and \cite{2015SoPh..290.3525L} recently discussed these lines as observed in a solar flare. They appeared as redshifted, single peaked profiles with a blue asymmetry at times of strongest redshift. 

RADYN uses the assumption of CRD when computing line profiles but it has been shown by \cite{2013ApJ...772...89L} that this is not valid for Mg~\textsc{ii}. Therefore we use the output RADYN atmospheres every 0.25~s as input to the RH radiative transfer code \citep{2001ApJ...557..389U}, which does use PRD, to synthesise the Mg~\textsc{ii} spectra, using a 22-level Mg~\textsc{ii} model atom. RH can perform the PRD radiative transfer with either the fast approximation/Hybrid PRD scheme of \cite{2012A&A...543A.109L} or with full angle-dependence. Full angle-dependent PRD takes into account that in an atmosphere with strong velocity gradients the radiation field is non-isotropic, and requires computing the angle-dependent redistribution function (costly both in time and memory). Instead it is possible to use the assumption of angle-averaged PRD and include transforms to/from the rest frame of particles (see \citealt{2012A&A...543A.109L} for details), to save on computational time whilst obtaining a good approximation to full angle-dependent PRD. We performed some tests of full angle-dependent PRD compared to the Hybrid PRD finding that angle-dependent computations took significantly longer with little difference in the emergent profile. We therefore used the Hybrid PRD. Figure~\ref{fig:mgii_k_lines}(a) shows the EB Mg~\textsc{ii} k profiles, and Figure~\ref{fig:mgii_k_lines}(b) shows the AW Mg~\textsc{ii} k profiles.

In the EB simulation the line profiles have an obvious central reversal at all times. This quickly becomes shallower, and the whole line becomes more intense. From $t=0.25-0.5$s k2r is stronger than k2v and the line reaches its maximum intensity. Over the next few seconds of the simulation the line core appears redshifted and kr2 decreases so that the k2r and k2v are largely symmetric. The blue wing develops an enhancement that moves steadily more blueward, making the line asymmetric. Between $t=4-5.5$~s k2r is again stronger than k2v, and the redshift of the line core becomes smaller. By $t=7$~s k2r and k2v have roughly equal intensity. By the end of the simulation a strong enhancement to the red between  0.15 and 0.5~\AA\ from line center has developed, and a weaker enhancement to the blue wing at $\sim$~0.75\AA . 

The line profiles in the AW simulation are very different from the EB profiles. Before $t=0.25$~s the profiles are very similar, though the AW profiles are more intense (AW k2r intensity is $I_{k2r} = 21.0\times10^7$~ergs~cm$^{-2}$~s$^{-1}$~sr$^{-1}$~\AA$^{-1}$, and for EB $17.0\times10^7$~ergs~cm$^{-2}$~s$^{-1}$~sr$^{-1}$~\AA$^{-1}$). By $t = 1.25$~s,  kr2 is very strong compared to k2v, and the line core is slightly blueshifted. In relation to k2r, k2v continues to decrease.  The maximum intensity of k2r is at $t=2.25$~s, with $I_{k2r} = 22.5\times10^7$~ergs~cm$^{-2}$~s$^{-1}$~sr$^{-1}$~\AA$^{-1}$ (compared to a maximum intensity in the EB simulation of $I = 19.0\times10^7$~ergs~cm$^{-2}$~s$^{-1}$~sr$^{-1}$~\AA$^{-1}$). At this point the central reversal is very shallow, the k2v is very weak, and k2r dominates. By $t=5$~s the k2v decreases so much that it becomes difficult to discern its presence. Instead the profile appears single peaked, with an extended blue wing or shoulder,  and a wider blue wing between 0.1\AA\ and 0.2\AA\ blueward of line centre, than the red wing.

\subsection{Mg~\textsc{ii} k Line Formation }\label{sec:mgii_k_formation}
\begin{figure*}
	\centering
	\vbox{
	\hbox{
		\subfloat[]{\includegraphics[width = 0.5\textwidth, clip = true, trim = 0.0cm 0cm 0cm 0.1cm]{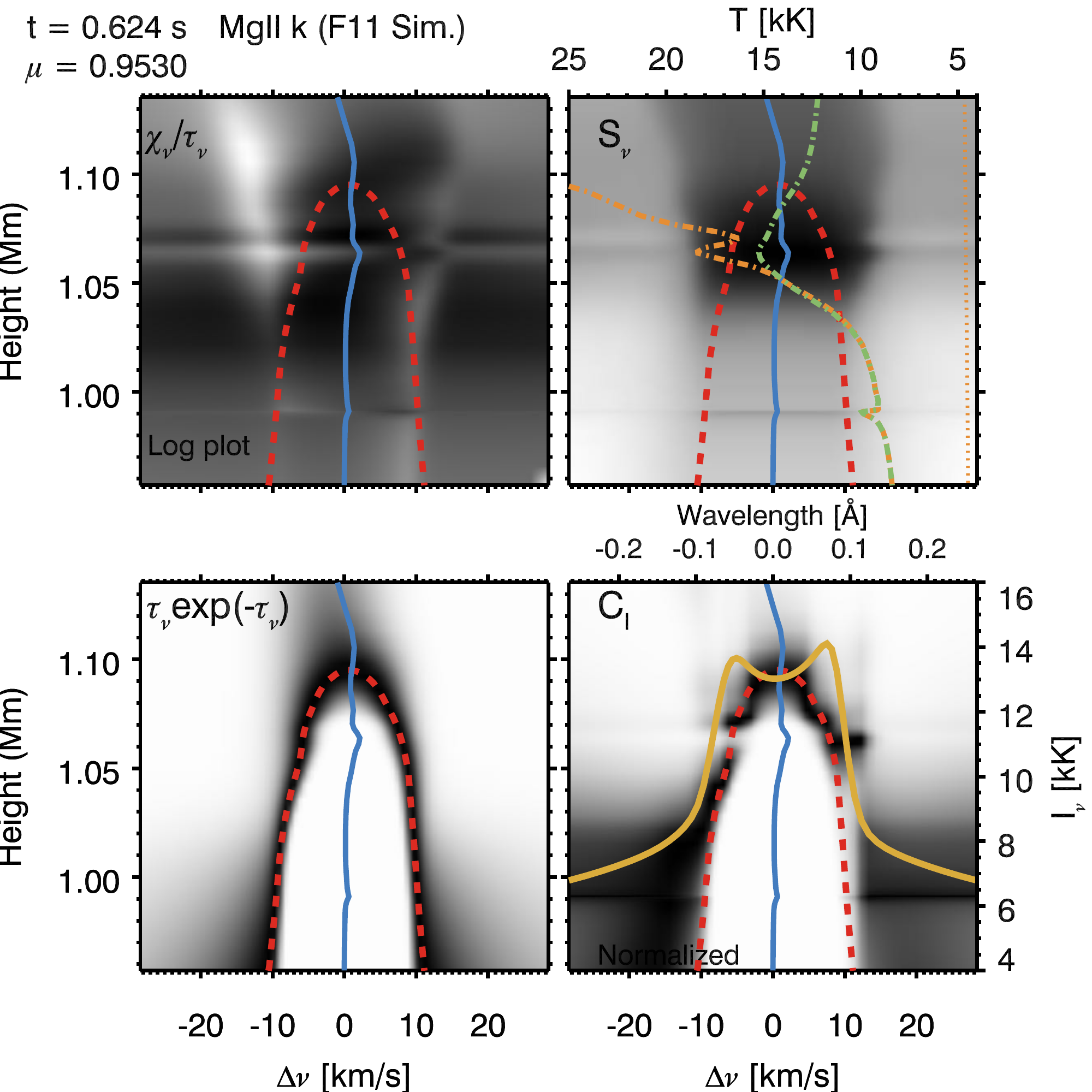}}		
		\subfloat[]{\includegraphics[width = 0.5\textwidth, clip = true, trim = 0.1cm 0cm 0cm 0.1cm]{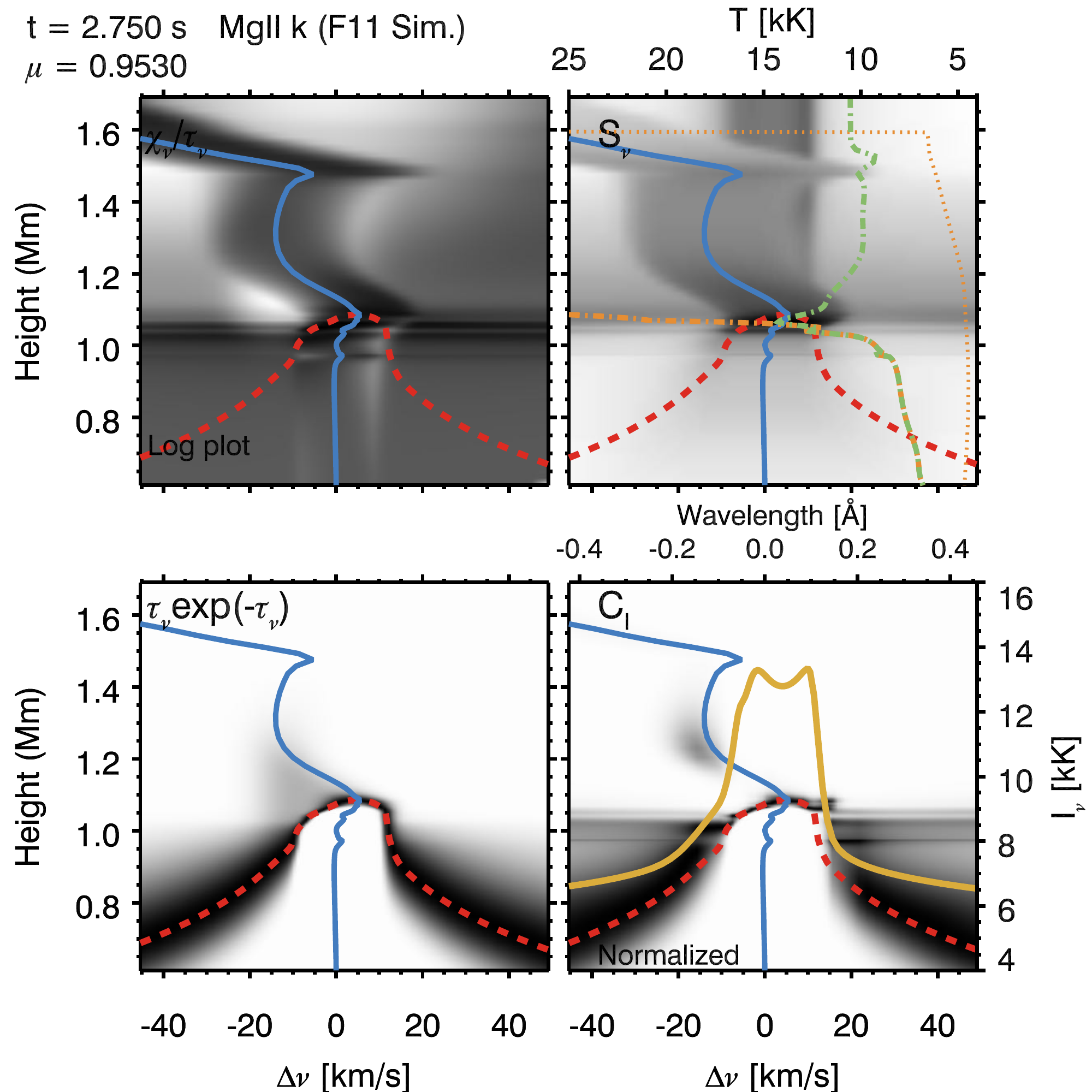}}		
		}
		}
		\vbox{
		\hbox{
		\subfloat[]{\includegraphics[width = 0.5\textwidth, clip = true, trim = 0.1cm 0cm 0cm 0.1cm]{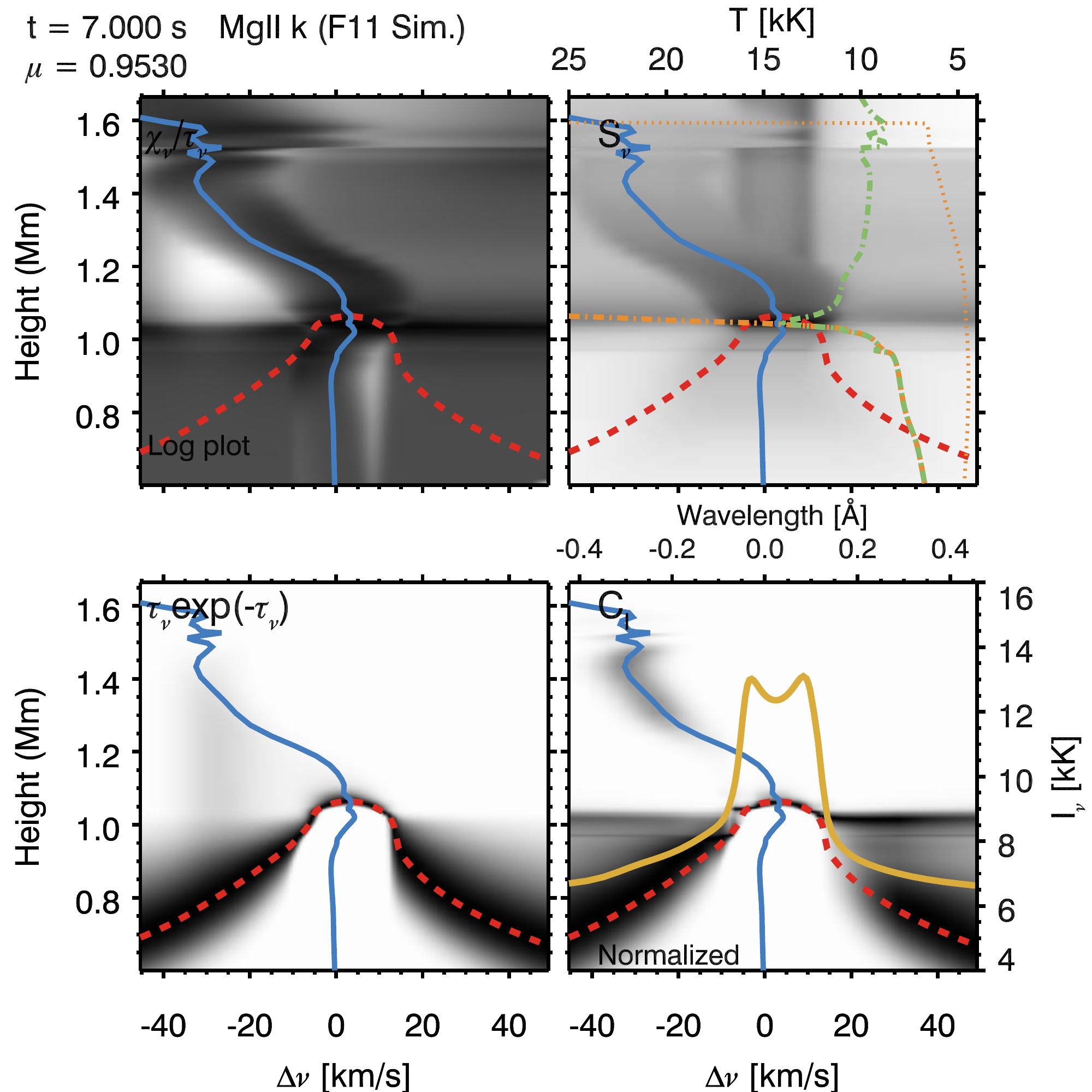}}		
		\subfloat[]{\includegraphics[width = 0.5\textwidth, clip = true, trim = 0.1cm 0cm 0cm 0.1cm]{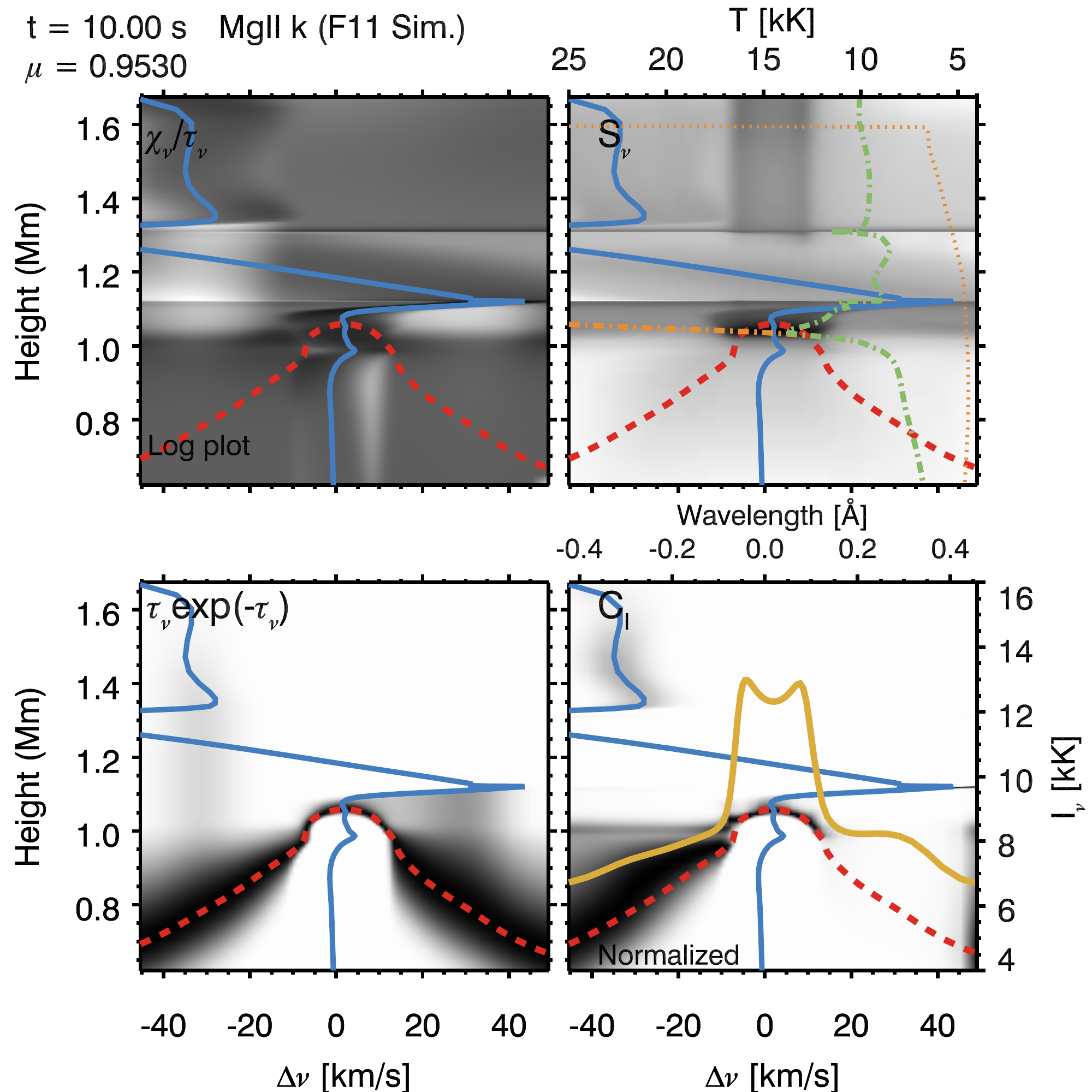}}		
		}
		}
			\caption{\textsl{The formation of the Mg~\textsc{ii} k line in the EB simulation at four timesteps, as indicated on each panel. The lines are as described in Figure~\ref{fig:caii_elec_contfn}. Note the different scales used in (a).}}
	\label{fig:mgiik_contfn_elec}
\end{figure*}

\begin{figure*}
	\centering
	\vbox{
	\hbox{
		\subfloat[]{\includegraphics[width = 0.5\textwidth, clip = true, trim = 0.1cm 0cm 0cm 0.1cm]{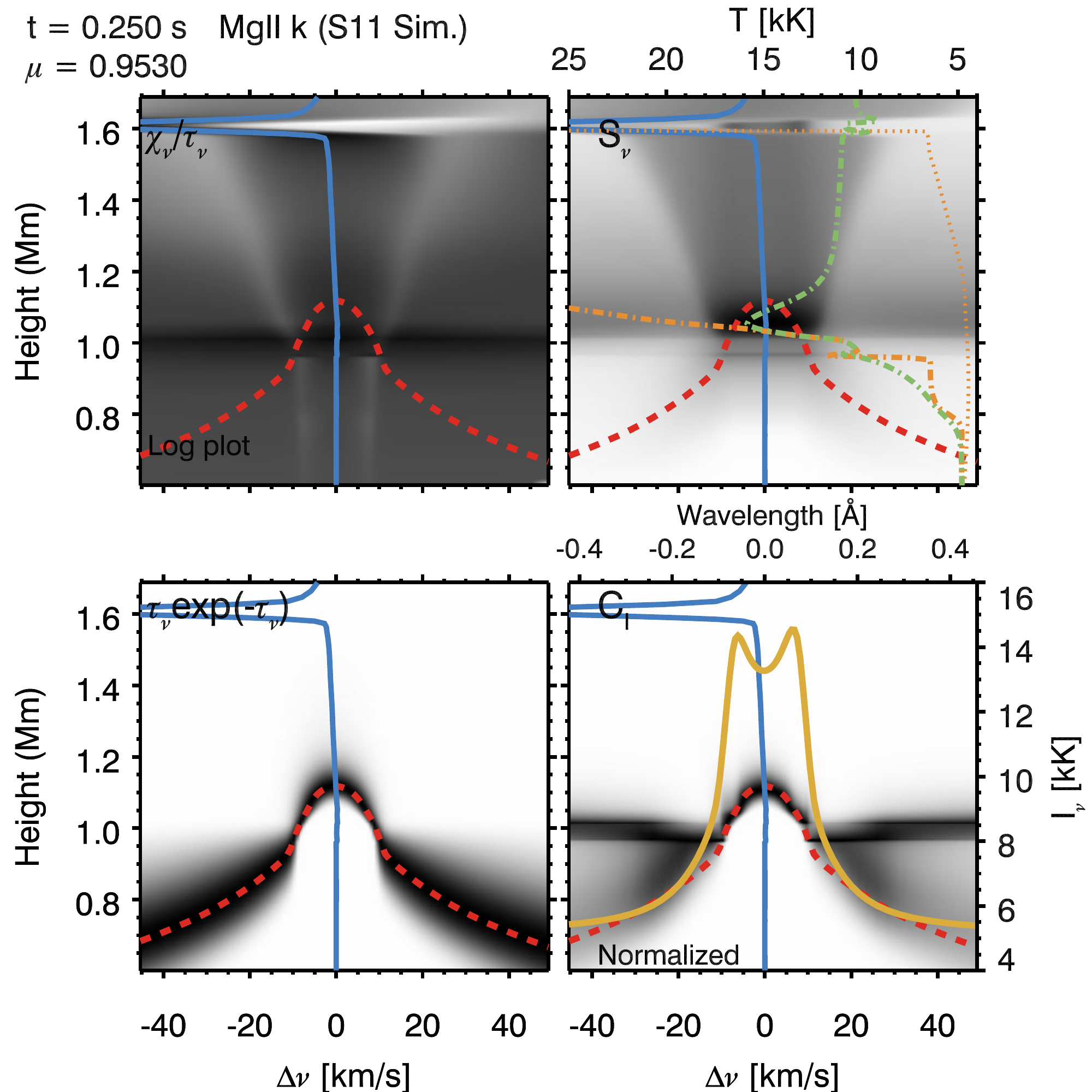}}		
		\subfloat[]{\includegraphics[width = 0.5\textwidth, clip = true, trim = 0.cm 0cm 0cm 0.1cm]{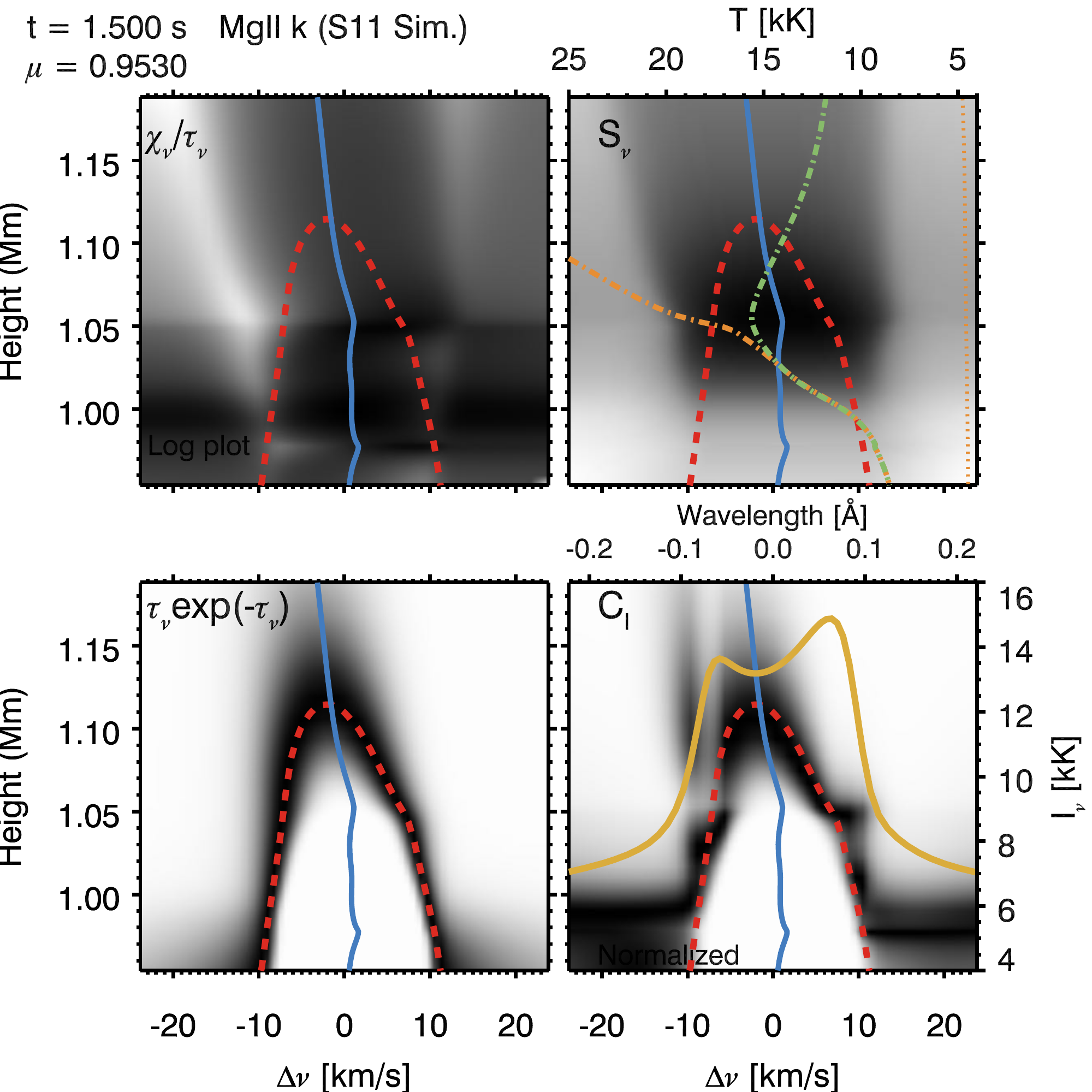}}		
		}
		}
		\vbox{
		\hbox{
		\subfloat[]{\includegraphics[width = 0.5\textwidth, clip = true, trim = 0.1cm 0cm 0cm 0.1cm]{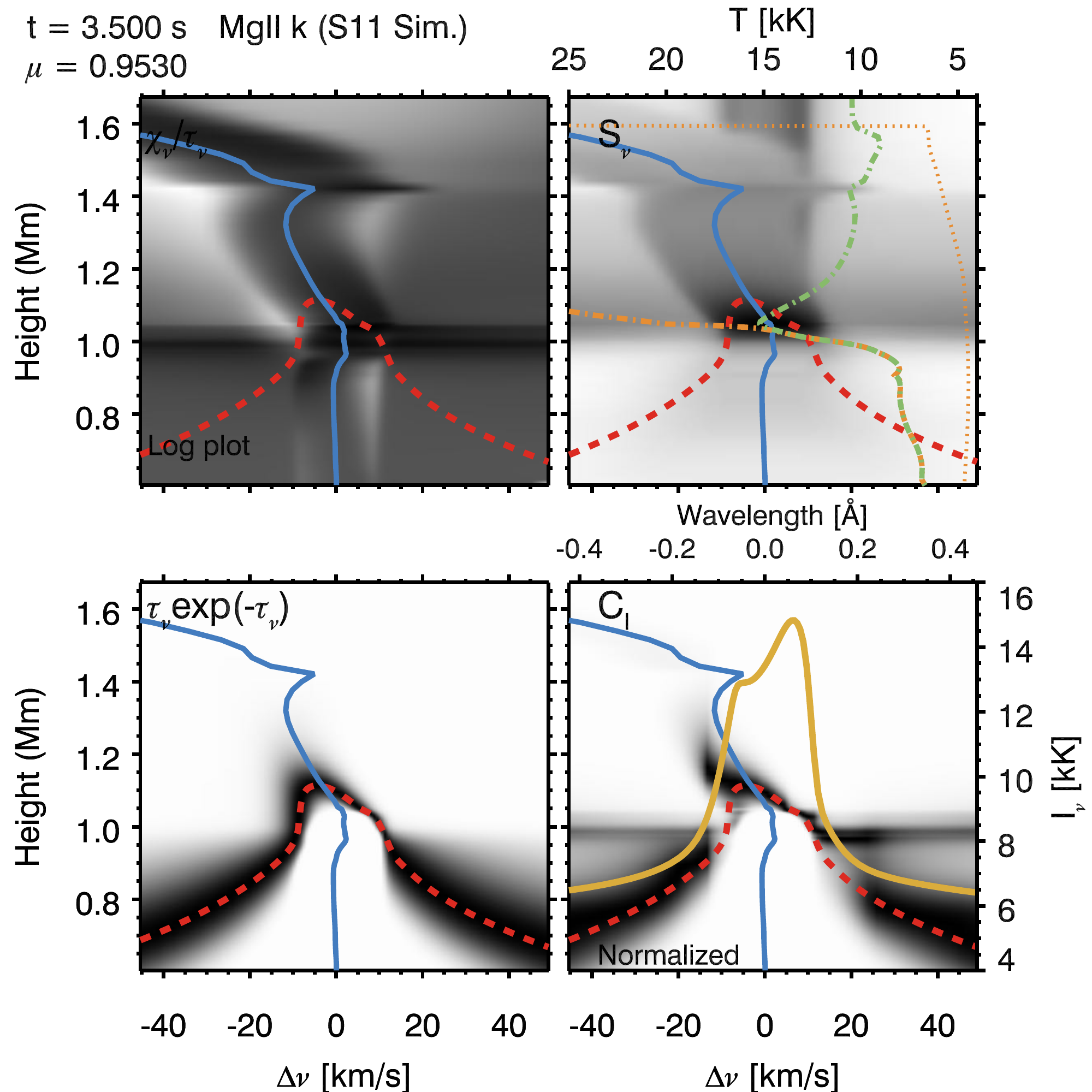}}		
		\subfloat[]{\includegraphics[width = 0.5\textwidth, clip = true, trim = 0.1cm 0cm 0cm 0.1cm]{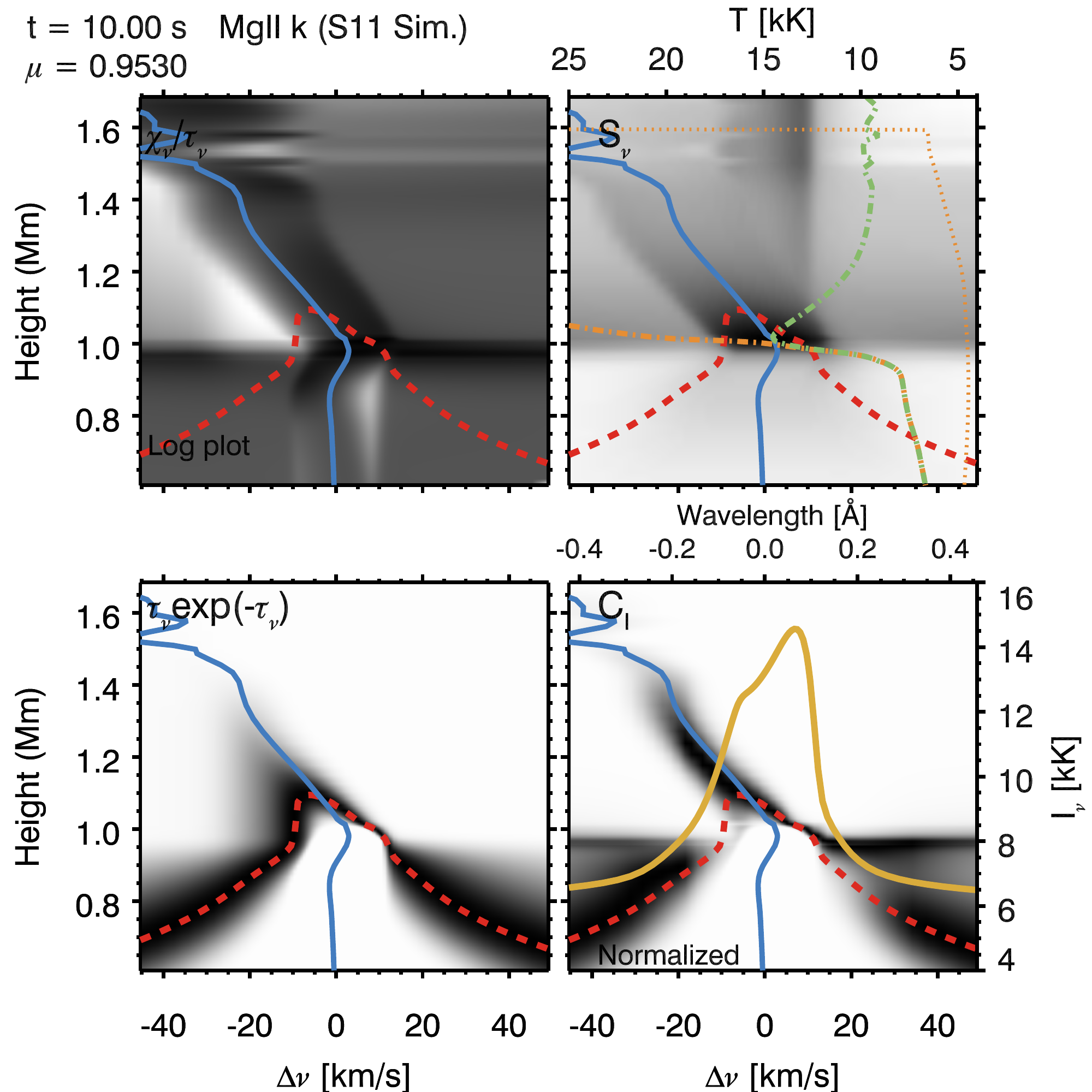}}		
		}
		}
			\caption{\textsl{The formation of the Mg~\textsc{ii} k line in the AW simulation at four timesteps, as indicated on each panel. The lines are as described in Figure~\ref{fig:caii_elec_contfn}. Note the different scales used in (b).}}
	\label{fig:mgiik_contfn_aw}
\end{figure*}

The computation of the Mg~\textsc{ii} k line formation using RH differs from Ca II 8542\AA\ using RADYN because using RH with PRD means that the source function is frequency-dependent and varies across the profile. Figures~\ref{fig:mgiik_contfn_elec} \& \ref{fig:mgiik_contfn_aw} show the formation of the line in the EB and AW simulation respectively. Four snapshots are shown, with the panels and lines as described in Section~\ref{sec:caii_8542_form}.\\

{\bf \textsc{EB Model}} 

{\bf t = 0.25~s:}  The line core formation height drops from its pre-flare location just below the  TR to $\sim1.1$~Mm, and k2r and k2v form slightly lower at  $\sim1.05$~Mm. This drop in formation height occurs because the Mg~\textsc{ii} k upper level becomes depopulated above $\sim1.15$~Mm, due to heating, while populations around 1.1~Mm increase. The central reversal occurs because at the k3 formation height, the source function has decoupled from the Planck function, and is decreasing with increasing height, whereas at the k2r and k2v peak formation height the source function is more strongly coupled to the Planck function, giving a higher intensity. However, the high temperatures and density produces conditions such that the difference in formation height between k3 and k2r,v components is fairly small, and so the depth of the reversal feature is very much smaller than it is in the quiet Sun. The line wings form deeper in the atmosphere, from $\sim0.7 - 1$~Mm in the wavelength range shown. 

{\bf t = 0.25-1~s:} see Figure~\ref{fig:mgiik_contfn_elec}(a). A small downflow has formed immediately above the core formation height, which increases the number of emitters contributing to the red wing relative to the blue wing. This widens the k2r peak and increases the intensity of k2r compared to k2v. 

{\bf t = 1-3~s:} see  Figure~\ref{fig:mgiik_contfn_elec}(b). The downflow develops and moves deeper, redshifting the core and the emission peaks. The peak of the opacity is also in the red (see upper left panels in Figure~\ref{fig:mgiik_contfn_elec}) so that red wing photons produced below the condensation are absorbed more than blue wing photons, steepening the extinction profile of the red wing. Between $\sim1.1-1.2$~Mm an upflow moves some emitters upwards to locations at smaller optical depths, so that extra blue-wing photons are emitted from 0.10 - 0.15~\AA~bluewards of the rest wavelength.   
 
{\bf t = 3-7~s:} see Figure~\ref{fig:mgiik_contfn_elec}(c). As the downflow moves deeper into the atmosphere it slows and the redshift of the core decreases. The k2r peak is sometimes formed slightly higher (and is slightly more intense) than the k2v peak, but when the overall redshift becomes smaller this height difference  also reduces. As the upflow speed increases the contribution function for optically thin blue-wing emission is pushed further out, to around 0.15 - 0.20~\AA~ from the core. It originates from a height of $\sim1.3-1.4$~Mm 

{\bf t = 8-10~s:} see Figure~\ref{fig:mgiik_contfn_elec}(d). A hot bubble has formed at $\sim1.2$~Mm, creating a large condensation that travels downwards. The flow associated with the bubble does not reach the core formation height, so has little effect on the k3 or k2 components, but it does result in peaks appearing in the red and blue wings. The condensation creates a very narrow layer of enhanced electron and mass density. The population of the Mg~\textsc{ii} k upper level increases at the condensation height, resulting in a strong red-shifted source function. The lower left panel of Figure~\ref{fig:mgiik_contfn_elec}(d) shows that the emitters in the condensation increase the attenuation of red wing photons, meaning that their contribution is almost exclusively from the condensation. The steep velocity gradient results in a bump in the red wing at  0.15 - 0.45~\AA\ from the rest wavelength. Similarly, the upflow results in emission between $\sim 0.60 - 0.80$~\AA\  blueward of the rest wavelength. \\

{\bf \textsc{AW Model}} 

{\bf t = 0.25~s:} see Figure~\ref{fig:mgiik_contfn_aw}(a). A shallow upflow is present in the chromosphere, from around the formation height of the k line at $\sim1.125$~Mm, to 1.55~Mm, resulting in a slightly blueshifted line core even at early times. Above this a much stronger upflow carries chromospheric material into the TR. The core is formed around $1.1-1.15$~Mm, and is centrally reversed, with a deeper reversal than the EB simulation as it is more decoupled from the Planck function. The k2 components are symmetric about the core, and are formed at $\sim1.05$~Mm. Since the AW atmosphere is hotter between 0.9-1.1~Mm, the k2 components are more intense than their EB counterparts. The core forms higher than in the EB simulation since the latter is hotter above 1.1~Mm, depopulating the upper level of the k line more than in the AW simulation. 

{\bf t = 0.25-2.0~s:} see Figure~\ref{fig:mgiik_contfn_aw}(b). The core becomes blueshifted as the  $\tau=1$ surface moves with the upflow. The k2v and k3 formation height move closer.  The k2r peak is formed slightly lower, and is more coupled to the Planck function so more intense that the k2v peak. The k2r peak appears wider than the k2v peak because the extinction on the blue side is higher - more photons are absorbed by the shifted opacity peak.   

{\bf t = 2.0-5.0~s:} see Figure~\ref{fig:mgiik_contfn_aw}(c). The upflow steepens between 1.1 and 1.3~Mm, and results in blueshifted optically thin emission between $\sim$ 0.10 - 0.15~\AA\ blueward of the rest wavelength. It is difficult to distinguish the core from the k2v peak. The theoretical core is defined as the maximum of the $\tau=1$ curve over the line, but by $t=5$~s this occurs at roughly the same formation height as the k2v peak. This may be due to the upflow, which extends over a wide height range above the core formation height. The opacity peak is shifted to the blue so that blue photons are preferentially absorbed relative to red, producing a red asymmetry. 

{\bf t = 5.0-10.0~s:} see Figure~\ref{fig:mgiik_contfn_aw}(d). The line profile does not change very much, but features become more extreme. The theoretical core position becomes more blueshifted, the optically thin component extends further into the blue wing, and the profile appears singly peaked with a blue `shoulder'. The k2r peak is very intense, formed in a region of high temperature and electron density, where the source function couples to the Planck function. Observationally k2r might be confused with the line core since the central reversal has all but vanished. A small condensation near the k2r formation height broadens this feature.


\section{Summary}\label{sec:discussion}
In the simulations presented, we find substantial differences in the atmospheric structure and flows in the AW simulations compared to the EB simulations. These differences result primarily from the different heating profiles.
\begin{enumerate}
\item In the EB simulation, the heating is strongly concentrated at the column depth corresponding to the stopping depth of the electrons at the beam cut-off energy. In the AW simulation the heating profile for the monochromatic wave is flatter. 
\item Both simulations result in fast low-density upflows in the upper chromosphere. For the energy flux used ($10^{11}\rm{ergs~cm^{-2}~s^{-1}}$) the localised, strong heating in the EB simulation means that helium as well as hydrogen becomes very highly ionised, and the plasma has a reduced ability to radiate. The resulting increase in temperature and pressure launches strong secondary, high-density, upflows and downflows in the mid-chromosphere. A similar but much less pronounced process results in a secondary upflow in the AW simulation, but the secondary flows are much weaker in the AW simulation for the same energy flux.
\item  Early in the EB simulation, the temperature in the lower chromosphere, from 0.6-0.9~Mm, is a few 100~K higher than in the AW simulation. After a short time this reverses and the AW simulation is hotter. The AW simulation is significantly hotter from $\sim0.9-1.1$~Mm, by $5,000-10,000$~K. 
\item In both models, H ionisation is complete above 1~Mm. In the EB model, H ionisation is higher between 1~Mm and 0.7~Mm compared to the AW model, due to non-thermal ionisation from the electron beam. After $\sim2$~s, ionisation at these depths in the AW simulation increases, and by the end of the simulation matches the EB simulation. The electron density in the EB simulation is mostly higher than in the AW simulation.
\end{enumerate}

These variations in the dynamic and thermodynamic properties of the atmosphere are reflected in the shape and variation of the lines in potentially distinguishable ways.
\begin{enumerate}
\setcounter{enumi}{4}
\item  Ca~\textsc{ii} 8542~\AA\  goes into emission almost immediately after the heating starts in the EB simulation, and rises quickly to its peak emission. In the AW simulation this takes a few seconds longer. In the case of Mg~\textsc{ii} k, the line peaks very quickly for both EB and AW heating.
\item For both Ca~\textsc{ii} 8542~\AA\  and Mg~\textsc{ii} k, the line intensity increases with time as the locations of the peaks of the contribution functions tend to move down as deeper layers of the atmosphere heat. Because the density in these deeper layers is higher, the emissivity is higher and the line source functions are more coupled to the Planck function. The shape and intensity of the wings is determined by the deeper atmosphere.
\item Ca~\textsc{ii} 8542~\AA\  does not show a reversal in either the EB or AW simulation. This is because the Ca~\textsc{ii} 8542~\AA\   source function and the Planck function are strongly coupled throughout the core formation region, so that the intensity increases with increasing temperature towards core formation heights. 
\item In the EB simulation small upflows at the location of Ca~\textsc{ii} 8542~\AA\  core formation lead to a small blueshift.  A weak, optically thin redshifted component contributes in the red wing. In the AW simulation the effect of redshifted, optically thin emission on the profile is more pronounced because the emission is stronger (higher temperature) and the downflow is larger.
\item In the EB simulation the core of Mg~\textsc{ii} k has a central reversal. The reversal occurs where the k3 source function has decoupled from the Planck function, leading to a drop in intensity compared to the k2v and k2r components which are formed where the source functions are more strongly coupled to the Planck function. 
\item In the AW simulation, there is initially a Mg~\textsc{ii} k central reversal, but over time the reversal becomes difficult to identify relative to the k2v peak. The profile becomes quite asymmetric because the source function and the Planck function are strongly coupled in a region with small downflows, leading to a strong red peak, whereas the blue side of the line core is dominated by weaker, optically thin emission from upflowing plasma. The line opacity is also primarily in the blue, increasing the asymmetry.
\item In the EB simulation there is very little optically thin Mg~\textsc{ii} k  emission near the line core, which remains quite symmetric. However we do see hints of the effect of absorption of red photons by downflowing material and blue photons by upflowing material. The far wings become enhanced by emission from the strong chromospheric condensation and upflow.

\end{enumerate}


\section{Conclusions and Next Steps}\label{sec:conclusion}
We have presented the results of radiation hydrodynamics simulations of a solar flare in which the atmosphere is heated by a monochromatic Alfv\'en wave or by an electron beam with the same total energy flux and duration. The electron beam calculation uses a well-established prescription for Coulomb heating and the treatment of Alfv\'en wave dissipation uses the WKB approximation which restricts it to waves of parallel wavelength less than the gradient scale length of the Alfv\'en speed. Line profiles from two important chromospheric lines were synthesized for both simulations.

Our results show that Alfv\'en wave dissipation is effective in heating the chromosphere and producing the high velocity flows observed in solar flares; indeed the temperature around 1~Mm in the AW simulation is ~5,000 - 10,000~K higher than in the EB simulation. In most regards the dynamic response of the atmosphere is similar in the two cases,  but the high-temperature shock in the mid-chromosphere that is a common feature of EB simulations is absent from the AW simulation. This has a significant impact on the line profiles.

The Ca~\textsc{ii} 8542\AA\ profiles produced in both the AW and EB simulation are similar to some recent flare observations \citep[e.g.][]{2015ApJ...804...56R} but neither show the blue asymmetries observed by \cite{2015ApJ...813..125K}. The AW simulation has a very small (0.05\AA) redshift. It takes slightly longer for the line intensity to increase in the AW simulation, but the delay we find here is likely to be an underestimate, as the transit time of the AW through the atmosphere is not captured in the simulation (the approximated form of AW energy input is a time average.) 

The Mg~\textsc{ii} k line profiles from the EB simulation do not provide a good match to observations, having a central reversal, and small amounts of broadening, redshift and asymmetry. Observed lines are very broad, usually do not have a central reversal  \citep[i.e. they are single peaked, e.g.][]{2015A&A...582A..50K, 2015SoPh..290.3525L} and have strong redshifts and asymmetries.  The Mg~\textsc{ii} spectra from the AW simulation, after a few seconds of energy input, appear single peaked,  redshifted and have an extended blue wing or shoulder, \citep[similar to observations by][]{2015A&A...582A..50K}.  The `theoretical core' (peak of the $\tau=1$ surface) is in fact formed in upflowing plasma, but more blue wing photons are absorbed than red wing photons, leading to the net red asymmetry \cite[see also][]{1994SoPh..152..393H}. This highlights the difficulty in interpreting observations of these optically thick lines.  

We have used a radiation hydrodynamics treatment to demonstrate that Alfv\'en waves can in principle heat the solar chromosphere and lead to emission of  important chromospheric flare lines. This follows the results that Alfv\'en waves can heat the TMR and upper chromosphere \citep{2013ApJ...765...81R,2016ApJ...818L..20R} including in flare simulations using a radiative loss function. Dissipation of Alfv\'en waves therefore looks like a viable candidate for generating many of the observed flare chromospheric signatures. We find that electron beam and Alfv\'en wave heating might be distinguishable via differences in the Mg~\textsc{ii} line profiles and their evolution. In our simulation of a single monochromatic wave, the more gradual heating profile of the Alfv\'en wave results in gentler flows and higher temperatures in the mid-chromosphere, producing single-peaked, asymmetric Mg~\textsc{ii} k line profiles more similar to current observations. However, we need to carry out a parameter study and also simulate the dissipation of a spectrum of waves. Simulations containing both electron beam and Alfv\'en wave heating will also be instructive.

The chromospheric response to intense heating, even in a 1-D model, is complicated. The shape of the emergent line profiles depends sensitively on the conditions - in particular the plasma flows - that arise at the line core formation heights. From various published electron beam simulations \citep[e.g.][]{2005ApJ...630..573A,2015A&A...578A..72K} strong upflows and downflows seem to be a common result of heating by a power-law beam with an energy flux in excess of $\sim 10^{10}\rm{ergs~cm^{-2}s^{-1}}$ because of the rapid ionisation of H and He in the chromosphere and the expansion that results.  It remains to be seen whether any such systematic behaviours arise from heating by  Alfv\'en waves, which could aid in distinguishing the contribution of each.\\

\textsc{Acknowledgments:} \small{}
The authors would like to thank Dr J. Reep for helpful discussions, and Dr J. Leenaarts for help with the RH code. GSK would like to acknowledge the financial support of a PhD scholarship from the College of Science and Engineering, Univeristy of Glasgow.  LF acknowledges support from STFC consolidated grant ST/L000741/1.  The research leading these results has received funding from the European Community's Seventh Framework Programme (FP7/2007-2013) under grant agreement no. 606862 (F-CHROMA). AJBR acknowledges support from STFC consolidated grant ST/K000993/1. AJBR and LF acknowledge support from ISSI (Switzerland) for the International Team on `Magnetic Waves in Solar Flares: Beyond the ``Standard" Flare Model'. JCA acknowledges funding support through the NASA Heliophysics Supporting Research and NASA Living With a Star programs.

\bibliographystyle{apj}
\bibliography{Kerr_etal_AlfvWave_2016}

\end{document}

%% file: rgb.tex
\definecolor{AliceBlue}{rgb}{0.94,0.97,1.00}
\definecolor{AntiqueWhite1}{rgb}{1.00,0.94,0.86}
\definecolor{AntiqueWhite2}{rgb}{0.93,0.87,0.80}
\definecolor{AntiqueWhite3}{rgb}{0.80,0.75,0.69}
\definecolor{AntiqueWhite4}{rgb}{0.55,0.51,0.47}
\definecolor{AntiqueWhite}{rgb}{0.98,0.92,0.84}
\definecolor{BlanchedAlmond}{rgb}{1.00,0.92,0.80}
\definecolor{BlueViolet}{rgb}{0.54,0.17,0.89}
\definecolor{CadetBlue1}{rgb}{0.60,0.96,1.00}
\definecolor{CadetBlue2}{rgb}{0.56,0.90,0.93}
\definecolor{CadetBlue3}{rgb}{0.48,0.77,0.80}
\definecolor{CadetBlue4}{rgb}{0.33,0.53,0.55}
\definecolor{CadetBlue}{rgb}{0.37,0.62,0.63}
\definecolor{CornflowerBlue}{rgb}{0.39,0.58,0.93}
\definecolor{DarkBlue}{rgb}{0.00,0.00,0.55}
\definecolor{DarkCyan}{rgb}{0.00,0.55,0.55}
\definecolor{DarkGoldenrod1}{rgb}{1.00,0.73,0.06}
\definecolor{DarkGoldenrod2}{rgb}{0.93,0.68,0.05}
\definecolor{DarkGoldenrod3}{rgb}{0.80,0.58,0.05}
\definecolor{DarkGoldenrod4}{rgb}{0.55,0.40,0.03}
\definecolor{DarkGoldenrod}{rgb}{0.72,0.53,0.04}
\definecolor{DarkGray}{rgb}{0.66,0.66,0.66}
\definecolor{DarkGreen}{rgb}{0.00,0.39,0.00}
\definecolor{DarkGrey}{rgb}{0.66,0.66,0.66}
\definecolor{DarkKhaki}{rgb}{0.74,0.72,0.42}
\definecolor{DarkMagenta}{rgb}{0.55,0.00,0.55}
\definecolor{DarkOliveGreen1}{rgb}{0.79,1.00,0.44}
\definecolor{DarkOliveGreen2}{rgb}{0.74,0.93,0.41}
\definecolor{DarkOliveGreen3}{rgb}{0.64,0.80,0.35}
\definecolor{DarkOliveGreen4}{rgb}{0.43,0.55,0.24}
\definecolor{DarkOliveGreen}{rgb}{0.33,0.42,0.18}
\definecolor{DarkOrange1}{rgb}{1.00,0.50,0.00}
\definecolor{DarkOrange2}{rgb}{0.93,0.46,0.00}
\definecolor{DarkOrange3}{rgb}{0.80,0.40,0.00}
\definecolor{DarkOrange4}{rgb}{0.55,0.27,0.00}
\definecolor{DarkOrange}{rgb}{1.00,0.55,0.00}
\definecolor{DarkOrchid1}{rgb}{0.75,0.24,1.00}
\definecolor{DarkOrchid2}{rgb}{0.70,0.23,0.93}
\definecolor{DarkOrchid3}{rgb}{0.60,0.20,0.80}
\definecolor{DarkOrchid4}{rgb}{0.41,0.13,0.55}
\definecolor{DarkOrchid}{rgb}{0.60,0.20,0.80}
\definecolor{DarkRed}{rgb}{0.55,0.00,0.00}
\definecolor{DarkSalmon}{rgb}{0.91,0.59,0.48}
\definecolor{DarkSeaGreen1}{rgb}{0.76,1.00,0.76}
\definecolor{DarkSeaGreen2}{rgb}{0.71,0.93,0.71}
\definecolor{DarkSeaGreen3}{rgb}{0.61,0.80,0.61}
\definecolor{DarkSeaGreen4}{rgb}{0.41,0.55,0.41}
\definecolor{DarkSeaGreen}{rgb}{0.56,0.74,0.56}
\definecolor{DarkSlateBlue}{rgb}{0.28,0.24,0.55}
\definecolor{DarkSlateGray1}{rgb}{0.59,1.00,1.00}
\definecolor{DarkSlateGray2}{rgb}{0.55,0.93,0.93}
\definecolor{DarkSlateGray3}{rgb}{0.47,0.80,0.80}
\definecolor{DarkSlateGray4}{rgb}{0.32,0.55,0.55}
\definecolor{DarkSlateGray}{rgb}{0.18,0.31,0.31}
\definecolor{DarkSlateGrey}{rgb}{0.18,0.31,0.31}
\definecolor{DarkTurquoise}{rgb}{0.00,0.81,0.82}
\definecolor{DarkViolet}{rgb}{0.58,0.00,0.83}
\definecolor{DeepPink1}{rgb}{1.00,0.08,0.58}
\definecolor{DeepPink2}{rgb}{0.93,0.07,0.54}
\definecolor{DeepPink3}{rgb}{0.80,0.06,0.46}
\definecolor{DeepPink4}{rgb}{0.55,0.04,0.31}
\definecolor{DeepPink}{rgb}{1.00,0.08,0.58}
\definecolor{DeepSkyBlue1}{rgb}{0.00,0.75,1.00}
\definecolor{DeepSkyBlue2}{rgb}{0.00,0.70,0.93}
\definecolor{DeepSkyBlue3}{rgb}{0.00,0.60,0.80}
\definecolor{DeepSkyBlue4}{rgb}{0.00,0.41,0.55}
\definecolor{DeepSkyBlue}{rgb}{0.00,0.75,1.00}
\definecolor{DimGray}{rgb}{0.41,0.41,0.41}
\definecolor{DimGrey}{rgb}{0.41,0.41,0.41}
\definecolor{DodgerBlue1}{rgb}{0.12,0.56,1.00}
\definecolor{DodgerBlue2}{rgb}{0.11,0.53,0.93}
\definecolor{DodgerBlue3}{rgb}{0.09,0.45,0.80}
\definecolor{DodgerBlue4}{rgb}{0.06,0.31,0.55}
\definecolor{DodgerBlue}{rgb}{0.12,0.56,1.00}
\definecolor{FloralWhite}{rgb}{1.00,0.98,0.94}
\definecolor{ForestGreen}{rgb}{0.13,0.55,0.13}
\definecolor{GhostWhite}{rgb}{0.97,0.97,1.00}
\definecolor{GreenYellow}{rgb}{0.68,1.00,0.18}
\definecolor{HotPink1}{rgb}{1.00,0.43,0.71}
\definecolor{HotPink2}{rgb}{0.93,0.42,0.65}
\definecolor{HotPink3}{rgb}{0.80,0.38,0.56}
\definecolor{HotPink4}{rgb}{0.55,0.23,0.38}
\definecolor{HotPink}{rgb}{1.00,0.41,0.71}
\definecolor{IndianRed1}{rgb}{1.00,0.42,0.42}
\definecolor{IndianRed2}{rgb}{0.93,0.39,0.39}
\definecolor{IndianRed3}{rgb}{0.80,0.33,0.33}
\definecolor{IndianRed4}{rgb}{0.55,0.23,0.23}
\definecolor{IndianRed}{rgb}{0.80,0.36,0.36}
\definecolor{LavenderBlush1}{rgb}{1.00,0.94,0.96}
\definecolor{LavenderBlush2}{rgb}{0.93,0.88,0.90}
\definecolor{LavenderBlush3}{rgb}{0.80,0.76,0.77}
\definecolor{LavenderBlush4}{rgb}{0.55,0.51,0.53}
\definecolor{LavenderBlush}{rgb}{1.00,0.94,0.96}
\definecolor{LawnGreen}{rgb}{0.49,0.99,0.00}
\definecolor{LemonChiffon1}{rgb}{1.00,0.98,0.80}
\definecolor{LemonChiffon2}{rgb}{0.93,0.91,0.75}
\definecolor{LemonChiffon3}{rgb}{0.80,0.79,0.65}
\definecolor{LemonChiffon4}{rgb}{0.55,0.54,0.44}
\definecolor{LemonChiffon}{rgb}{1.00,0.98,0.80}
\definecolor{LightBlue1}{rgb}{0.75,0.94,1.00}
\definecolor{LightBlue2}{rgb}{0.70,0.87,0.93}
\definecolor{LightBlue3}{rgb}{0.60,0.75,0.80}
\definecolor{LightBlue4}{rgb}{0.41,0.51,0.55}
\definecolor{LightBlue}{rgb}{0.68,0.85,0.90}
\definecolor{LightCoral}{rgb}{0.94,0.50,0.50}
\definecolor{LightCyan1}{rgb}{0.88,1.00,1.00}
\definecolor{LightCyan2}{rgb}{0.82,0.93,0.93}
\definecolor{LightCyan3}{rgb}{0.71,0.80,0.80}
\definecolor{LightCyan4}{rgb}{0.48,0.55,0.55}
\definecolor{LightCyan}{rgb}{0.88,1.00,1.00}
\definecolor{LightGoldenrod1}{rgb}{1.00,0.93,0.55}
\definecolor{LightGoldenrod2}{rgb}{0.93,0.86,0.51}
\definecolor{LightGoldenrod3}{rgb}{0.80,0.75,0.44}
\definecolor{LightGoldenrod4}{rgb}{0.55,0.51,0.30}
\definecolor{LightGoldenrodYellow}{rgb}{0.98,0.98,0.82}
\definecolor{LightGoldenrod}{rgb}{0.93,0.87,0.51}
\definecolor{LightGray}{rgb}{0.83,0.83,0.83}
\definecolor{LightGreen}{rgb}{0.56,0.93,0.56}
\definecolor{LightGrey}{rgb}{0.83,0.83,0.83}
\definecolor{LightPink1}{rgb}{1.00,0.68,0.73}
\definecolor{LightPink2}{rgb}{0.93,0.64,0.68}
\definecolor{LightPink3}{rgb}{0.80,0.55,0.58}
\definecolor{LightPink4}{rgb}{0.55,0.37,0.40}
\definecolor{LightPink}{rgb}{1.00,0.71,0.76}
\definecolor{LightSalmon1}{rgb}{1.00,0.63,0.48}
\definecolor{LightSalmon2}{rgb}{0.93,0.58,0.45}
\definecolor{LightSalmon3}{rgb}{0.80,0.51,0.38}
\definecolor{LightSalmon4}{rgb}{0.55,0.34,0.26}
\definecolor{LightSalmon}{rgb}{1.00,0.63,0.48}
\definecolor{LightSeaGreen}{rgb}{0.13,0.70,0.67}
\definecolor{LightSkyBlue1}{rgb}{0.69,0.89,1.00}
\definecolor{LightSkyBlue2}{rgb}{0.64,0.83,0.93}
\definecolor{LightSkyBlue3}{rgb}{0.55,0.71,0.80}
\definecolor{LightSkyBlue4}{rgb}{0.38,0.48,0.55}
\definecolor{LightSkyBlue}{rgb}{0.53,0.81,0.98}
\definecolor{LightSlateBlue}{rgb}{0.52,0.44,1.00}
\definecolor{LightSlateGray}{rgb}{0.47,0.53,0.60}
\definecolor{LightSlateGrey}{rgb}{0.47,0.53,0.60}
\definecolor{LightSteelBlue1}{rgb}{0.79,0.88,1.00}
\definecolor{LightSteelBlue2}{rgb}{0.74,0.82,0.93}
\definecolor{LightSteelBlue3}{rgb}{0.64,0.71,0.80}
\definecolor{LightSteelBlue4}{rgb}{0.43,0.48,0.55}
\definecolor{LightSteelBlue}{rgb}{0.69,0.77,0.87}
\definecolor{LightYellow1}{rgb}{1.00,1.00,0.88}
\definecolor{LightYellow2}{rgb}{0.93,0.93,0.82}
\definecolor{LightYellow3}{rgb}{0.80,0.80,0.71}
\definecolor{LightYellow4}{rgb}{0.55,0.55,0.48}
\definecolor{LightYellow}{rgb}{1.00,1.00,0.88}
\definecolor{LimeGreen}{rgb}{0.20,0.80,0.20}
\definecolor{MediumAquamarine}{rgb}{0.40,0.80,0.67}
\definecolor{MediumBlue}{rgb}{0.00,0.00,0.80}
\definecolor{MediumOrchid1}{rgb}{0.88,0.40,1.00}
\definecolor{MediumOrchid2}{rgb}{0.82,0.37,0.93}
\definecolor{MediumOrchid3}{rgb}{0.71,0.32,0.80}
\definecolor{MediumOrchid4}{rgb}{0.48,0.22,0.55}
\definecolor{MediumOrchid}{rgb}{0.73,0.33,0.83}
\definecolor{MediumPurple1}{rgb}{0.67,0.51,1.00}
\definecolor{MediumPurple2}{rgb}{0.62,0.47,0.93}
\definecolor{MediumPurple3}{rgb}{0.54,0.41,0.80}
\definecolor{MediumPurple4}{rgb}{0.36,0.28,0.55}
\definecolor{MediumPurple}{rgb}{0.58,0.44,0.86}
\definecolor{MediumSeaGreen}{rgb}{0.24,0.70,0.44}
\definecolor{MediumSlateBlue}{rgb}{0.48,0.41,0.93}
\definecolor{MediumSpringGreen}{rgb}{0.00,0.98,0.60}
\definecolor{MediumTurquoise}{rgb}{0.28,0.82,0.80}
\definecolor{MediumVioletRed}{rgb}{0.78,0.08,0.52}
\definecolor{MidnightBlue}{rgb}{0.10,0.10,0.44}
\definecolor{MintCream}{rgb}{0.96,1.00,0.98}
\definecolor{MistyRose1}{rgb}{1.00,0.89,0.88}
\definecolor{MistyRose2}{rgb}{0.93,0.84,0.82}
\definecolor{MistyRose3}{rgb}{0.80,0.72,0.71}
\definecolor{MistyRose4}{rgb}{0.55,0.49,0.48}
\definecolor{MistyRose}{rgb}{1.00,0.89,0.88}
\definecolor{NavajoWhite1}{rgb}{1.00,0.87,0.68}
\definecolor{NavajoWhite2}{rgb}{0.93,0.81,0.63}
\definecolor{NavajoWhite3}{rgb}{0.80,0.70,0.55}
\definecolor{NavajoWhite4}{rgb}{0.55,0.47,0.37}
\definecolor{NavajoWhite}{rgb}{1.00,0.87,0.68}
\definecolor{NavyBlue}{rgb}{0.00,0.00,0.50}
\definecolor{OldLace}{rgb}{0.99,0.96,0.90}
\definecolor{OliveDrab1}{rgb}{0.75,1.00,0.24}
\definecolor{OliveDrab2}{rgb}{0.70,0.93,0.23}
\definecolor{OliveDrab3}{rgb}{0.60,0.80,0.20}
\definecolor{OliveDrab4}{rgb}{0.41,0.55,0.13}
\definecolor{OliveDrab}{rgb}{0.42,0.56,0.14}
\definecolor{OrangeRed1}{rgb}{1.00,0.27,0.00}
\definecolor{OrangeRed2}{rgb}{0.93,0.25,0.00}
\definecolor{OrangeRed3}{rgb}{0.80,0.22,0.00}
\definecolor{OrangeRed4}{rgb}{0.55,0.15,0.00}
\definecolor{OrangeRed}{rgb}{1.00,0.27,0.00}
\definecolor{PaleGoldenrod}{rgb}{0.93,0.91,0.67}
\definecolor{PaleGreen1}{rgb}{0.60,1.00,0.60}
\definecolor{PaleGreen2}{rgb}{0.56,0.93,0.56}
\definecolor{PaleGreen3}{rgb}{0.49,0.80,0.49}
\definecolor{PaleGreen4}{rgb}{0.33,0.55,0.33}
\definecolor{PaleGreen}{rgb}{0.60,0.98,0.60}
\definecolor{PaleTurquoise1}{rgb}{0.73,1.00,1.00}
\definecolor{PaleTurquoise2}{rgb}{0.68,0.93,0.93}
\definecolor{PaleTurquoise3}{rgb}{0.59,0.80,0.80}
\definecolor{PaleTurquoise4}{rgb}{0.40,0.55,0.55}
\definecolor{PaleTurquoise}{rgb}{0.69,0.93,0.93}
\definecolor{PaleVioletRed1}{rgb}{1.00,0.51,0.67}
\definecolor{PaleVioletRed2}{rgb}{0.93,0.47,0.62}
\definecolor{PaleVioletRed3}{rgb}{0.80,0.41,0.54}
\definecolor{PaleVioletRed4}{rgb}{0.55,0.28,0.36}
\definecolor{PaleVioletRed}{rgb}{0.86,0.44,0.58}
\definecolor{PapayaWhip}{rgb}{1.00,0.94,0.84}
\definecolor{PeachPuff1}{rgb}{1.00,0.85,0.73}
\definecolor{PeachPuff2}{rgb}{0.93,0.80,0.68}
\definecolor{PeachPuff3}{rgb}{0.80,0.69,0.58}
\definecolor{PeachPuff4}{rgb}{0.55,0.47,0.40}
\definecolor{PeachPuff}{rgb}{1.00,0.85,0.73}
\definecolor{PowderBlue}{rgb}{0.69,0.88,0.90}
\definecolor{RosyBrown1}{rgb}{1.00,0.76,0.76}
\definecolor{RosyBrown2}{rgb}{0.93,0.71,0.71}
\definecolor{RosyBrown3}{rgb}{0.80,0.61,0.61}
\definecolor{RosyBrown4}{rgb}{0.55,0.41,0.41}
\definecolor{RosyBrown}{rgb}{0.74,0.56,0.56}
\definecolor{RoyalBlue1}{rgb}{0.28,0.46,1.00}
\definecolor{RoyalBlue2}{rgb}{0.26,0.43,0.93}
\definecolor{RoyalBlue3}{rgb}{0.23,0.37,0.80}
\definecolor{RoyalBlue4}{rgb}{0.15,0.25,0.55}
\definecolor{RoyalBlue}{rgb}{0.25,0.41,0.88}
\definecolor{SaddleBrown}{rgb}{0.55,0.27,0.07}
\definecolor{SandyBrown}{rgb}{0.96,0.64,0.38}
\definecolor{SeaGreen1}{rgb}{0.33,1.00,0.62}
\definecolor{SeaGreen2}{rgb}{0.31,0.93,0.58}
\definecolor{SeaGreen3}{rgb}{0.26,0.80,0.50}
\definecolor{SeaGreen4}{rgb}{0.18,0.55,0.34}
\definecolor{SeaGreen}{rgb}{0.18,0.55,0.34}
\definecolor{SkyBlue1}{rgb}{0.53,0.81,1.00}
\definecolor{SkyBlue2}{rgb}{0.49,0.75,0.93}
\definecolor{SkyBlue3}{rgb}{0.42,0.65,0.80}
\definecolor{SkyBlue4}{rgb}{0.29,0.44,0.55}
\definecolor{SkyBlue}{rgb}{0.53,0.81,0.92}
\definecolor{SlateBlue1}{rgb}{0.51,0.44,1.00}
\definecolor{SlateBlue2}{rgb}{0.48,0.40,0.93}
\definecolor{SlateBlue3}{rgb}{0.41,0.35,0.80}
\definecolor{SlateBlue4}{rgb}{0.28,0.24,0.55}
\definecolor{SlateBlue}{rgb}{0.42,0.35,0.80}
\definecolor{SlateGray1}{rgb}{0.78,0.89,1.00}
\definecolor{SlateGray2}{rgb}{0.73,0.83,0.93}
\definecolor{SlateGray3}{rgb}{0.62,0.71,0.80}
\definecolor{SlateGray4}{rgb}{0.42,0.48,0.55}
\definecolor{SlateGray}{rgb}{0.44,0.50,0.56}
\definecolor{SlateGrey}{rgb}{0.44,0.50,0.56}
\definecolor{SpringGreen1}{rgb}{0.00,1.00,0.50}
\definecolor{SpringGreen2}{rgb}{0.00,0.93,0.46}
\definecolor{SpringGreen3}{rgb}{0.00,0.80,0.40}
\definecolor{SpringGreen4}{rgb}{0.00,0.55,0.27}
\definecolor{SpringGreen}{rgb}{0.00,1.00,0.50}
\definecolor{SteelBlue1}{rgb}{0.39,0.72,1.00}
\definecolor{SteelBlue2}{rgb}{0.36,0.67,0.93}
\definecolor{SteelBlue3}{rgb}{0.31,0.58,0.80}
\definecolor{SteelBlue4}{rgb}{0.21,0.39,0.55}
\definecolor{SteelBlue}{rgb}{0.27,0.51,0.71}
\definecolor{VioletRed1}{rgb}{1.00,0.24,0.59}
\definecolor{VioletRed2}{rgb}{0.93,0.23,0.55}
\definecolor{VioletRed3}{rgb}{0.80,0.20,0.47}
\definecolor{VioletRed4}{rgb}{0.55,0.13,0.32}
\definecolor{VioletRed}{rgb}{0.82,0.13,0.56}
\definecolor{WhiteSmoke}{rgb}{0.96,0.96,0.96}
\definecolor{YellowGreen}{rgb}{0.60,0.80,0.20}
\definecolor{aliceblue}{rgb}{0.94,0.97,1.00}
\definecolor{antiquewhite}{rgb}{0.98,0.92,0.84}
\definecolor{aquamarine1}{rgb}{0.50,1.00,0.83}
\definecolor{aquamarine2}{rgb}{0.46,0.93,0.78}
\definecolor{aquamarine3}{rgb}{0.40,0.80,0.67}
\definecolor{aquamarine4}{rgb}{0.27,0.55,0.45}
\definecolor{aquamarine}{rgb}{0.50,1.00,0.83}
\definecolor{azure1}{rgb}{0.94,1.00,1.00}
\definecolor{azure2}{rgb}{0.88,0.93,0.93}
\definecolor{azure3}{rgb}{0.76,0.80,0.80}
\definecolor{azure4}{rgb}{0.51,0.55,0.55}
\definecolor{azure}{rgb}{0.94,1.00,1.00}
\definecolor{beige}{rgb}{0.96,0.96,0.86}
\definecolor{bisque1}{rgb}{1.00,0.89,0.77}
\definecolor{bisque2}{rgb}{0.93,0.84,0.72}
\definecolor{bisque3}{rgb}{0.80,0.72,0.62}
\definecolor{bisque4}{rgb}{0.55,0.49,0.42}
\definecolor{bisque}{rgb}{1.00,0.89,0.77}
\definecolor{black}{rgb}{0.00,0.00,0.00}
\definecolor{blanchedalmond}{rgb}{1.00,0.92,0.80}
\definecolor{blue1}{rgb}{0.00,0.00,1.00}
\definecolor{blue2}{rgb}{0.00,0.00,0.93}
\definecolor{blue3}{rgb}{0.00,0.00,0.80}
\definecolor{blue4}{rgb}{0.00,0.00,0.55}
\definecolor{blueviolet}{rgb}{0.54,0.17,0.89}
\definecolor{blue}{rgb}{0.00,0.00,1.00}
\definecolor{brown1}{rgb}{1.00,0.25,0.25}
\definecolor{brown2}{rgb}{0.93,0.23,0.23}
\definecolor{brown3}{rgb}{0.80,0.20,0.20}
\definecolor{brown4}{rgb}{0.55,0.14,0.14}
\definecolor{brown}{rgb}{0.65,0.16,0.16}
\definecolor{burlywood1}{rgb}{1.00,0.83,0.61}
\definecolor{burlywood2}{rgb}{0.93,0.77,0.57}
\definecolor{burlywood3}{rgb}{0.80,0.67,0.49}
\definecolor{burlywood4}{rgb}{0.55,0.45,0.33}
\definecolor{burlywood}{rgb}{0.87,0.72,0.53}
\definecolor{cadetblue}{rgb}{0.37,0.62,0.63}
\definecolor{chartreuse1}{rgb}{0.50,1.00,0.00}
\definecolor{chartreuse2}{rgb}{0.46,0.93,0.00}
\definecolor{chartreuse3}{rgb}{0.40,0.80,0.00}
\definecolor{chartreuse4}{rgb}{0.27,0.55,0.00}
\definecolor{chartreuse}{rgb}{0.50,1.00,0.00}
\definecolor{chocolate1}{rgb}{1.00,0.50,0.14}
\definecolor{chocolate2}{rgb}{0.93,0.46,0.13}
\definecolor{chocolate3}{rgb}{0.80,0.40,0.11}
\definecolor{chocolate4}{rgb}{0.55,0.27,0.07}
\definecolor{chocolate}{rgb}{0.82,0.41,0.12}
\definecolor{coral1}{rgb}{1.00,0.45,0.34}
\definecolor{coral2}{rgb}{0.93,0.42,0.31}
\definecolor{coral3}{rgb}{0.80,0.36,0.27}
\definecolor{coral4}{rgb}{0.55,0.24,0.18}
\definecolor{coral}{rgb}{1.00,0.50,0.31}
\definecolor{cornflowerblue}{rgb}{0.39,0.58,0.93}
\definecolor{cornsilk1}{rgb}{1.00,0.97,0.86}
\definecolor{cornsilk2}{rgb}{0.93,0.91,0.80}
\definecolor{cornsilk3}{rgb}{0.80,0.78,0.69}
\definecolor{cornsilk4}{rgb}{0.55,0.53,0.47}
\definecolor{cornsilk}{rgb}{1.00,0.97,0.86}
\definecolor{cyan1}{rgb}{0.00,1.00,1.00}
\definecolor{cyan2}{rgb}{0.00,0.93,0.93}
\definecolor{cyan3}{rgb}{0.00,0.80,0.80}
\definecolor{cyan4}{rgb}{0.00,0.55,0.55}
\definecolor{cyan}{rgb}{0.00,1.00,1.00}
\definecolor{darkblue}{rgb}{0.00,0.00,0.55}
\definecolor{darkcyan}{rgb}{0.00,0.55,0.55}
\definecolor{darkgoldenrod}{rgb}{0.72,0.53,0.04}
\definecolor{darkgray}{rgb}{0.66,0.66,0.66}
\definecolor{darkgreen}{rgb}{0.00,0.39,0.00}
\definecolor{darkgrey}{rgb}{0.66,0.66,0.66}
\definecolor{darkkhaki}{rgb}{0.74,0.72,0.42}
\definecolor{darkmagenta}{rgb}{0.55,0.00,0.55}
\definecolor{darkolive}{rgb}{0.33,0.42,0.18}
\definecolor{darkorange}{rgb}{1.00,0.55,0.00}
\definecolor{darkorchid}{rgb}{0.60,0.20,0.80}
\definecolor{darkred}{rgb}{0.55,0.00,0.00}
\definecolor{darksalmon}{rgb}{0.91,0.59,0.48}
\definecolor{darksea}{rgb}{0.56,0.74,0.56}
\definecolor{darkslate}{rgb}{0.18,0.31,0.31}
\definecolor{darkslate}{rgb}{0.18,0.31,0.31}
\definecolor{darkslate}{rgb}{0.28,0.24,0.55}
\definecolor{darkturquoise}{rgb}{0.00,0.81,0.82}
\definecolor{darkviolet}{rgb}{0.58,0.00,0.83}
\definecolor{deeppink}{rgb}{1.00,0.08,0.58}
\definecolor{deepsky}{rgb}{0.00,0.75,1.00}
\definecolor{dimgray}{rgb}{0.41,0.41,0.41}
\definecolor{dimgrey}{rgb}{0.41,0.41,0.41}
\definecolor{dodgerblue}{rgb}{0.12,0.56,1.00}
\definecolor{firebrick1}{rgb}{1.00,0.19,0.19}
\definecolor{firebrick2}{rgb}{0.93,0.17,0.17}
\definecolor{firebrick3}{rgb}{0.80,0.15,0.15}
\definecolor{firebrick4}{rgb}{0.55,0.10,0.10}
\definecolor{firebrick}{rgb}{0.70,0.13,0.13}
\definecolor{floralwhite}{rgb}{1.00,0.98,0.94}
\definecolor{forestgreen}{rgb}{0.13,0.55,0.13}
\definecolor{gainsboro}{rgb}{0.86,0.86,0.86}
\definecolor{ghostwhite}{rgb}{0.97,0.97,1.00}
\definecolor{gold1}{rgb}{1.00,0.84,0.00}
\definecolor{gold2}{rgb}{0.93,0.79,0.00}
\definecolor{gold3}{rgb}{0.80,0.68,0.00}
\definecolor{gold4}{rgb}{0.55,0.46,0.00}
\definecolor{goldenrod1}{rgb}{1.00,0.76,0.15}
\definecolor{goldenrod2}{rgb}{0.93,0.71,0.13}
\definecolor{goldenrod3}{rgb}{0.80,0.61,0.11}
\definecolor{goldenrod4}{rgb}{0.55,0.41,0.08}
\definecolor{goldenrod}{rgb}{0.85,0.65,0.13}
\definecolor{gold}{rgb}{1.00,0.84,0.00}
\definecolor{gray0}{rgb}{0.00,0.00,0.00}
\definecolor{gray100}{rgb}{1.00,1.00,1.00}
\definecolor{gray10}{rgb}{0.10,0.10,0.10}
\definecolor{gray11}{rgb}{0.11,0.11,0.11}
\definecolor{gray12}{rgb}{0.12,0.12,0.12}
\definecolor{gray13}{rgb}{0.13,0.13,0.13}
\definecolor{gray14}{rgb}{0.14,0.14,0.14}
\definecolor{gray15}{rgb}{0.15,0.15,0.15}
\definecolor{gray16}{rgb}{0.16,0.16,0.16}
\definecolor{gray17}{rgb}{0.17,0.17,0.17}
\definecolor{gray18}{rgb}{0.18,0.18,0.18}
\definecolor{gray19}{rgb}{0.19,0.19,0.19}
\definecolor{gray1}{rgb}{0.01,0.01,0.01}
\definecolor{gray20}{rgb}{0.20,0.20,0.20}
\definecolor{gray21}{rgb}{0.21,0.21,0.21}
\definecolor{gray22}{rgb}{0.22,0.22,0.22}
\definecolor{gray23}{rgb}{0.23,0.23,0.23}
\definecolor{gray24}{rgb}{0.24,0.24,0.24}
\definecolor{gray25}{rgb}{0.25,0.25,0.25}
\definecolor{gray26}{rgb}{0.26,0.26,0.26}
\definecolor{gray27}{rgb}{0.27,0.27,0.27}
\definecolor{gray28}{rgb}{0.28,0.28,0.28}
\definecolor{gray29}{rgb}{0.29,0.29,0.29}
\definecolor{gray2}{rgb}{0.02,0.02,0.02}
\definecolor{gray30}{rgb}{0.30,0.30,0.30}
\definecolor{gray31}{rgb}{0.31,0.31,0.31}
\definecolor{gray32}{rgb}{0.32,0.32,0.32}
\definecolor{gray33}{rgb}{0.33,0.33,0.33}
\definecolor{gray34}{rgb}{0.34,0.34,0.34}
\definecolor{gray35}{rgb}{0.35,0.35,0.35}
\definecolor{gray36}{rgb}{0.36,0.36,0.36}
\definecolor{gray37}{rgb}{0.37,0.37,0.37}
\definecolor{gray38}{rgb}{0.38,0.38,0.38}
\definecolor{gray39}{rgb}{0.39,0.39,0.39}
\definecolor{gray3}{rgb}{0.03,0.03,0.03}
\definecolor{gray40}{rgb}{0.40,0.40,0.40}
\definecolor{gray41}{rgb}{0.41,0.41,0.41}
\definecolor{gray42}{rgb}{0.42,0.42,0.42}
\definecolor{gray43}{rgb}{0.43,0.43,0.43}
\definecolor{gray44}{rgb}{0.44,0.44,0.44}
\definecolor{gray45}{rgb}{0.45,0.45,0.45}
\definecolor{gray46}{rgb}{0.46,0.46,0.46}
\definecolor{gray47}{rgb}{0.47,0.47,0.47}
\definecolor{gray48}{rgb}{0.48,0.48,0.48}
\definecolor{gray49}{rgb}{0.49,0.49,0.49}
\definecolor{gray4}{rgb}{0.04,0.04,0.04}
\definecolor{gray50}{rgb}{0.50,0.50,0.50}
\definecolor{gray51}{rgb}{0.51,0.51,0.51}
\definecolor{gray52}{rgb}{0.52,0.52,0.52}
\definecolor{gray53}{rgb}{0.53,0.53,0.53}
\definecolor{gray54}{rgb}{0.54,0.54,0.54}
\definecolor{gray55}{rgb}{0.55,0.55,0.55}
\definecolor{gray56}{rgb}{0.56,0.56,0.56}
\definecolor{gray57}{rgb}{0.57,0.57,0.57}
\definecolor{gray58}{rgb}{0.58,0.58,0.58}
\definecolor{gray59}{rgb}{0.59,0.59,0.59}
\definecolor{gray5}{rgb}{0.05,0.05,0.05}
\definecolor{gray60}{rgb}{0.60,0.60,0.60}
\definecolor{gray61}{rgb}{0.61,0.61,0.61}
\definecolor{gray62}{rgb}{0.62,0.62,0.62}
\definecolor{gray63}{rgb}{0.63,0.63,0.63}
\definecolor{gray64}{rgb}{0.64,0.64,0.64}
\definecolor{gray65}{rgb}{0.65,0.65,0.65}
\definecolor{gray66}{rgb}{0.66,0.66,0.66}
\definecolor{gray67}{rgb}{0.67,0.67,0.67}
\definecolor{gray68}{rgb}{0.68,0.68,0.68}
\definecolor{gray69}{rgb}{0.69,0.69,0.69}
\definecolor{gray6}{rgb}{0.06,0.06,0.06}
\definecolor{gray70}{rgb}{0.70,0.70,0.70}
\definecolor{gray71}{rgb}{0.71,0.71,0.71}
\definecolor{gray72}{rgb}{0.72,0.72,0.72}
\definecolor{gray73}{rgb}{0.73,0.73,0.73}
\definecolor{gray74}{rgb}{0.74,0.74,0.74}
\definecolor{gray75}{rgb}{0.75,0.75,0.75}
\definecolor{gray76}{rgb}{0.76,0.76,0.76}
\definecolor{gray77}{rgb}{0.77,0.77,0.77}
\definecolor{gray78}{rgb}{0.78,0.78,0.78}
\definecolor{gray79}{rgb}{0.79,0.79,0.79}
\definecolor{gray7}{rgb}{0.07,0.07,0.07}
\definecolor{gray80}{rgb}{0.80,0.80,0.80}
\definecolor{gray81}{rgb}{0.81,0.81,0.81}
\definecolor{gray82}{rgb}{0.82,0.82,0.82}
\definecolor{gray83}{rgb}{0.83,0.83,0.83}
\definecolor{gray84}{rgb}{0.84,0.84,0.84}
\definecolor{gray85}{rgb}{0.85,0.85,0.85}
\definecolor{gray86}{rgb}{0.86,0.86,0.86}
\definecolor{gray87}{rgb}{0.87,0.87,0.87}
\definecolor{gray88}{rgb}{0.88,0.88,0.88}
\definecolor{gray89}{rgb}{0.89,0.89,0.89}
\definecolor{gray8}{rgb}{0.08,0.08,0.08}
\definecolor{gray90}{rgb}{0.90,0.90,0.90}
\definecolor{gray91}{rgb}{0.91,0.91,0.91}
\definecolor{gray92}{rgb}{0.92,0.92,0.92}
\definecolor{gray93}{rgb}{0.93,0.93,0.93}
\definecolor{gray94}{rgb}{0.94,0.94,0.94}
\definecolor{gray95}{rgb}{0.95,0.95,0.95}
\definecolor{gray96}{rgb}{0.96,0.96,0.96}
\definecolor{gray97}{rgb}{0.97,0.97,0.97}
\definecolor{gray98}{rgb}{0.98,0.98,0.98}
\definecolor{gray99}{rgb}{0.99,0.99,0.99}
\definecolor{gray9}{rgb}{0.09,0.09,0.09}
\definecolor{gray}{rgb}{0.75,0.75,0.75}
\definecolor{green1}{rgb}{0.00,1.00,0.00}
\definecolor{green2}{rgb}{0.00,0.93,0.00}
\definecolor{green3}{rgb}{0.00,0.80,0.00}
\definecolor{green4}{rgb}{0.00,0.55,0.00}
\definecolor{greenyellow}{rgb}{0.68,1.00,0.18}
\definecolor{green}{rgb}{0.00,1.00,0.00}
\definecolor{grey0}{rgb}{0.00,0.00,0.00}
\definecolor{grey100}{rgb}{1.00,1.00,1.00}
\definecolor{grey10}{rgb}{0.10,0.10,0.10}
\definecolor{grey11}{rgb}{0.11,0.11,0.11}
\definecolor{grey12}{rgb}{0.12,0.12,0.12}
\definecolor{grey13}{rgb}{0.13,0.13,0.13}
\definecolor{grey14}{rgb}{0.14,0.14,0.14}
\definecolor{grey15}{rgb}{0.15,0.15,0.15}
\definecolor{grey16}{rgb}{0.16,0.16,0.16}
\definecolor{grey17}{rgb}{0.17,0.17,0.17}
\definecolor{grey18}{rgb}{0.18,0.18,0.18}
\definecolor{grey19}{rgb}{0.19,0.19,0.19}
\definecolor{grey1}{rgb}{0.01,0.01,0.01}
\definecolor{grey20}{rgb}{0.20,0.20,0.20}
\definecolor{grey21}{rgb}{0.21,0.21,0.21}
\definecolor{grey22}{rgb}{0.22,0.22,0.22}
\definecolor{grey23}{rgb}{0.23,0.23,0.23}
\definecolor{grey24}{rgb}{0.24,0.24,0.24}
\definecolor{grey25}{rgb}{0.25,0.25,0.25}
\definecolor{grey26}{rgb}{0.26,0.26,0.26}
\definecolor{grey27}{rgb}{0.27,0.27,0.27}
\definecolor{grey28}{rgb}{0.28,0.28,0.28}
\definecolor{grey29}{rgb}{0.29,0.29,0.29}
\definecolor{grey2}{rgb}{0.02,0.02,0.02}
\definecolor{grey30}{rgb}{0.30,0.30,0.30}
\definecolor{grey31}{rgb}{0.31,0.31,0.31}
\definecolor{grey32}{rgb}{0.32,0.32,0.32}
\definecolor{grey33}{rgb}{0.33,0.33,0.33}
\definecolor{grey34}{rgb}{0.34,0.34,0.34}
\definecolor{grey35}{rgb}{0.35,0.35,0.35}
\definecolor{grey36}{rgb}{0.36,0.36,0.36}
\definecolor{grey37}{rgb}{0.37,0.37,0.37}
\definecolor{grey38}{rgb}{0.38,0.38,0.38}
\definecolor{grey39}{rgb}{0.39,0.39,0.39}
\definecolor{grey3}{rgb}{0.03,0.03,0.03}
\definecolor{grey40}{rgb}{0.40,0.40,0.40}
\definecolor{grey41}{rgb}{0.41,0.41,0.41}
\definecolor{grey42}{rgb}{0.42,0.42,0.42}
\definecolor{grey43}{rgb}{0.43,0.43,0.43}
\definecolor{grey44}{rgb}{0.44,0.44,0.44}
\definecolor{grey45}{rgb}{0.45,0.45,0.45}
\definecolor{grey46}{rgb}{0.46,0.46,0.46}
\definecolor{grey47}{rgb}{0.47,0.47,0.47}
\definecolor{grey48}{rgb}{0.48,0.48,0.48}
\definecolor{grey49}{rgb}{0.49,0.49,0.49}
\definecolor{grey4}{rgb}{0.04,0.04,0.04}
\definecolor{grey50}{rgb}{0.50,0.50,0.50}
\definecolor{grey51}{rgb}{0.51,0.51,0.51}
\definecolor{grey52}{rgb}{0.52,0.52,0.52}
\definecolor{grey53}{rgb}{0.53,0.53,0.53}
\definecolor{grey54}{rgb}{0.54,0.54,0.54}
\definecolor{grey55}{rgb}{0.55,0.55,0.55}
\definecolor{grey56}{rgb}{0.56,0.56,0.56}
\definecolor{grey57}{rgb}{0.57,0.57,0.57}
\definecolor{grey58}{rgb}{0.58,0.58,0.58}
\definecolor{grey59}{rgb}{0.59,0.59,0.59}
\definecolor{grey5}{rgb}{0.05,0.05,0.05}
\definecolor{grey60}{rgb}{0.60,0.60,0.60}
\definecolor{grey61}{rgb}{0.61,0.61,0.61}
\definecolor{grey62}{rgb}{0.62,0.62,0.62}
\definecolor{grey63}{rgb}{0.63,0.63,0.63}
\definecolor{grey64}{rgb}{0.64,0.64,0.64}
\definecolor{grey65}{rgb}{0.65,0.65,0.65}
\definecolor{grey66}{rgb}{0.66,0.66,0.66}
\definecolor{grey67}{rgb}{0.67,0.67,0.67}
\definecolor{grey68}{rgb}{0.68,0.68,0.68}
\definecolor{grey69}{rgb}{0.69,0.69,0.69}
\definecolor{grey6}{rgb}{0.06,0.06,0.06}
\definecolor{grey70}{rgb}{0.70,0.70,0.70}
\definecolor{grey71}{rgb}{0.71,0.71,0.71}
\definecolor{grey72}{rgb}{0.72,0.72,0.72}
\definecolor{grey73}{rgb}{0.73,0.73,0.73}
\definecolor{grey74}{rgb}{0.74,0.74,0.74}
\definecolor{grey75}{rgb}{0.75,0.75,0.75}
\definecolor{grey76}{rgb}{0.76,0.76,0.76}
\definecolor{grey77}{rgb}{0.77,0.77,0.77}
\definecolor{grey78}{rgb}{0.78,0.78,0.78}
\definecolor{grey79}{rgb}{0.79,0.79,0.79}
\definecolor{grey7}{rgb}{0.07,0.07,0.07}
\definecolor{grey80}{rgb}{0.80,0.80,0.80}
\definecolor{grey81}{rgb}{0.81,0.81,0.81}
\definecolor{grey82}{rgb}{0.82,0.82,0.82}
\definecolor{grey83}{rgb}{0.83,0.83,0.83}
\definecolor{grey84}{rgb}{0.84,0.84,0.84}
\definecolor{grey85}{rgb}{0.85,0.85,0.85}
\definecolor{grey86}{rgb}{0.86,0.86,0.86}
\definecolor{grey87}{rgb}{0.87,0.87,0.87}
\definecolor{grey88}{rgb}{0.88,0.88,0.88}
\definecolor{grey89}{rgb}{0.89,0.89,0.89}
\definecolor{grey8}{rgb}{0.08,0.08,0.08}
\definecolor{grey90}{rgb}{0.90,0.90,0.90}
\definecolor{grey91}{rgb}{0.91,0.91,0.91}
\definecolor{grey92}{rgb}{0.92,0.92,0.92}
\definecolor{grey93}{rgb}{0.93,0.93,0.93}
\definecolor{grey94}{rgb}{0.94,0.94,0.94}
\definecolor{grey95}{rgb}{0.95,0.95,0.95}
\definecolor{grey96}{rgb}{0.96,0.96,0.96}
\definecolor{grey97}{rgb}{0.97,0.97,0.97}
\definecolor{grey98}{rgb}{0.98,0.98,0.98}
\definecolor{grey99}{rgb}{0.99,0.99,0.99}
\definecolor{grey9}{rgb}{0.09,0.09,0.09}
\definecolor{grey}{rgb}{0.75,0.75,0.75}
\definecolor{honeydew1}{rgb}{0.94,1.00,0.94}
\definecolor{honeydew2}{rgb}{0.88,0.93,0.88}
\definecolor{honeydew3}{rgb}{0.76,0.80,0.76}
\definecolor{honeydew4}{rgb}{0.51,0.55,0.51}
\definecolor{honeydew}{rgb}{0.94,1.00,0.94}
\definecolor{hotpink}{rgb}{1.00,0.41,0.71}
\definecolor{indianred}{rgb}{0.80,0.36,0.36}
\definecolor{ivory1}{rgb}{1.00,1.00,0.94}
\definecolor{ivory2}{rgb}{0.93,0.93,0.88}
\definecolor{ivory3}{rgb}{0.80,0.80,0.76}
\definecolor{ivory4}{rgb}{0.55,0.55,0.51}
\definecolor{ivory}{rgb}{1.00,1.00,0.94}
\definecolor{khaki1}{rgb}{1.00,0.96,0.56}
\definecolor{khaki2}{rgb}{0.93,0.90,0.52}
\definecolor{khaki3}{rgb}{0.80,0.78,0.45}
\definecolor{khaki4}{rgb}{0.55,0.53,0.31}
\definecolor{khaki}{rgb}{0.94,0.90,0.55}
\definecolor{lavenderblush}{rgb}{1.00,0.94,0.96}
\definecolor{lavender}{rgb}{0.90,0.90,0.98}
\definecolor{lawngreen}{rgb}{0.49,0.99,0.00}
\definecolor{lemonchiffon}{rgb}{1.00,0.98,0.80}
\definecolor{lightblue}{rgb}{0.68,0.85,0.90}
\definecolor{lightcoral}{rgb}{0.94,0.50,0.50}
\definecolor{lightcyan}{rgb}{0.88,1.00,1.00}
\definecolor{lightgoldenrod}{rgb}{0.93,0.87,0.51}
\definecolor{lightgoldenrod}{rgb}{0.98,0.98,0.82}
\definecolor{lightgray}{rgb}{0.83,0.83,0.83}
\definecolor{lightgreen}{rgb}{0.56,0.93,0.56}
\definecolor{lightgrey}{rgb}{0.83,0.83,0.83}
\definecolor{lightpink}{rgb}{1.00,0.71,0.76}
\definecolor{lightsalmon}{rgb}{1.00,0.63,0.48}
\definecolor{lightsea}{rgb}{0.13,0.70,0.67}
\definecolor{lightsky}{rgb}{0.53,0.81,0.98}
\definecolor{lightslate}{rgb}{0.47,0.53,0.60}
\definecolor{lightslate}{rgb}{0.47,0.53,0.60}
\definecolor{lightslate}{rgb}{0.52,0.44,1.00}
\definecolor{lightsteel}{rgb}{0.69,0.77,0.87}
\definecolor{lightyellow}{rgb}{1.00,1.00,0.88}
\definecolor{limegreen}{rgb}{0.20,0.80,0.20}
\definecolor{linen}{rgb}{0.98,0.94,0.90}
\definecolor{magenta1}{rgb}{1.00,0.00,1.00}
\definecolor{magenta2}{rgb}{0.93,0.00,0.93}
\definecolor{magenta3}{rgb}{0.80,0.00,0.80}
\definecolor{magenta4}{rgb}{0.55,0.00,0.55}
\definecolor{magenta}{rgb}{1.00,0.00,1.00}
\definecolor{maroon1}{rgb}{1.00,0.20,0.70}
\definecolor{maroon2}{rgb}{0.93,0.19,0.65}
\definecolor{maroon3}{rgb}{0.80,0.16,0.56}
\definecolor{maroon4}{rgb}{0.55,0.11,0.38}
\definecolor{maroon}{rgb}{0.69,0.19,0.38}
\definecolor{mediumaquamarine}{rgb}{0.40,0.80,0.67}
\definecolor{mediumblue}{rgb}{0.00,0.00,0.80}
\definecolor{mediumorchid}{rgb}{0.73,0.33,0.83}
\definecolor{mediumpurple}{rgb}{0.58,0.44,0.86}
\definecolor{mediumsea}{rgb}{0.24,0.70,0.44}
\definecolor{mediumslate}{rgb}{0.48,0.41,0.93}
\definecolor{mediumspring}{rgb}{0.00,0.98,0.60}
\definecolor{mediumturquoise}{rgb}{0.28,0.82,0.80}
\definecolor{mediumviolet}{rgb}{0.78,0.08,0.52}
\definecolor{midnightblue}{rgb}{0.10,0.10,0.44}
\definecolor{mintcream}{rgb}{0.96,1.00,0.98}
\definecolor{mistyrose}{rgb}{1.00,0.89,0.88}
\definecolor{moccasin}{rgb}{1.00,0.89,0.71}
\definecolor{navajowhite}{rgb}{1.00,0.87,0.68}
\definecolor{navyblue}{rgb}{0.00,0.00,0.50}
\definecolor{navy}{rgb}{0.00,0.00,0.50}
\definecolor{oldlace}{rgb}{0.99,0.96,0.90}
\definecolor{olivedrab}{rgb}{0.42,0.56,0.14}
\definecolor{orange1}{rgb}{1.00,0.65,0.00}
\definecolor{orange2}{rgb}{0.93,0.60,0.00}
\definecolor{orange3}{rgb}{0.80,0.52,0.00}
\definecolor{orange4}{rgb}{0.55,0.35,0.00}
\definecolor{orangered}{rgb}{1.00,0.27,0.00}
\definecolor{orange}{rgb}{1.00,0.65,0.00}
\definecolor{orchid1}{rgb}{1.00,0.51,0.98}
\definecolor{orchid2}{rgb}{0.93,0.48,0.91}
\definecolor{orchid3}{rgb}{0.80,0.41,0.79}
\definecolor{orchid4}{rgb}{0.55,0.28,0.54}
\definecolor{orchid}{rgb}{0.85,0.44,0.84}
\definecolor{palegoldenrod}{rgb}{0.93,0.91,0.67}
\definecolor{palegreen}{rgb}{0.60,0.98,0.60}
\definecolor{paleturquoise}{rgb}{0.69,0.93,0.93}
\definecolor{paleviolet}{rgb}{0.86,0.44,0.58}
\definecolor{papayawhip}{rgb}{1.00,0.94,0.84}
\definecolor{peachpuff}{rgb}{1.00,0.85,0.73}
\definecolor{peru}{rgb}{0.80,0.52,0.25}
\definecolor{pink1}{rgb}{1.00,0.71,0.77}
\definecolor{pink2}{rgb}{0.93,0.66,0.72}
\definecolor{pink3}{rgb}{0.80,0.57,0.62}
\definecolor{pink4}{rgb}{0.55,0.39,0.42}
\definecolor{pink}{rgb}{1.00,0.75,0.80}
\definecolor{plum1}{rgb}{1.00,0.73,1.00}
\definecolor{plum2}{rgb}{0.93,0.68,0.93}
\definecolor{plum3}{rgb}{0.80,0.59,0.80}
\definecolor{plum4}{rgb}{0.55,0.40,0.55}
\definecolor{plum}{rgb}{0.87,0.63,0.87}
\definecolor{powderblue}{rgb}{0.69,0.88,0.90}
\definecolor{purple1}{rgb}{0.61,0.19,1.00}
\definecolor{purple2}{rgb}{0.57,0.17,0.93}
\definecolor{purple3}{rgb}{0.49,0.15,0.80}
\definecolor{purple4}{rgb}{0.33,0.10,0.55}
\definecolor{purple}{rgb}{0.63,0.13,0.94}
\definecolor{red1}{rgb}{1.00,0.00,0.00}
\definecolor{red2}{rgb}{0.93,0.00,0.00}
\definecolor{red3}{rgb}{0.80,0.00,0.00}
\definecolor{red4}{rgb}{0.55,0.00,0.00}
\definecolor{red}{rgb}{1.00,0.00,0.00}
\definecolor{rosybrown}{rgb}{0.74,0.56,0.56}
\definecolor{royalblue}{rgb}{0.25,0.41,0.88}
\definecolor{saddlebrown}{rgb}{0.55,0.27,0.07}
\definecolor{salmon1}{rgb}{1.00,0.55,0.41}
\definecolor{salmon2}{rgb}{0.93,0.51,0.38}
\definecolor{salmon3}{rgb}{0.80,0.44,0.33}
\definecolor{salmon4}{rgb}{0.55,0.30,0.22}
\definecolor{salmon}{rgb}{0.98,0.50,0.45}
\definecolor{sandybrown}{rgb}{0.96,0.64,0.38}
\definecolor{seagreen}{rgb}{0.18,0.55,0.34}
\definecolor{seashell1}{rgb}{1.00,0.96,0.93}
\definecolor{seashell2}{rgb}{0.93,0.90,0.87}
\definecolor{seashell3}{rgb}{0.80,0.77,0.75}
\definecolor{seashell4}{rgb}{0.55,0.53,0.51}
\definecolor{seashell}{rgb}{1.00,0.96,0.93}
\definecolor{sienna1}{rgb}{1.00,0.51,0.28}
\definecolor{sienna2}{rgb}{0.93,0.47,0.26}
\definecolor{sienna3}{rgb}{0.80,0.41,0.22}
\definecolor{sienna4}{rgb}{0.55,0.28,0.15}
\definecolor{sienna}{rgb}{0.63,0.32,0.18}
\definecolor{skyblue}{rgb}{0.53,0.81,0.92}
\definecolor{slateblue}{rgb}{0.42,0.35,0.80}
\definecolor{slategray}{rgb}{0.44,0.50,0.56}
\definecolor{slategrey}{rgb}{0.44,0.50,0.56}
\definecolor{snow1}{rgb}{1.00,0.98,0.98}
\definecolor{snow2}{rgb}{0.93,0.91,0.91}
\definecolor{snow3}{rgb}{0.80,0.79,0.79}
\definecolor{snow4}{rgb}{0.55,0.54,0.54}
\definecolor{snow}{rgb}{1.00,0.98,0.98}
\definecolor{springgreen}{rgb}{0.00,1.00,0.50}
\definecolor{steelblue}{rgb}{0.27,0.51,0.71}
\definecolor{tan1}{rgb}{1.00,0.65,0.31}
\definecolor{tan2}{rgb}{0.93,0.60,0.29}
\definecolor{tan3}{rgb}{0.80,0.52,0.25}
\definecolor{tan4}{rgb}{0.55,0.35,0.17}
\definecolor{tan}{rgb}{0.82,0.71,0.55}
\definecolor{thistle1}{rgb}{1.00,0.88,1.00}
\definecolor{thistle2}{rgb}{0.93,0.82,0.93}
\definecolor{thistle3}{rgb}{0.80,0.71,0.80}
\definecolor{thistle4}{rgb}{0.55,0.48,0.55}
\definecolor{thistle}{rgb}{0.85,0.75,0.85}
\definecolor{tomato1}{rgb}{1.00,0.39,0.28}
\definecolor{tomato2}{rgb}{0.93,0.36,0.26}
\definecolor{tomato3}{rgb}{0.80,0.31,0.22}
\definecolor{tomato4}{rgb}{0.55,0.21,0.15}
\definecolor{tomato}{rgb}{1.00,0.39,0.28}
\definecolor{turquoise1}{rgb}{0.00,0.96,1.00}
\definecolor{turquoise2}{rgb}{0.00,0.90,0.93}
\definecolor{turquoise3}{rgb}{0.00,0.77,0.80}
\definecolor{turquoise4}{rgb}{0.00,0.53,0.55}
\definecolor{turquoise}{rgb}{0.25,0.88,0.82}
\definecolor{violetred}{rgb}{0.82,0.13,0.56}
\definecolor{violet}{rgb}{0.93,0.51,0.93}
\definecolor{wheat1}{rgb}{1.00,0.91,0.73}
\definecolor{wheat2}{rgb}{0.93,0.85,0.68}
\definecolor{wheat3}{rgb}{0.80,0.73,0.59}
\definecolor{wheat4}{rgb}{0.55,0.49,0.40}
\definecolor{wheat}{rgb}{0.96,0.87,0.70}
\definecolor{whitesmoke}{rgb}{0.96,0.96,0.96}
\definecolor{white}{rgb}{1.00,1.00,1.00}
\definecolor{yellow1}{rgb}{1.00,1.00,0.00}
\definecolor{yellow2}{rgb}{0.93,0.93,0.00}
\definecolor{yellow3}{rgb}{0.80,0.80,0.00}
\definecolor{yellow4}{rgb}{0.55,0.55,0.00}
\definecolor{yellowgreen}{rgb}{0.60,0.80,0.20}
\definecolor{yellow}{rgb}{1.00,1.00,0.00}